%% file: paper.tex
\documentclass[format=acmsmall,manuscript,screen]{acmart}
\pdfoutput=1

\acmJournal{TOMS}
\setcopyright{none}

\newif\ifarxiv
\arxivtrue

\ifarxiv
\acmDOI{}
\acmPrice{}
\acmArticle{}
\setkeys{acmart.cls}{nonacm=true}
\fi

\usepackage[utf8]{inputenc}
\usepackage{array}
\usepackage{algorithm}
\usepackage{algpseudocode}
\usepackage{amsmath}
\usepackage{amssymb}
\usepackage{booktabs}
\usepackage[english]{babel}
\usepackage{enumitem}
\usepackage{hyperref}
\usepackage{cleveref}
\usepackage{tikz}
\usetikzlibrary{calc}
\usepackage{pgfplots}
\usepackage{pgfplotstable}
\usepackage{verbatim}
\usepackage{subcaption}
\usepackage{xspace}
\newcommand{\tsfc}{\texttt{tsfc}\xspace}
\newcommand{\jump}[1]{\ensuremath[\![#1]\!]}
\newcommand{\avg}[1]{\ensuremath\{#1\}}
\newcommand{\dx}{\ensuremath\,\text{d}x}
\newcommand{\ds}{\ensuremath\,\text{d}s}
\usetikzlibrary{patterns}

\pgfplotstableset{create on use/Build/.style={create
    col/expr={\thisrow{SNESFunctionEval} + \thisrow{SNESJacobianEval}}}}
\usepackage{listings}
\definecolor{DarkBlue}{rgb}{0.00,0.00,0.55}
\definecolor{DarkRed}{rgb}{0.55,0.00,0.00}
\definecolor{DarkGreen}{rgb}{0.00,0.55,0.00}
\definecolor{Bittersweet}{rgb}{1.0, 0.44, 0.37}
\definecolor{Purple}{rgb}{0.5, 0.0, 0.5}

\graphicspath{{code/pictures/}}

\title{Code generation for generally mapped finite elements}

\author{Robert C. Kirby}
\affiliation{%
  \institution{Baylor University}
  \department{Department of Mathematics}
  \streetaddress{One Bear Place}
  \city{Waco}
  \state{TX}
  \country{USA}}
\email{robert_kirby@baylor.edu}

\author{Lawrence Mitchell}
\affiliation{%
  \institution{Durham University}
  \department{Department of Computer Science}
  \streetaddress{Lower Mountjoy}
  \city{Durham}
  \postcode{DH1 3LE}
  \country{UK}}
\email{lawrence.mitchell@durham.ac.uk}

\numberwithin{equation}{section}
\crefname{algorithm}{Algorithm}{Algorithms}
\crefname{figure}{Fig.}{Figs.}
\crefname{table}{Table}{Tables}

\begin{abstract}
Many classical finite elements such as the Argyris and Bell elements
have long been absent from high-level PDE
software.  Building on recent theoretical work, we describe how
to implement very general finite element transformations in FInAT and
hence into the Firedrake finite element system.  Numerical results evaluate the
new elements, comparing them to existing methods for classical
problems.  For a second order model problem, we find that new elements
give smooth solutions at a mild increase in cost over standard
Lagrange elements.  For fourth-order problems, however, the
newly-enabled methods significantly outperform interior penalty
formulations.  We also give some advanced use cases, solving the
nonlinear Cahn-Hilliard equation and some
biharmonic eigenvalue problems (including Chladni plates) using $C^1$
discretizations.  
\end{abstract}


\begin{document}
\maketitle

\section{Introduction}
The FIAT\footnote{The FInite element Automatic Tabulator} project~\citep{Kirby:2004} has provided a new generation of finite
element codes such as FEniCS~\citep{Logg:2012} and Firedrake~\citep{Rathgeber:2016} with a
diverse set of finite element basis functions.  Internally, FIAT
represents the nodal bases for finite elements in terms of orthogonal
polynomials and dense linear algebra, and it is capable of tabulating
reference basis functions and their derivatives at any desired points.  Wrapped behind
a high-level declarative interface, many elements generally regarded as
difficult to implement can be rapidly deployed in a flexible,
performant environment.  In the case of FEniCS and Firedrake, FIAT's
usage is hidden behind a \emph{form compiler}~\citep{Kirby2006,homolya2018tsfc} that translates
UFL~\citep{Alnaes:2014}, a domain-specific language for variational forms, into
efficient lower-level code for constructing matrices and vectors.  As
an example of how this works, we refer to the code listing in \cref{fig:listing}.
\begin{figure}[htbp]
  \centering
  \begin{lstlisting}
    from firedrake import *
    mesh = UnitSquareMesh(10, 10)
    V = FunctionSpace(mesh, "CG", 3)
    u = TrialFunction(V)
    v = TestFunction(V)
    f = ...
    a = (dot(grad(v), grad(u)) + v * u) * dx
    L = f * v * dx
    u0 = Function(V)
    solve(a == L, u0)
\end{lstlisting}
\caption{Sample code listing for a model problem in
  Firedrake using cubic Lagrange basis functions.}
\label{fig:listing}
\end{figure}

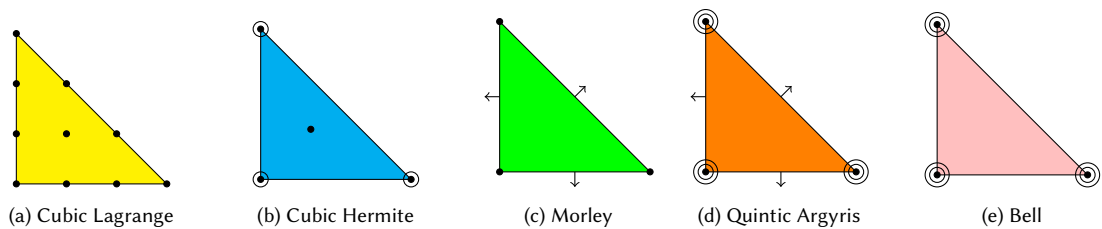
\begin{figure}[htbp]
  \begin{subfigure}[t]{0.2\textwidth}
    \centering
    \begin{tikzpicture}[scale=2.0] % lagrange
    \draw[fill=yellow] (0,0) -- (1, 0) -- (0, 1) -- cycle;
    \foreach \i in {0, 1, 2, 3} {
      \foreach \j in {0,...,\i}{
      \draw[fill=black] (1-\i/3, \j/3) circle (0.02);
      }
      }
    \end{tikzpicture}
  \caption{Cubic Lagrange}
  \label{lag3}
  \end{subfigure}
  \hfill
  \begin{subfigure}[t]{0.18\textwidth}
    \centering
  \begin{tikzpicture}[scale=2.0] % hermite
    \draw[fill=cyan] (0,0) -- (1, 0) -- (0, 1) -- cycle;
    \foreach \i/\j in {0/0, 1/0, 0/1}{
      \draw[fill=black] (\i, \j) circle (0.02);
      \draw (\i, \j) circle (0.05);
    }
    \draw[fill=black] (1/3, 1/3) circle (0.02);
  \end{tikzpicture}
  \caption{Cubic Hermite}
  \label{herm3}
  \end{subfigure}
  \hfill
  \begin{subfigure}[t]{0.18\textwidth} % Morley
    \centering
    \begin{tikzpicture}[scale=2.0]
    \draw[fill=green] (0,0) -- (1, 0) -- (0, 1) -- cycle;
    \foreach \i/\j in {0/0, 1/0, 0/1}{
      \draw[fill=black] (\i, \j) circle (0.02);
    }
    \foreach \i/\j/\n/\t in {0.5/0.0/0.0/-1, 0.5/0.5/0.707/0.707, 0.0/0.5/-1/0}{
      \draw[->] (\i, \j) -- (\i+\n/10, \j+\t/10);
    }
    \end{tikzpicture}
    \caption{Morley}
  \label{morley}
    \end{subfigure}
  \begin{subfigure}[t]{0.18\textwidth}
    \centering
  \begin{tikzpicture}[scale=2.0] % Argyris
    \draw[fill=orange] (0,0) -- (1, 0) -- (0, 1) -- cycle;
    \foreach \i/\j in {0/0, 1/0, 0/1}{
      \draw[fill=black] (\i, \j) circle (0.02);
      \draw (\i, \j) circle (0.05);
      \draw (\i, \j) circle (0.08);
    }
    \foreach \i/\j/\n/\t in {0.5/0.0/0.0/-1, 0.5/0.5/0.707/0.707, 0.0/0.5/-1/0}{
      \draw[->] (\i, \j) -- (\i+\n/10, \j+\t/10);
      }
  \end{tikzpicture}
  \caption{Quintic Argyris}
  \label{arg}
  \end{subfigure}
  \hfill
  \begin{subfigure}[t]{0.18\textwidth}
    \centering
  \begin{tikzpicture}[scale=2.0] % Bell
    \draw[fill=pink] (0,0) -- (1, 0) -- (0, 1) -- cycle;
    \foreach \i/\j in {0/0, 1/0, 0/1}{
      \draw[fill=black] (\i, \j) circle (0.02);
      \draw (\i, \j) circle (0.05);
      \draw (\i, \j) circle (0.08);
    }
  \end{tikzpicture}
  \caption{Bell}
  \label{bell}
  \end{subfigure}
  \caption{Some famous triangular elements.  Solid dots represent
    point value degrees of freedom, smaller circles represent
    gradients, and larger circles represent the collection of second
    derivatives.  The arrows indicate directional derivatives
    evaluated at the tail of the arrow.}
    \label{fig:els}
\end{figure}

However, time and usage have revealed certain limitations of FIAT.
Its Python implementation makes it most suitable to ``run once'' in a
given simulation, tabulating basis functions at reference element
quadrature points and then letting the rest of the finite element code
map these to each cell in the mesh.  Moreover, the dense tables of
arrays that FIAT produces make it difficult to exploit tensor-product
or other structure that might
lead to more efficient algorithms.  As a third issue, it is difficult
for FIAT to communicate how bases should be mapped, leaving clients to
handle their own pullbacks.  This works naturally when reference and
physical elements are equivalent under affine or Piola pullback,
but implementing elements (see \cref{fig:els}) such as
Hermite~\citep{ciarlet1972general}, Morley~\citep{morley1971constant},
Bell~\citep{bell1969refined}, and Argyris~\citep{argyris1968tuba} that
require more complex
transformations~\citep{dominguez2008algorithm, kirby-zany} easily leads
to a proliferation of special
cases.  It also requires encoding basis tabulation and
transformation in completely different code components.
Just as the dense tabulation historically limited the ability of
FEniCS and Firedrake to employ sum-factorization and other fast
algorithms, lack of general transformations has limited their ability
to address higher-order PDEs and other applications benefitting from
smooth approximations.

Recent work on FInAT~\citep{Homolya2017finat} has addressed many of
these issues. Unlike FIAT, FInAT is not a tabulator. Instead, it
constructs abstract syntax using GEM~\citep{homolya2018tsfc} for basis
function tabulation. By emitting GEM, which is the tensor-based
intermediate representation of the \tsfc form compiler, FInAT provides
symbolic, as well as numerical, tabulations. For example, as well as
including syntax for table lookup from FIAT-generated values, FInAT
also provides a kind of element calculus by which one may construct
vector- or tensor-valued elements; use tensor-products of
lower-dimensional bases to build structured bases on quadrilaterals,
prisms, or hexahedra; or indicate that basis evaluation at particular
points satisfies a Kronecker delta property. In
\citet{Homolya2017finat}, we describe these features of FInAT and
extensions of \tsfc to make use of them in generating efficient code.
Our goal is for FInAT to become a ``single source of truth'' for
finite element bases. This includes not only basis function
evaluation, like FIAT, but also structural and algorithmic
considerations as well as reference element transformations.

In this work, we present further developments in FInAT to enable the
transformations of Argyris and other such elements described
in \citet{kirby-zany}.  In particular, we equip FInAT with abstract syntax to
provide basis transformations.  FInAT's routines for basis evaluation
now assume that the caller will provide an object capable of providing
(abstract syntax for) the required geometric quantities such as Jacobians,
normals, and tangents required by the mapping.  By extending \tsfc to
provide these, we have been able to deploy these extensions within
Firedrake, making the utilization of such bases (nearly) seamless from
the user perspective.  Essentially, we enable users to replace the line
\lstinline[basicstyle=\normalsize\ttfamily]{V = FunctionSpace(mesh, "CG", 3)}
in \cref{fig:listing} with \lstinline[basicstyle=\normalsize\ttfamily]{V = FunctionSpace(mesh, "Hermite", 3)}
or any of the other elements such as those depicted in \cref{fig:els}.

Although our examples are two-dimensional, our work extends in
principle (and code infrastructure supports) tetrahedral elements as
well. The code already has Hermite
tetrahedra~\cite{ciarlet1972general}, and implementing Morley-like
elements~\cite{wang2013minimal} would be straightforward.
Three-dimensional $C^1$ analogs of the Argyris triangle, however,
require very high polynomial degree. This high degree and the
associated high cost of direct methods limits their practical use. An
alternative is to use numerical approaches such as isogeometric
analysis~\cite{hughes2005isogeometric} or maximum-entropy methods
\cite{arroyo2006local} which naturally offer high order continuity,
potentially at reduced cost. Of these methods, simplicial splines
\cite{LaiSch07} are of future interest.  Such splines would require additional 
infrastructure in FInAT to reason about macro elements, but would
bring Firedrake and other FInAT clients a step closer to the
flexibility to change order and continuity available on (patches of)
mapped logically uniform meshes in isogeometric analysis.

Among existing high-level PDE codes like FEniCS, Firedrake,
deal.II~\citep{Bangerth:2007}, Sundance~\citep{Long:2003}, and
Feel++~\citep{prud2012feel++} we have only found a general support for
the kinds of elements we consider here in
GetFEM++~\citep{fournie2010cfd,perduta2013enhancing}, which implements
the Morley, and Argyris elements on triangles, and the Hermite element
on triangles and tetrahedra.  Except for
FEniCS and Firedrake, these codes are all C++ libraries with varying
degrees of abstraction in the problem description.  While there is no
impediment in such libraries to implementing the transformation theory
described in \citet{kirby-zany}, we believe UFL affords a more
succinct user interface, and code generation more opportunities for
optimization~\citep{kirby2005optimizing,kirby2006topological,Luporini2017,Homolya2017finat},

In the rest of the paper, we first review the theory developed
in \citet{kirby-zany} in \cref{sec:transformtheory}.
In \cref{sec:code} we describe the modifications to
FInAT and the rest of the Firedrake code stack that are necessary to
implement and automate the generalized mappings for these new elements.
Strongly-enforced boundary conditions are the one limitation we face
when implementing these elements.  In \cref{sec:bcbad}, we
describe some of the mathematical issues that have to date prevented a
general approach, so (like GetFEM++) we defer to weak enforcement of
essential boundary conditions through Nitsche's method~\citep{Nitsche}.
To test our
implementation and the practical use of these elements, we present a
series of numerical experiments in \cref{sec:ex}.  These include an
attempt to quantify the practical effect in terms of the cost of our general
transformations.

\section{Overview of transformation theory}
\label{sec:transformtheory}
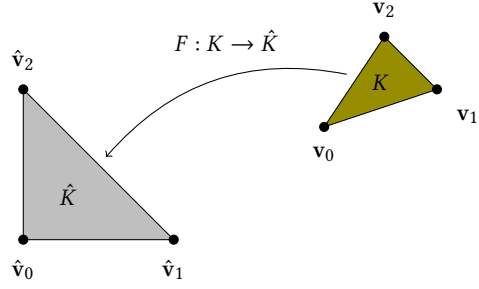
\begin{figure}[htbp]
  \centering
  \begin{tikzpicture}
    \draw[fill=lightgray] (0,0) coordinate (vhat1)
    -- (2,0) coordinate(vhat2)
    -- (0,2) coordinate(vhat3)--cycle;
    \foreach \pt\labpos\lab in {vhat1/below/\hat{\mathbf{v}}_0, vhat2/below/\hat{\mathbf{v}}_1, vhat3/above/\hat{\mathbf{v}}_2}{
      \filldraw (\pt) circle(.6mm) node[\labpos=1.5mm, fill=white]{$\lab$};
    }
    \draw[fill=olive] (4.0, 1.5) coordinate (v1)
    -- (5.5, 2.0) coordinate (v2)
    -- (4.8, 2.7) coordinate (v3) -- cycle;
    \foreach \pt\labpos\lab in {v1/below/\mathbf{v}_0, v2/below right/\mathbf{v}
_1, v3/above/\mathbf{v}_2}{
      \filldraw (\pt) circle(.6mm) node[\labpos=1.5mm, fill=white]{$\lab$};
    }
    \draw[<-] (1.1, 1.1) to[bend left] (4.3, 2.2);
    \node at (2.7, 2.65) {$F:K\rightarrow\hat{K}$};
    \node at (0.6,0.6) {$\hat{K}$};
    \node at (4.75, 2.1) {$K$};
  \end{tikzpicture}
    \caption{Affine mapping to a reference cell \(\hat{K}\) from a
    typical cell \( K \).}
    \label{fig:affmap}
\end{figure}

Let $\hat{K}$ be a reference domain with vertices
$\{\hat{\mathbf{v}}_i\}_{i=0}^{d}$ and let $K$ be a typical element
with vertices $\{\mathbf{v}_i\}_{i=0}^{d}$.  We let the affine
mapping $F: K \rightarrow \hat{K}$ be as shown in
\cref{fig:affmap}.  Given any function $\hat{\phi}$ defined on
$\hat{K}$, we define its \emph{pullback} by
\begin{equation}
  \label{eq:pullback}
  \phi = F^*(\hat{\phi}) = \hat{\phi} \circ F.
\end{equation}
To evaluate the new function $\phi$ at a point $\mathbf{x}$
that is the image of some $\hat{\mathbf{x}}$ under $F$, we have
\begin{equation}
  \label{eq:pullbackx}
  \phi(\mathbf{x}) = \hat{\phi}(\hat{\mathbf{x}}).
\end{equation}
Similarly, if $J$ is the Jacobian matrix of the mapping $F$, we can
differentiate $\phi$ by the chain rule
\begin{equation}
  \label{eq:chainrule}
  \nabla \phi = J^{T} \hat{\nabla} \hat{\phi},
\end{equation}
where $\nabla$ denotes differentiation with respect to the coordinate
system on $K$ and $\hat{\nabla}$ to that on $\hat{K}$.

We shall also need the \emph{push forward} of linear
functionals, which is defined simply by composition with the pullback.
If $P$ is a set of functions mapping $K$ into $\mathbb{R}$ (or some other vector space), then
\begin{equation}
  F_*(n) = n \circ F^* \quad \forall n \in P.
\end{equation}

Lagrange finite elements in simplicial domains form
\emph{affine-equivalent} families.  One can find the Lagrange
basis on $\hat{K}$ and then obtain the
Lagrange basis on any other domain $K$ by means of the affine pullback.
From \cref{eq:pullbackx} and \cref{eq:chainrule}, it is straightforward to tabulate the
basis functions and their derivatives at reference domain quadrature
points and map the results to any cell in the mesh.  A similar
property holds for vector-valued Raviart-Thomas~\citep{Raviart1977} and
N\'ed\'elec elements~\citep{nedelec1980mixed}, which can be defined on
a reference domain and mapped via Piola transforms.
However, other elements, such as scalar-valued elements with
derivative degrees of freedom, do not typically satisfy affine
equivalence, and this complicates the use of a reference element.

\begin{figure}[htbp]
  \begin{subfigure}[t]{0.4\textwidth}
    \centering
  \begin{tikzpicture}[scale=2.0] % hermite
    \draw[fill=cyan] (0,0) -- (1, 0) -- (0, 1) -- cycle;
    \foreach \i/\j in {0/0, 1/0, 0/1}{
      \draw[fill=black] (\i, \j) circle (0.02);
      \draw[->] (\i, \j) -- (\i, \j + 0.1);
      \draw[->] (\i, \j) -- (\i+0.1, \j);
    }
    \draw[fill=black] (1/3, 1/3) circle (0.02);
  \end{tikzpicture}
    \caption{Reference Hermite element}
  \label{refherm}
  \end{subfigure}
  \hfill
  \begin{subfigure}[t]{0.4\textwidth}
    \centering
      \begin{tikzpicture}[scale=2.0]
    \draw[fill=cyan] (0,0) -- (1.5, 0.5) -- (0.8, 1.2) -- cycle;
    \foreach \i/\j in {0/0, 1.5/0.5, 0.8/1.2}{
      \draw[fill=black] (\i, \j) circle (0.02);
      \draw[->] (\i, \j) -- (\i, \j + 0.1);
      \draw[->] (\i, \j) -- (\i+0.1, \j);
    }
    \draw[fill=black] (0.7667, 0.5667) circle (0.02);
    \end{tikzpicture}
        \caption{Physical Hermite element}
    \label{physherm}
  \end{subfigure}
  \caption{Reference and physical cubic Hermite elements with
    gradient degrees of freedom expressed in terms of local Cartesian
    directional derivatives.}
  \label{fig:hermrefandphys}
\end{figure}
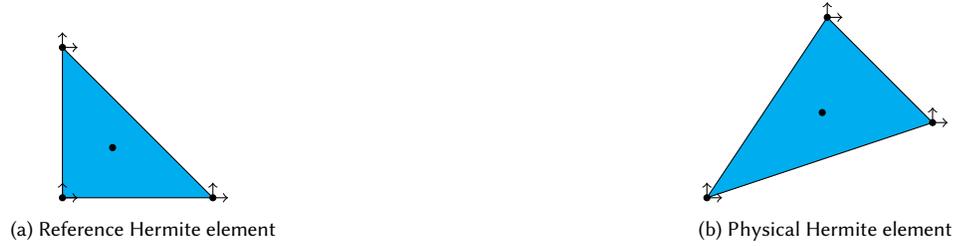

The cubic Hermite triangle~\citep{ciarlet1972general} provides a simple example.
The ten degrees of freedom parametrizing a Hermite triangle are the
function values and gradients at the vertices and the function value
at the barycenter.   We will specify the
gradient in terms of the partial derivatives in the local Cartesian
coordinate directions as shown in \cref{fig:hermrefandphys}.
Now, let $\hat{\psi}(\hat{\mathbf{x}})$ be the reference Hermite
polynomial having unit horizontal derivative
at $\hat{\mathbf{v}}_0$, with all other degrees of freedom vanishing.
The chain rule tells us that
\begin{equation}
\nabla \psi (\mathbf{v}_0) =
J^{T} \hat{\nabla} \hat{\psi}(\hat{\mathbf{v}}_0),
\end{equation}
which is not equal to $\left[\begin{smallmatrix} 1
    \\ 0 \end{smallmatrix} \right]$ except in very special geometry.
It is clear that $\psi(\mathbf{v}_i) = 0$ and that the
gradient vanishes at the other two vertices.

As we show in \citet{kirby-zany}, we can take a simple linear combination of
the pullbacks of the Hermite basis functions on $\hat{K}$ to obtain
the Hermite basis functions on $K$.  Let $\{\psi_{3i}\}_{i=0}^2$ be the
basis functions taking unit value at vertex $\mathbf{v}_i$, $\{
\psi_{3i+1}\}_{i=0}^2$ having unit $x$-derivative at $\mathbf{v}_i$, $\{
\psi_{3i+2}\}_{i=0}^2$ having unit $y$-derivative at $\mathbf{v}_i$,
and
$\psi_9$ having unit value at the barycenter and the corresponding
numbering of $\{\hat{\psi}_i\}_{i=0}^9$ on $\hat{K}$. Then, our theory
gives that
\begin{equation}
  \begin{bmatrix}
    \psi_0 \\ \psi_1 \\ \psi_2 \\ \psi_3 \\ \psi_4 \\ \psi_5 \\ \psi_6
    \\ \psi_7 \\ \psi_8 \\ \psi_9 \end{bmatrix}
  =
  \begin{bmatrix}
    1 & 0 & 0 & 0 & 0 & 0 & 0 & 0 & 0 & 0 \\
    0 & \tfrac{\partial \hat{x}}{\partial x}& \tfrac{\partial
      \hat{x}}{\partial y}& 0 & 0 & 0 & 0 & 0 & 0 & 0 \\
    0 & \tfrac{\partial \hat{y}}{\partial x} & \tfrac{\partial
      \hat{y}}{\partial y}& 0 & 0 & 0 & 0 & 0 & 0 & 0 \\
    0 & 0 & 0 & 1 & 0 & 0 & 0 & 0 & 0 & 0 \\
    0 & 0 & 0 & 0 & \tfrac{\partial \hat{x}}{\partial x} &  \tfrac{\partial
      \hat{x}}{\partial y}& 0 & 0 & 0 & 0 \\
    0 & 0 & 0 & 0 & \tfrac{\partial \hat{y}}{\partial x} & \tfrac{\partial
      \hat{y}}{\partial y} & 0 & 0 & 0 &0  \\
    0 & 0 & 0 & 0 & 0 & 0 &1 & 0& 0 & 0 \\
    0 & 0 & 0 & 0 & 0 & 0 & 0 & \tfrac{\partial \hat{x}}{\partial x} & \tfrac{\partial
      \hat{x}}{\partial y} & 0 \\
    0 & 0 & 0 & 0 &0 & 0 & 0 & \tfrac{\partial \hat{y}}{\partial x} & \tfrac{\partial
      \hat{y}}{\partial y} & 0 \\
    0 &0 & 0 & 0 & 0 & 0 & 0 & 0 & 0 & 1
  \end{bmatrix}
  \begin{bmatrix}
    F^*(\hat{\psi}_0) \\ F^*(\hat{\psi}_1) \\ F^*(\hat{\psi}_2) \\ F^*(\hat{\psi}_3) \\ F^*(\hat{\psi}_4) \\ F^*(\hat{\psi}_5) \\ F^*(\hat{\psi}_6)
    \\ F^*(\hat{\psi}_7) \\ F^*(\hat{\psi}_8)
    \\ F^*(\hat{\psi}_9) \end{bmatrix}.
  \label{eq:MforHerm}
\end{equation}
Note that the gradient basis-functions are mapped in pairs, while the
pull-back in fact maps the basis functions for point values correctly.

More generally, if $F^*$ maps not just Hermite but some other finite
element function space on $\hat{K}$ onto that on $K$, then there
exists a matrix $M$ such that
\begin{equation}
  \label{eq:M}
\psi_i = \sum_{j=1}^N M_{ij} F^*(\hat{\psi}_j).
\end{equation}
It is frequently much easier to construct $V=M^T$, which relates the
\emph{push-forward} of the physical finite element nodes to the reference element nodes.

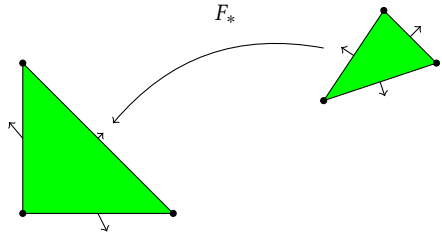
\begin{figure}[htbp]
  \centering
  \begin{tikzpicture}
    \draw[fill=green] (0,0) -- (2,0) -- (0,2) -- cycle;
    \foreach \x\y in {0/0, 2/0, 0/2}{
      \filldraw (\x,\y) circle(0.04);
    }
    \foreach \x\y\nx\ny in {1/1/.3536/.3536,0/1/-.9113/1.070,1/0/.6010/-1.1971}{
      \draw[->] (\x, \y) -- (\x+0.2*\nx, \y+0.2*\ny);
      }
    \draw[fill=green] (4.0, 1.5) -- (5.5, 2.0) -- (4.8, 2.7) -- cycle;
    \foreach \x\y in {4/1.5, 5.5/2, 4.8/2.7}{
      \filldraw (\x,\y) circle(.04);
    }
    \foreach \x\y\nx\ny in {5.15/2.35/.707/.707,4.4/2.1/-0.8321/0.5547,4.75/1.75/.3163/-.9487}{
      \draw[->] (\x,\y) -- (\x+0.2*\nx,\y+0.2*\ny);
      }
    \draw[<-] (1.2, 1.2) to[bend left] (4.0, 2.2);
    \node at (2.7, 2.65) {$F_*$};
    \end{tikzpicture}
    \caption{Pushing forward the Morley derivative nodes in physical
    space does \emph{not} produce the reference derivative nodes.}
\label{fig:morleypushforward}
\end{figure}

While the assumption behind \cref{eq:M} applies to Morley and Argyris
elements, the corresponding $M$ is more complicated because they do
not form affine-interpolation equivalent families of elements.
Equivalently \citep{Brenner:2008}, the spans of the reference nodes and
pushed-forward physical nodes do not coincide.  To see why this
presents difficulty, consider \cref{fig:morleypushforward},
which shows the physical nodes pushed forward to the reference
element.  Since there is only a single directional derivative on each
edge, pairs of basis function corresponding to a gradient (in some
coordinates) cannot be adjusted by the Jacobian like in the
Hermite case.

In \citet{kirby-zany}, we develop a three-step mapping technique that
generalizes the approach in \citet{dominguez2008algorithm}.  First,
one extends the reference and 
physical nodal sets with additional nodes such that their spans do
coincide.  The
transformation between these sets can be constructed in a method
similar to Hermite.  Second, the action of the new nodes on the finite
element space must be constructed in terms of the given ones.  This
can typically be done with interpolation theory.  Finally, one
extracts the nodes of the mapped finite element as a subset of the
enriched set.  Each stage can be expressed as a matrix, and we have
that
\begin{equation}
  V = E V^C D.
\end{equation}
Here, $D$, which has more rows than columns, maps the finite element nodes
to the extended nodes.  The square matrix $V^C$ relates the
push-forward of the extended  nodes in physical space to the extended
reference nodes.  $E$ has a single nonzero per row, selecting out the
subset of extended reference nodes belonging to the reference finite
element.

\begin{figure}[htbp]
    \centering
  \begin{tikzpicture}
    \draw[fill=green] (0,0) -- (2,0) -- (0,2) -- cycle;
    \foreach \x\y in {0/0, 2/0, 0/2}{
      \filldraw (\x,\y) circle(0.04);
    }
    \foreach \x\y\nx\ny\tx\ty in {1/1/.7071/.7071/-.7071/.7071,
      0/1/-1/0/0/1,
      1/0/0/-1/1/0}
    {
      \draw[->] (\x, \y) -- (\x+0.2*\nx, \y+0.2*\ny);
      \draw[->] (\x, \y) -- (\x+0.2*\tx, \y+0.2*\ty);
      }
    \draw[fill=green] (4.0, 1.5) -- (5.5, 2.0) -- (4.8, 2.7) -- cycle;
    \foreach \x\y in {4/1.5, 5.5/2, 4.8/2.7}{
      \filldraw (\x,\y) circle(.04);
    }
    \foreach \x\y\nx\ny\tx\ty in {5.15/2.35/.707/.707/-.707/.707,4.4/2.1/-0.8321/0.5547/-.5547/-.8321,4.75/1.75/.3163/-.9487/.9487/.3163}{
      \draw[->] (\x,\y) -- (\x+0.2*\nx,\y+0.2*\ny);
      \draw[->] (\x,\y) -- (\x+0.2*\tx,\y+0.2*\ty);
      }
    %% \draw[<-] (1.2, 1.2) to[bend left] (4.0, 2.2);
    %% \node at (2.7, 2.65) {$F_*$};
    \end{tikzpicture}
    \caption{Extended nodal sets for Morley
    reference (left) and typical (right) element are formed by
    including tangential derivatives along with normal derivatives at
    each edge midpoint.}
\label{fig:morleybridge}
\end{figure}
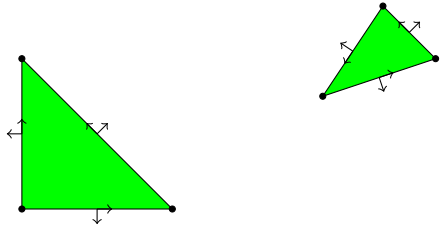

In the case of the Morley element, we extend the vertex values and
normal derivatives at edge midpoints with the tangential derivatives
at the same, as shown in \cref{fig:morleybridge}.
The tangential derivative of a quadratic at the edge midpoint can
actually be computed by differencing the associated vertex values, so
the $D$ matrix is relatively easy to construct.  The matrix $V^C$ is
similar to Hermite, although it is slightly more involved since it
transforms gradients in normal and tangential rather than Cartesian
coordinates.  Let $\mathbf{n}_i$ and $\mathbf{t}_i$ denote the normal
and tangent to edge $i$ of $K$ and $\hat{\mathbf{n}}_i$ and
$\hat{\mathbf{t}}_i$ those for $\hat{K}$.  Then, we put
$G_i = \left[\begin{smallmatrix} \mathbf{n}_i & \mathbf{t}_i \end{smallmatrix}\right]^T$
with an analagous definition for $\hat{G}_i$ and hence define
\begin{equation}
  B^i = \hat{G}_i J^{-T} G_i^T.
\end{equation}
The matrix $V^C$ is logically block-diagonal with unit values
corresponding to the vertex degrees of freedom and $B^i$ for the
derivative nodes on edge $i$.  Putting this together, we have
\begin{equation}
  M^T = V =
  \begin{bmatrix}
    1 & 0 & 0 & 0 & 0 & 0 \\
    0 & 1 & 0 & 0 & 0 & 0 \\
    0 & 0 & 1 & 0 & 0 & 0 \\
    0 & \tfrac{-B_{01}^0}{\ell_0} & \tfrac{B_{01}^0}{\ell_0} & B_{00}^0 & 0 & 0 \\
    \tfrac{-B_{01}^1}{\ell_1} & 0 & \tfrac{B_{01}^1}{\ell_1} & 0 & B_{00}^1 & 0 \\
    \tfrac{-B_{01}^2}{\ell_2} & \tfrac{B_{01}^2}{\ell_2} & 0 & 0 & 0 & B_{00}^2
  \end{bmatrix}.
\end{equation}

Owing to the higher degree and second-order derivatives, the Argyris
triangle (\cref{arg}) transformation is more complicated, but
follows the same kind of process.  The Bell element
(\cref{bell}), however, requires an extension of this theory.  It is
billed as a ``reduced'' or ``simplified'' element because it lacks the
Argyris element's normal derivatives at edge midpoints.  However,
these are removed by constraining the function space to have cubic normal
derivatives rather than the quartic ones typical to a quintic
polynomial.  This means that the affine pullback does not preserve the
function space.  We deal with this by
constructing and mapping a full quintic finite element, a subset of
whose basis functions are the Bell basis functions.  We refer the
reader to \citet{kirby-zany} for further details.

We address one further wrinkle in \citet{kirby-zany}.  In particular,
global basis functions with a unit derivative value may differ in size
considerably from those corresponding to a unit point value.  The
resulting scale separation between basis functions leads to a
considerable increase in the condition number that degrades as $h
\searrow 0$.  To compensate for this, we can adjust $M$ by composing
with another linear transform that scales the derivative basis
functions by a suitable, locally-defined, function of the mesh spacing
$h$.

\section{Incorporating transformations in the Firedrake stack}
\label{sec:code}

Making these elements work seamlessly from the end user perspective requires careful
modification of several components of Firedrake.  Throughout this
process, we have been driven by two main design goals.  First, we seek
to provide all information related to basis function evaluation
through FInAT (internally using FIAT for reference element tabulation
as needed).  Second, we seek to avoid making FInAT dependent on other
Firedrake components so that it may be used by projects besides
Firedrake.  To address these two concerns we have developed an
abstract interface, which a form compiler must implement, that allows
FInAT to request expressions for the evaluation of geometric
information on physical cells.

To facilitate these callbacks between the element library on the one
hand, and the form compiler on the other, they need to agree on a
common language for exchanging information.  Where FIAT communicated
with the form compiler through numerical arrays, FInAT communicates
with the form compiler by exchanging GEM
expressions~\citep{homolya2018tsfc}. GEM is the intermediate language used in both
\tsfc and FInAT to describe tensor algebra.  It is the natural
choice for the exchange of geometric information, since any form
compiler using FInAT must already understand GEM.  Since it is a
fairly basic language describing array indexing and arithmetic, it
should be possible to convert from GEM to other representations
such as pymbolic (\url{https://mathema.tician.de/software/pymbolic/})
or sympy~\citep{joyner2012open}.  Our new elements then overload the
\texttt{basis\_evaluation} method to call FIAT for reference values
and then provide abstract syntax for constructing and applying $M$ to
those values.

By isolating the transformations at the level of basis functions in
this way, we required almost no changes in the rest of the Firedrake
code base.  In UFL, we made trivial changes to ``register'' the new
elements: adding them to a list of available elements in the
language.  In \tsfc, we bind the UFL elements to concrete FInAT
classes and provide the necessary implementation of geometric mapping
information.  Most of the 
expressions are readily obtainable by translation of appropriate
UFL, which has symbolic objects representing Jacobians, facet normals,
and so forth.  The mesh cell size at
vertices, used for scaling of derivative basis functions, presents
certain difficulties.  The form
compiler only makes reference to a single element, but the
transformation theory requires a characteristic mesh size that is available at
vertices and agreed upon by all cells sharing a given vertex.
We therefore provide the mesh cell size as a normal
coefficient field, and require that Firedrake provide it when
assembling the local element tensor.  This marginally changes the
interface between \tsfc and Firedrake: element kernels may now
have one extra argument.  However, this is handled internally and
unbeknownst to the user.

\subsection{Inserting the transformation matrix}
\label{sec:transformation-matrix-choice}

Given the ability to construct the transformation $M$, it can be
inserted into a finite element code in (at least) two distinct places.
For one, the transformation can be applied to
reference values at each quadrature point before the local integration
is carried out on each cell.  Alternatively, one could compute a local
vector or matrix with the untransformed basis values and
post-multiply by the resulting transformations as needed.  Similarly,
one can either evaluate members of the function space by transforming the
basis functions or by using the associative property to transform the
expansion coefficients and then using the unmodified basis functions.

In the first, case, suppose that a bilinear form over some
$\Omega \subset \mathbb{R}^2$ requires both basis
values and gradients.  If we have a basis with $N_f$ members and a
quadrature rule with $N_q$ points, then transforming all
the basis values and gradients will require 3 $N_q$ matrix-vector
products with $M$.  Forming a load vector or matrix action will
also require similar additional effort.

On the other hand, suppose that one forms an element matrix with
untransformed basis functions and applies the transformation
afterward.  Essentially, one must compute the congruence transform
\begin{equation}
  \label{eq:cong}
  A = M \tilde{A} M^T,
\end{equation}
where $\tilde{A}$ is the `wrong' element matrix.  Beyond computing
$\tilde{A}$, one must act on each row of $\tilde{A}$ with $M$ and then
on each column of the result.  This amounts to $2 N_f$ matrix-vector
products with $M$.  Unless $N_q < \tfrac{2}{3} N_f$, using \cref{eq:cong}
will be moderately less expensive.  At any rate, as we point out in
\citet{kirby-zany}, either are relatively small additional costs compared
to the overall cost of forming the element matrix.

Although the cost of matrix formation is not significantly different
with these two approaches, the difference is somewhat more substantial
with load vectors or matrix actions.  Considering the matrix
action via \cref{eq:cong}:
\begin{equation}
A u = M \tilde{A} M^T u,
\end{equation}
where $u$ is vector of degrees of freedom on an element, this can be
factorized so that it requires only two applications of $M$ by using
the associative property in the correct way, reusing a kernel for
computing the mapping $u \mapsto \tilde{A} u$.

In our implementation, we have opted simply to map the basis functions
at each quadrature point.  This required somewhat less invasive
modifications to the form compiler and, as we see in
\cref{sec:element-matrix-cost}, does not seem to create any
significant performance penalties.  Future work could include
additional refactoring of \tsfc to support and compare either mode and
select based on the particular use case or user-specified settings.

\subsection{Boundary conditions}
\label{sec:bcbad}

Strongly enforcing boundary conditions with most of these elements
appears to be rather difficult.  Instead, we use Nitsche-type
penalty methods for essential boundary conditions.
Similar techniques are used in GetFEM++ and have also
been adopted in spline-based finite elements~\citep{embar2010imposing}
for similar reasons.  Fortunately, expressing the additional
terms in the variational problem in UFL is not difficult.  Here, we
briefly demonstrate some of the issues in strongly constraining
boundary values.

With Lagrange elements, (approximately) setting
$u|_{\partial \Omega}= g$ for some continuous function $g$ by
interpolation is relatively
straightforward, fixing $u$ to agree with $g$ at the boundary nodes.
With Hermite elements, for example, this becomes more difficult.
Since $g$ lives only on $\partial \Omega$, one would have
to provide a differentiable extension of $g$ into $\Omega$ and,
moreover, a code interface to obtaining the derivatives of $g$ to
enforce boundary conditions.

Further difficulties for Hermite elements are apparent, even
just considering the simpler case of the homogeneous condition
$u|_{\partial \Omega} = 0$.  On a line segment, a cubic polynomial
vanishes iff its endpoint values and derivatives do.  So, on a
triangle containing a boundary edge, we must constrain the correct
values and tangential derivatives on the
boundary vertices to force $u=0$ on the edge.  

\begin{figure}[htbp]
  \begin{subfigure}[t]{0.4\textwidth}
    \centering
      \begin{tikzpicture}
        \draw[fill] (0,0) circle (0.05) node (0) {} node[anchor=east] {0};
        \draw (0) circle (0.075);
        \draw[fill] (0,1) circle (0.05) node (1) {} node[anchor=east] {1};
        \draw (1) circle (0.075);
        \draw[fill] (0,2) circle (0.05) node (2) {} node[anchor=east] {2};
        \draw (2) circle (0.075);
        \draw[very thick] (0.center)--(1.center);
        \draw[very thick] (1.center)--(2.center);
        \draw[fill] (1, 0.5) circle (0.05) node (3) {} node[anchor=west] {3};
        \draw[fill] (1, 1.5) circle (0.05) node (4) {} node[anchor=west] {4};
        \draw[thin, dashed] (0.center)--(3.center);
        \draw[thin] (1.center)--(3.center);
        \draw[thin] (1.center)--(4.center);
        \draw[thin, dashed] (2.center)--(4.center);
        \draw[thin, dashed] (3.center)--(4.center);
      \end{tikzpicture}
        \caption{Setting $u=0$ on the solid boundary
      requires setting vertex values and $y-$ derivatives at nodes 0, 1,
      and 2 to be zero. No $x-$ derivatives are modified in this case.}
    \label{fig:vertbc}
  \end{subfigure}
  \hfill
  \begin{subfigure}[t]{0.4\textwidth}
       \centering
      \begin{tikzpicture}
        \draw[fill] (0,0) circle (0.05) node (0) {} node[anchor=east] {0};
        \draw (0) circle (0.075);
        \draw[fill] (-0.2,1) circle (0.05) node (1) {} node[anchor=east] {1};
        \draw (1) circle (0.075);
        \draw[fill] (0,2) circle (0.05) node (2) {} node[anchor=east] {2};
        \draw (2) circle (0.075);
        \draw[very thick] (0.center)--(1.center);
        \draw[very thick] (1.center)--(2.center);
        \draw[fill] (1, 0.5) circle (0.05) node (3) {} node[anchor=west] {3};
        \draw[fill] (1, 1.5) circle (0.05) node (4) {} node[anchor=west] {4};
        \draw[thin, dashed] (0.center)--(3.center);
        \draw[thin] (1.center)--(3.center);
        \draw[thin] (1.center)--(4.center);
        \draw[thin, dashed] (2.center)--(4.center);
        \draw[thin, dashed] (3.center)--(4.center);
      \end{tikzpicture}
           \caption{Setting $u=0$ on the solid boundary
         requires setting vertex values at nodes 0, 1, and 2 to be
         zero as well as the entire gradient at vertex 1 and linear
         combinations of the directional derivatives at vertices 0 and
       2.}
    \label{fig:novertbc}
  \end{subfigure}
  \caption{Enforcing homogeneous Dirichlet boundary conditions with
    Hermite elements.}
  \label{fig:badbc}
\end{figure}
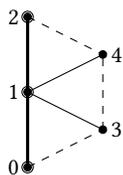
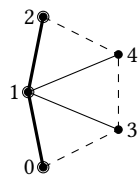

On one hand, consider a portion of a domain with a vertical
boundary, as shown in \cref{fig:vertbc}.  To enforce $u=0$ on
the boundary segments connecting vertices 0 to 1 and 1 to 2, one can
set the nodal values of $u$ at vertices 0, 1, and 2 to zero.  Then, if
one also sets the $y$ derivative of $u$ at these points to zero, $u$
will vanish along both of these segments.  This amounts to six constraints.
In the more general case of colinear edges that are not
aligned with a coordinate axis, one still has six constraints.
However, these set linear combinations of the gradient components
rather than particular degrees of freedom.

On the other hand, consider the case in
\cref{fig:novertbc}, where the boundary is not straight.
Clearly, we must set $u=0$ at vertices 0,
1, and 2.  Moreover, we need to set the tangential derivatives in the
directions running along the edges at vertices 0 and 1.  Setting these
tangential derivatives requires constraining a linear combination of
the gradient components.  We must set derivatives
in two distinct directions at vertex 1 to force $u=0$ along both of
the edges.  This, then, forces the entire gradient of $u$ to vanish
at vertex 1.  Two immediate problems arise.  First, the homogeneous
Dirichlet condition on the boundary cannot be enforced without
imposing an additional homogeneous Neumann condition at the vertices.
Also, the number of constraints changes as soon as the boundary is not
straight as we now require \emph{seven} constraints: three function
values, a linear combinations of derivatives at two vertices, and the
full gradient at a third.   As an additional technical problem,
interpolation of inhomogeneous boundary conditions in spaces with
derivative degrees of freedom requires higher regularity of the data.

\section{Examples}
\label{sec:ex}
In this section, we turn to demonstrating and evaluating the new
capabilities of Firedrake.  In the first three subsections, we explore
various computational aspects of these newly-enabled elements for some
model problems.  We consider the accuracy and computational costs of
various discretizations of the Poisson and biharmonic problems.  In
particular, we can give per-element FLOP estimates for the generated
code for various discretizations and also compare run time to build
and solve the global sparse finite element matrices.  Later, we also
present some more advanced examples of usage.  All timing results were
obtained a on single core of an Intel E5-2640v3 (Haswell) processor.
The measured single core STREAM triad \citep{McCalpin:1995} bandwidth
is $13 \operatorname{GB/s}$, and the peak floating point performance
of a single core is $41.6 \operatorname{Gflop/s}$.  

Dealing with high-order polynomials can lead to difficulties with
linear solvers, especially for fourth-order problems such as the
biharmonic equation.  We do not explore optimal preconditioners or
other aspects of iterative methods~\citep{brenner1989optimal,bramble1995multigrid,xu1996auxiliary}, instead using sparse LU
factorization provided by PaStiX~\citep{Henon:2002}, accessed through
the PETSc library~\citep{Balay:1997,Balay:2018}.  In all examples
where we study the convergence, we have also included one iteration of
iterative refinement, which adds only a small extra cost, to improve
the roundoff error that can become rather large at high degree or on
fine meshes.

When using Lagrange elements for Poisson's equation (or more generally,
conforming discretizations of second-order elliptic operators),
\emph{static condensation} is also a reasonable technique to perform
before sparse factorization.  This is implemented through
SLATE~\citep{gibson2018domain}, a domain-specific language for
expressing localized linear algebra on finite element tensors.  While
the experiments in \citet{gibson2018domain} demonstrate its
effectiveness for hybridizing mixed and HDG-type methods,
here we use its facilities for static condensation for continuous
Galerkin methods.  Since SLATE interfaces with the rest of Firedrake
as a preconditioner acting on the implicit matrices described
in \citet{kirby2018solver}, we are able to factor either the original
or condensed global matrix by a change of solver options.
Except for Hermite, our newly-enabled elements have no internal
degrees of freedom, so static condensation gives no assistance in
these cases.  Similarly, it does not work for interior penalty biharmonic
discretizations since internal degrees of freedom couple across cell
boundaries through jump terms on the derivatives.

Because scalable solvers are quite difficult and somewhat distinct
from the innovations in code generation required for this work, we
have focused on single-process performance. These new developments in
no way clash with Firedrake's existing parallel abilities, but absent
effective preconditioners one may be limited to sparse direct parallel
solvers such as PaStiX or MUMPS~\cite{amestoy2000mumps}.

\subsection{Convergence on model problems}
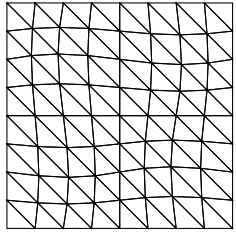
\begin{figure}[htbp]
  \centering
  \input{code/pictures/tikzmesh}
  \caption{Coarsest mesh for convergence studies.}
  \label{fig:mesh}
\end{figure}

As a first example, we demonstrate that we obtain
theoretically-predicted convergence rates for some model problems on
the unit square, labelled $\Omega$.  Our coarsest mesh is shown in
\cref{fig:mesh}, which has internal vertices of a regular
$8 \times 8$ mesh sinusoidally perturbed to introduce geometric
nonuniformity.

First, we validate our new elements for Poisson's equation
\begin{equation}
  -\Delta u = f,
\end{equation}
subject to Dirichlet boundary conditions $u=0$ on $\partial \Omega$.
One multiplies by a
test function and integrates by parts in the typical way to obtain a
variational formulation.  In line with our observation in \cref{sec:bcbad}, we will
enforce the boundary condition through a consistent modification of
the bilinear form, as introduced by \citet{Nitsche}.
\begin{equation}
a_h(u, v) = \int_\Omega \nabla u \cdot \nabla v \dx
- \int_{\partial \Omega} (\nabla u \cdot n) v \ds
- \int_{\partial \Omega} u (\nabla v \cdot n) \ds
+ \frac{\alpha}{h} \int_{\partial \Omega} u v \ds.
\label{eq:apoisson-nitsche}
\end{equation}
Here, $\alpha$ is a positive constant large enough to render the
bilinear form coercive in a perturbed $H^1$ norm, and $h$ is
the characteristic mesh size.  For
$V_h \subset H^1$ consisting of one of the finite element spaces
under consideration, we solve the discrete variational problem
\begin{equation}
  a_h(u_h, v_h) = \left( f , v_h \right) \equiv \int_\Omega f v_h \dx
\end{equation}
although additional terms would appear on the right-hand side with inhomogeneous
Dirichlet boundary conditions.  Implementing this in
Firedrake requires straightforward modifications of the listing in \cref{fig:listing}.

\begin{figure}[htbp]
  \begin{subfigure}[l]{0.475\textwidth}
    \begin{tikzpicture}[scale=0.88]
      \begin{loglogaxis}[xlabel={$N$}, ylabel={$\|u-u_h\|_{L^2}$},
        log basis x=2,
        ylabel near ticks,
        legend cell align=left,
        legend style={at={(0.5, 1.05)}, anchor=south},
        legend columns=3]
        \addplot[dotted,mark=*,mark options={solid,fill=gray}] table [x=N, y=Error, col sep=comma]
        {code/data/poisson.Lagrange.3.csv};
        \addlegendentry{$P^3$}
        \addplot[dotted,mark=*,mark options={solid,fill}] table [x=N, y=Error, col sep=comma]
        {code/data/poisson.Lagrange.4.csv};
        \addlegendentry{$P^4$}
        \addplot[dotted,mark=o, mark options={solid}] table [x=N, y=Error, col sep=comma]
        {code/data/poisson.Lagrange.5.csv};
        \addlegendentry{$P^5$}
        \addplot[dashdotted,mark=diamond*,mark options={solid,fill=gray}] table [x=N, y=Error, col sep=comma]
        {code/data/poisson.Hermite.3.csv};
        \addlegendentry{Hermite}
        \addplot[dashdotted,mark=diamond*,mark options={solid,fill}] table [x=N, y=Error, col sep=comma]
        {code/data/poisson.Bell.5.csv};
        \addlegendentry{Bell}
        \addplot[dashdotted,mark=diamond, mark options={solid}] table [x=N, y=Error, col sep=comma]
        {code/data/poisson.Argyris.5.csv};
        \addlegendentry{Argyris}
        \addplot [domain=32:128] {10/pow(x,4)} node[above,midway, yshift=-1pt, anchor=south west] {$h^{-4}$};
        \addplot [domain=32:128] {2/pow(x,5)} node[above, midway, yshift=-1pt, anchor=south west] {$h^{-5}$};
        \addplot [domain=32:128] {0.1/pow(x,6)} node[below, yshift=-2pt, midway] {$h^{-6}$};
      \end{loglogaxis}
    \end{tikzpicture}
    \caption{$L^2$ error in solving Poisson's equation on a perturbed $N
      \times N$ mesh.}
    \label{fig:poissonconv}
  \end{subfigure}
  \hspace{0.04\textwidth}
  \begin{subfigure}[r]{0.475\textwidth}
    \begin{tikzpicture}[scale=0.88]
      \begin{loglogaxis}[xlabel={$N$}, ylabel={$\|u-u_h\|_{L^2}$},
        log basis x=2,
        ylabel near ticks,
        legend cell align=left,
        legend style={at={(0.5, 1.05)}, anchor=south},
        legend columns=4]
        \addplot[dotted,mark=otimes,mark options={solid, fill=gray}] table [x=N, y=Error, col sep=comma]
        {code/data/biharmonicdg.Lagrange.2.csv};
        \addlegendentry{$P^2$}
        \addplot[dotted,mark=*,mark options={solid,fill=gray}] table [x=N, y=Error, col sep=comma]
        {code/data/biharmonicdg.Lagrange.3.csv};
        \addlegendentry{$P^3$}
        \addplot[dotted,mark=*,mark options={solid,fill}] table [x=N, y=Error, col sep=comma]
        {code/data/biharmonicdg.Lagrange.4.csv};
        \addlegendentry{$P^4$}
        \addplot[dotted,mark=o, mark options={solid}] table [x=N, y=Error, col sep=comma]
        {code/data/biharmonicdg.Lagrange.5.csv};
        \addlegendentry{$P^5$}
        \addplot[dashdotted,mark=diamond*,mark options={solid,fill=gray}] table [x=N, y=Error, col sep=comma]
        {code/data/biharmonic.Morley.2.csv};
        \addlegendentry{Morley}
        \addplot[dashdotted,mark=diamond*,mark options={solid,fill}] table [x=N, y=Error, col sep=comma]
        {code/data/biharmonic.Bell.5.csv};
        \addlegendentry{Bell}
        \addplot[dashdotted,mark=diamond, mark options={solid}] table [x=N, y=Error, col sep=comma]
        {code/data/biharmonic.Argyris.5.csv};
        \addlegendentry{Argyris}
        \addplot [domain=32:128] {20/pow(x,2)} node[above, yshift=-1pt, midway, anchor=south west] {$h^{-2}$};
        \addplot [domain=32:128] {10/pow(x,4)} node[above, yshift=-1pt, midway, anchor=south west] {$h^{-4}$};
        \addplot [domain=32:128] {1/pow(x,5)} node[above, yshift=-1pt,  midway, anchor=south west] {$h^{-5}$};
        \addplot [domain=32:128] {0.1/pow(x,6)} node[below, yshift=-2pt, midway] {$h^{-6}$};
      \end{loglogaxis}
    \end{tikzpicture}
    \caption{$L^2$ error in solving biharmonic equation on a perturbed $N
      \times N$ mesh.}
    \label{fig:biharmconv}
  \end{subfigure}
  \caption{$L^2$ errors solving prototypical second- and fourth-order
    PDEs using various higher-order finite elements.}
  \label{fig:convergence}
\end{figure}
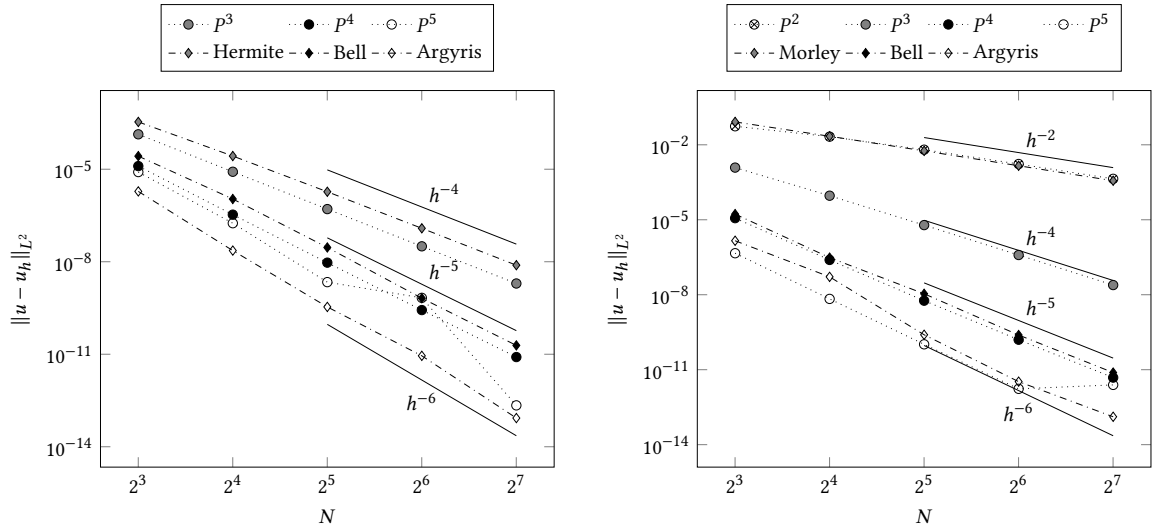

\Cref{fig:poissonconv} plots the $L^2$ error versus mesh
refinement on uniform refinements of the mesh in
\cref{fig:mesh}.  $P^3$ appears slightly more accurate, but of
the same order, as cubic Hermite.  Since $P^3$ is a strictly larger
space than Hermite, its $H^1$ best approximation result (which affects the $L^2$
estimates through Aubin-Nitsche duality) should also be slightly better.
The same comparison holds for $P^4$ and Bell, while we do not observe
the same effect for $P^5$ and Argyris elements.

While some situations could call for smooth approximations of second
order equations, many of our new elements come into their own
for the plate-bending biharmonic equation
\begin{equation}
  \Delta^2 u = f
\end{equation}
on $\Omega$.   In our examples, we consider clamped boundary
conditions $u = \tfrac{\partial u}{\partial n} = 0$ on $\partial \Omega$.
Following \citet{Brenner:2008}, we employ the bilinear form
\begin{equation}
  a(u, v) = \int_{\Omega} \Delta u \Delta v - \left(1-\nu \right)
  \left( 2 u_{xx} v_{yy} + 2 u_{yy} v_{xx} - 4 u_{xy} v_{xy} \right)
  \dx,
  \label{eq:aplate}
\end{equation}
where $0 \leq \nu \leq \tfrac{1}{2}$ is the Poisson ratio for the
plate.  The terms multiplied by $(1-\nu)$ may separately be integrated
by parts to give $u_{xxyy}v$ times zero plus terms for
incorporating strongly-supported boundary conditions.
Subject to clamped boundary conditions, or on any subspace of
$H^2$ not containing linear polynomials over $\Omega$, the bilinear
form $a$ is coercive.  Argyris and Bell elements give
conforming, optimal-order approximations.  Morley elements give a
suitable, albeit suboptimal, nonconforming method.  

As an alternative to discretizing the $H^2$ bilinear form with Morley,
Bell, or Argyris elements, it is possible to adapt interior penalty
techniques for $C^0$ elements~\citep{engel2002continuous,wells2004discontinuous}.  This allows use of
$P^k$ elements for $k \geq 2$ by means of
penalizing jumps in derivatives.  These methods do not require
the extra terms scaled by $(1-\nu)$ in \cref{eq:aplate}, instead using the bilinear form
\begin{equation}
  a_h(u, v) = \sum_{K \in \mathcal{T}} \int_{K} \Delta u \Delta v \dx
+ \sum_{E \in \mathcal{E}_h^{\text{int}}}\left(\int_{E} \frac{\alpha}{h_E} \jump{\nabla u}\jump{\nabla v} \ds
- \int_{E} \avg{\Delta u}\jump{\nabla v}\ds
- \int_{E} \jump{\nabla u}  \avg{\Delta v} \ds\right).
\label{eq:aip}
\end{equation}
Here, $K$ are the cells in a triangulation $\mathcal{T}$ and
$\mathcal{E}_h^{\text{int}}$ are the interior edges. The operators
$\jump{\cdot}$ and $\avg{\cdot}$ are the standard jump and average
operators used in discontinuous Galerkin methods. This method is known
to give optimal convergence rates in $L^2$ for $k > 2$ (but only
second order for quadratics), and requires only Lagrange or other
standard $H^1$-conforming elements. In particular, it is frequently
used with Lagrange elements. Although we will not include an
exhaustive comparison, it is also worth mentioning other alternatives
such as the Reissner-Mindlin formulation. This leads to a system of
singularly-perturbed $H^1$
equations~\cite{duran1992mixed,hale2018simple} and allows direct
enforcement of Dirichlet boundary conditions for rotations at the cost
of a mixed system. An alternative approach, using a mixed formulation
is given by Li~\cite{lithesis} using generalized Regge-type elements.

\Cref{fig:biharmconv} plots the error versus mesh refinement on uniform
refinements of the mesh in \cref{fig:mesh} for the two
biharmonic models.  Unlike the Poisson equation, we are not directly
comparing the same discretization with different elements, but an
$H^2$ discretization (nonconforming for Morley) versus an interior
penalty method.  Still, we see that $P^2$ with interior penalty gives
slightly lower error than Morley elements, and $P^4$ is slightly more
accurate than Bell until the finest meshes.  On the other hand,
Argyris is more accurate than $P^5$ on all the meshes, for which we
observe suboptimal convergence rates, even after iterative refinement.

\subsection{Cost of forming element matrices}
\label{sec:element-matrix-cost}
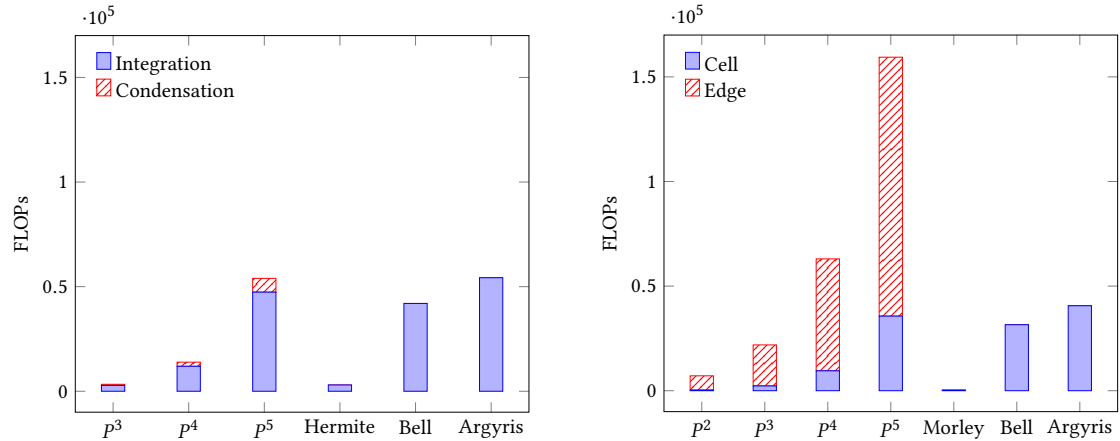
\begin{figure}[htbp]
  \centering
  \pgfplotstableread[col sep=comma,]{code/data/poissonflops.csv}\datatable
  \begin{subfigure}[t]{0.475\textwidth}
    \centering
    \begin{tikzpicture}[scale=0.88]
      \begin{axis}[ybar stacked,
        ylabel near ticks,
        ymin=-1e4, ymax=1.7e5,
        legend pos=north west,
        legend cell align=left,
        legend style={draw=none},
        xtick=data,
        xticklabels from table={\datatable}{Name}, ylabel={FLOPs}]
        \addplot+ table [x expr=\coordindex, y={A Flops}]{\datatable};
        \addplot+[pattern=north east lines, pattern color=red] table
        [x expr=\coordindex, y={S Flops}, pattern=north east lines, fill=none]{\datatable}; \legend{Integration, Condensation}
      \end{axis}
    \end{tikzpicture}
    \caption{Laplace operator of
      \cref{eq:apoisson-nitsche}.  Stacked above $P^k$ elements are
      the additional FLOPs required to form the local Schur complement
      in static condensation}
    \label{fig:poisson-flops}
  \end{subfigure}
  \hspace{0.04\textwidth}
  \begin{subfigure}[t]{0.475\linewidth}
    \begin{tikzpicture}[scale=0.88]
      \pgfplotstableread[col sep=comma,]{code/data/biharmonicflops.csv}\datatable
      \begin{axis}[ ybar stacked,
        ylabel near ticks,
        legend pos=north west,
        ymin=-1e4, ymax=1.7e5,
        legend cell align=left,
        legend style={draw=none},
        xtick=data,
        xticklabels from table={\datatable}{Name}, ylabel={FLOPs}]
        \addplot+ table [x expr=\coordindex, y={Cell Flops}]{\datatable};
        \addplot+[pattern=north east lines, pattern color=red] table
        [x expr=\coordindex, y={Edge Flops}]{\datatable};
        \legend{Cell, Edge}
      \end{axis}
    \end{tikzpicture}
    \caption{Biharmonic operator.  The
      last three entries relate to the bilinear form
      in \cref{eq:aplate} and include no edge terms, while the $P^k$
      entries are from the DG form \cref{eq:aip}.  The per-cell cost
      of the edge integrals seems to drive the overall FLOP count for
      interior penalty methods.}
    \label{fig:biharmflops}
  \end{subfigure}
  \caption{Per-element FLOP counts for assembling Laplace and biharmonic operators
    using various elements.}
  \label{fig:flop-counts}
\end{figure}

While our newly-enabled elements provide effective solutions, we must
also consider the computational cost associated with using them.
Clearly, when the element transformation $M \ne \mathbb{I}$,
extra work is required to perform element-level integrals as
compared to Lagrange and other affine-equivalent elements.  Here, we
document the additional FLOPs required and the effect on time to
assemble matrices for the problems considered above.
\cref{fig:poisson-flops} shows
the \tsfc-reported FLOP counts for building the element
stiffness matrix for the Laplacian using Lagrange elements of degrees
3 through 5 as well as Hermite, Bell, and Argyris elements.  These
numbers include constructing the element Jacobian from the physical
coordinates, transforming basis function gradients using $M$ and the
chain rule, and the loop over
quadrature points and pairs of basis functions.  They do not include
any interaction with the global sparse matrix.  In addition, we report the
per-element cost of performing static condensation for Lagrange
elements with a separate bar above that of forming the element matrices.

We notice that forming the element integral with Hermite elements
requires slightly more operations than
with $P^3$ Lagrange elements.  Since the transformation \cref{eq:MforHerm}
requires only modifying some of the basis functions with the
(already-computed) Jacobian, the actual up-tick in operation count is
very small.  On the other hand, the gap between $P^5$ Lagrange and
Argyris is a bit larger.  Two factors contribute to this.  First, the
Argyris transformation requires the evaluation of several geometric
quantities (tangents, normals, edge lengths, Hessians) not otherwise
required.  Second, there are relatively more nonzeros in $M$ for
Argyris than Hermite.  Comparing Bell elements to $P^4$ shows them to
be quite a bit more expensive for the same order of approximation.
Bell elements have nearly as many degrees of freedom as Argyris,
require a complicated transformation matrix $M$, and also require a
more accurate quadrature rule owing to the presence of quintic terms.
\Cref{fig:poisson-flops} shows that the per-element
cost is indeed closer to (but less than) the Argyris element than it
is to $P^4$.  Beyond this, static condensation adds a small
per-element cost to $P^k$ elements, making $P^3$ and Hermite roughly
comparable as well as $P^5$ and Argyris.

We now turn to the biharmonic operator.  Although the bilinear
form \cref{eq:aplate} has a more complex cell integral
and requires the transformation $M$, the cost of the edge terms in the
bilinear form \cref{eq:aip} is quite large.  Since
there are about 1.5 times as many edges as triangles in our meshes, we
report the FLOP count of the kernel to perform the element integral
plus 1.5 times the per-facet FLOP count.  Observing
\cref{fig:biharmflops}, the $H^2$ methods have considerably
lower per-element operation counts.

\subsection{Building and solving global systems}
Per-element FLOP counts are not the only factor affecting the overall
cost of assembling linear systems.  In fact, the
concentration of degrees of freedom at vertices for Hermite, Bell, and
Argyris elements leads to overall smaller systems with markedly
different costs of assembly and sparsity patterns than Lagrange
elements.

\begin{figure}[htbp]
  \begin{subfigure}[t]{0.475\textwidth}
    \centering
    \begin{tikzpicture}[scale=0.88]
      \begin{loglogaxis}[xlabel={$N$}, ylabel={Time ($s$)},
        ylabel near ticks,
        log basis x=2,
        legend pos=north west,
        legend cell align=left,
        legend style={draw=none}]
        \addplot[dotted,mark=*,mark options={solid,fill=gray}] table [x=N, y=SNESJacobianEval, col sep=comma]
        {code/data/poisson.Lagrange.3.csv};
        \addlegendentry{$P^3$}
        \addplot[dotted,mark=*,mark options={solid,fill}] table [x=N, y=SNESJacobianEval, col sep=comma]
        {code/data/poisson.Lagrange.4.csv};
        \addlegendentry{$P^4$}
        \addplot[dotted,mark=o, mark options={solid}] table [x=N, y=SNESJacobianEval, col sep=comma]
        {code/data/poisson.Lagrange.5.csv};
        \addlegendentry{$P^5$}
        \addplot[dashdotted,mark=diamond*,mark options={solid,fill=gray}] table [x=N, y=SNESJacobianEval, col sep=comma]
        {code/data/poisson.Hermite.3.csv};
        \addlegendentry{Hermite}
        \addplot[dashdotted,mark=diamond*,mark options={solid,fill}] table [x=N, y=SNESJacobianEval, col sep=comma]
        {code/data/poisson.Bell.5.csv};
        \addlegendentry{Bell}
        \addplot[dashdotted,mark=diamond, mark options={solid}] table [x=N, y=SNESJacobianEval, col sep=comma]
        {code/data/poisson.Argyris.5.csv};
        \addlegendentry{Argyris}
      \end{loglogaxis}
    \end{tikzpicture}
    \caption{Laplace assembly time.  We see that the higher FLOP
      counts for $C^1$ elements (c.f.~\cref{fig:poisson-flops})
      only have a noticeable effect on assembly time for Bell elements
      (relative to $P^4$).}
    \label{fig:poissonbuild}
  \end{subfigure}
  \hspace{0.04\textwidth}
  \begin{subfigure}[t]{0.475\textwidth}
    \begin{tikzpicture}[scale=0.88]
      \begin{loglogaxis}[xlabel={$N$}, ylabel={Time ($s$)},
        ylabel near ticks,
        log basis x=2,
        legend pos=north west,
        legend cell align=left,
        legend style={draw=none},
        legend columns=2]
        \addplot[dotted,mark=otimes,mark options={solid,fill=gray}] table [x=N, y=SNESJacobianEval, col sep=comma]
        {code/data/biharmonicdg.Lagrange.2.csv};
        \addlegendentry{$P^2$}
        \addplot[dotted,mark=*,mark options={solid,fill=gray}] table [x=N, y=SNESJacobianEval, col sep=comma]
        {code/data/biharmonicdg.Lagrange.3.csv};
        \addlegendentry{$P^3$}
        \addplot[dotted,mark=*,mark options={solid,fill}] table [x=N, y=SNESJacobianEval, col sep=comma]
        {code/data/biharmonicdg.Lagrange.4.csv};
        \addlegendentry{$P^4$}
        \addplot[dotted,mark=o, mark options={solid}] table [x=N, y=SNESJacobianEval, col sep=comma]
        {code/data/biharmonicdg.Lagrange.5.csv};
        \addlegendentry{$P^5$}
        \addplot[dashdotted,mark=diamond*,mark options={solid,fill=gray}] table [x=N, y=SNESJacobianEval, col sep=comma]
        {code/data/biharmonic.Morley.2.csv};
        \addlegendentry{Morley}
        \addplot[dashdotted,mark=diamond*,mark options={solid,fill}] table [x=N, y=SNESJacobianEval, col sep=comma]
        {code/data/biharmonic.Bell.5.csv};
        \addlegendentry{Bell}
        \addplot[dashdotted,mark=diamond, mark options={solid}] table [x=N, y=SNESJacobianEval, col sep=comma]
        {code/data/biharmonic.Argyris.5.csv};
        \addlegendentry{Argyris}
      \end{loglogaxis}
    \end{tikzpicture}
    \caption{Biharmonic assembly time.  The lower FLOP counts for
      $C^1$ discretisations translate into significantly lower
      assembly times relative to the appropriate $C^0$ interior
      penalty scheme.}
    \label{fig:buildbiharm}
  \end{subfigure}
  \caption{Time to assemble the Laplace and biharmonic operators on a
    perturbed $N \times N$ mesh.  The timings include preallocation of
    the matrix sparsity (which is amortizable over solves), as well as
    building element tensors, and matrix insertion.}
  \label{fig:build-time}
\end{figure}
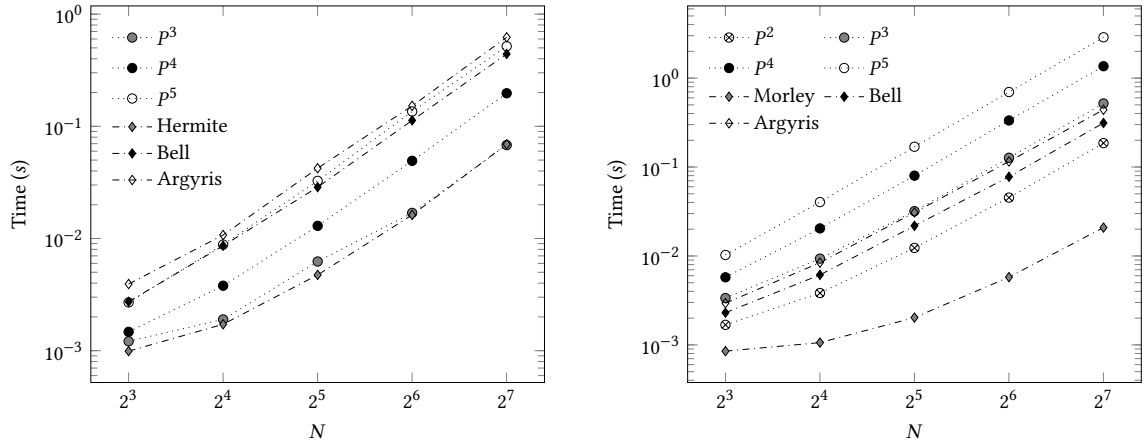

\Cref{fig:poissonbuild} gives the time to build the discrete
Poisson system on the meshes considered in our convergence study
above.  We observe that Hermite and $P^3$, and Argyris and $P^5$
elements require very similar assembly time, with Bell somewhat higher
than $P^4$.  By sharing vertex degrees of freedom between many
elements, we may be achieving slightly better insertion patterns when
summing element matrices into the global system for $H^2$ elements.
We speculate that this, combined with the overall smaller system size
help offset the higher per-element FLOP counts.
\Cref{fig:buildbiharm} shows a clear win for $H^2$ elements.
With fewer global degrees of freedom and less expensive element-level
computation, we see a much lower cost to assemble than for the
interior penalty method with $P^k$ elements.

\begin{figure}[htbp]
  \centering
  \begin{subfigure}[t]{0.15\textwidth}
    \centering
    \includegraphics[height=0.95\textwidth]{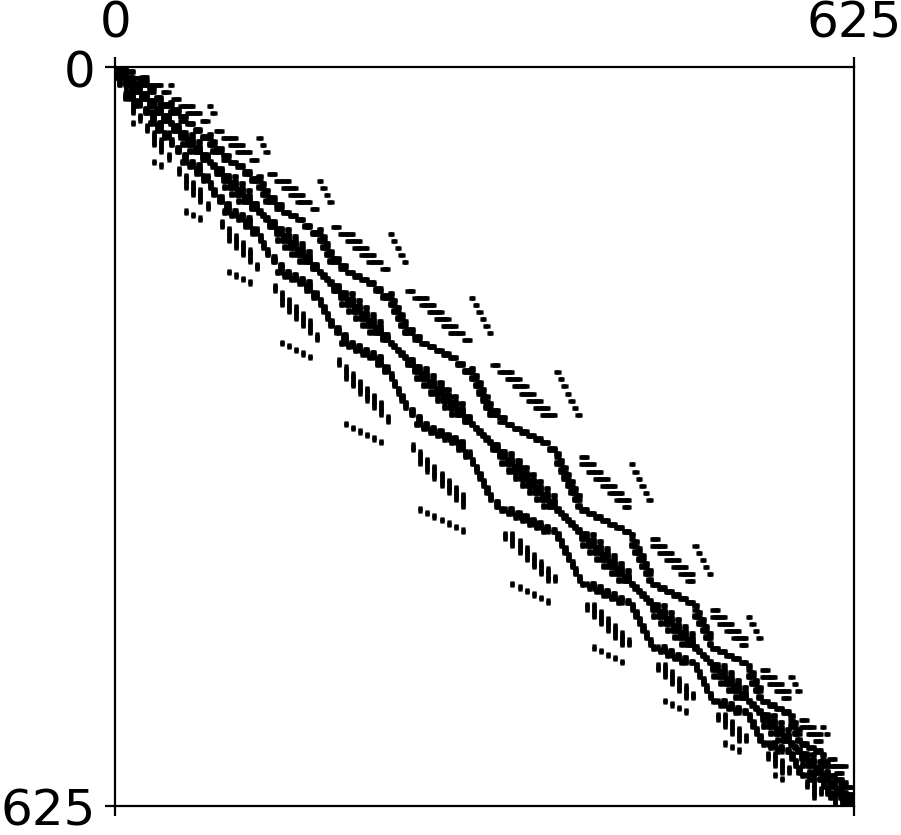}
    \caption{$P^3$}
  \end{subfigure}
  \hspace{0.002\textwidth}
  \begin{subfigure}[t]{0.15\textwidth}
    \centering
    \includegraphics[height=0.95\textwidth]{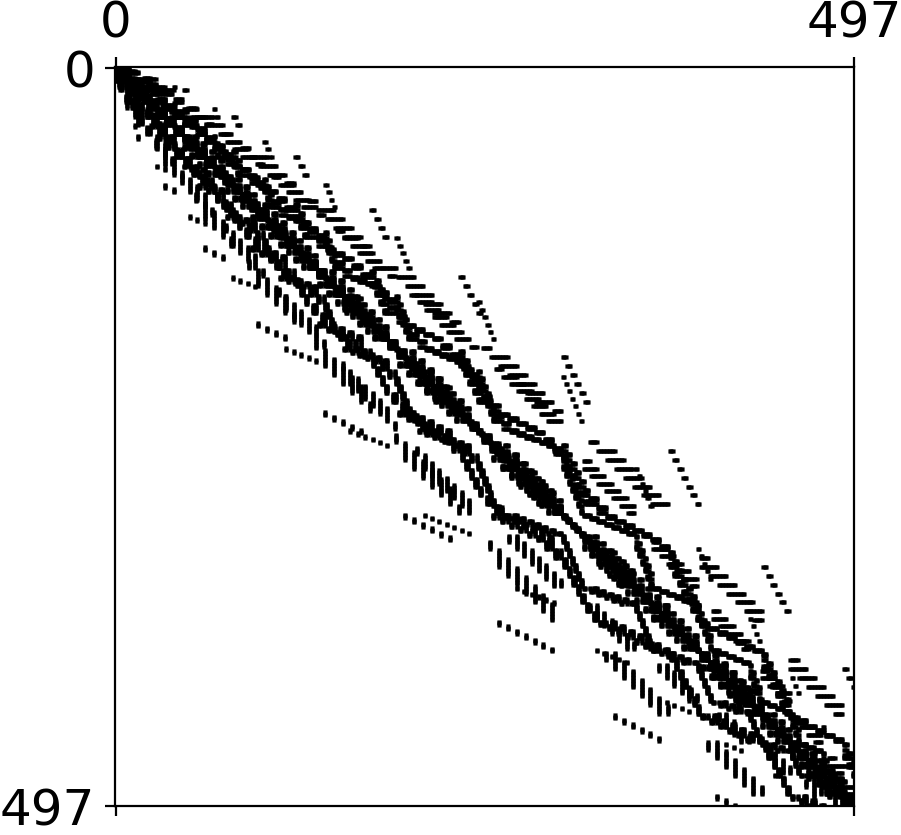}
    \caption{Condensed $P^3$}
  \end{subfigure}
  \hspace{0.02\textwidth}
  \begin{subfigure}[t]{0.15\textwidth}
    \centering
    \includegraphics[height=0.95\textwidth]{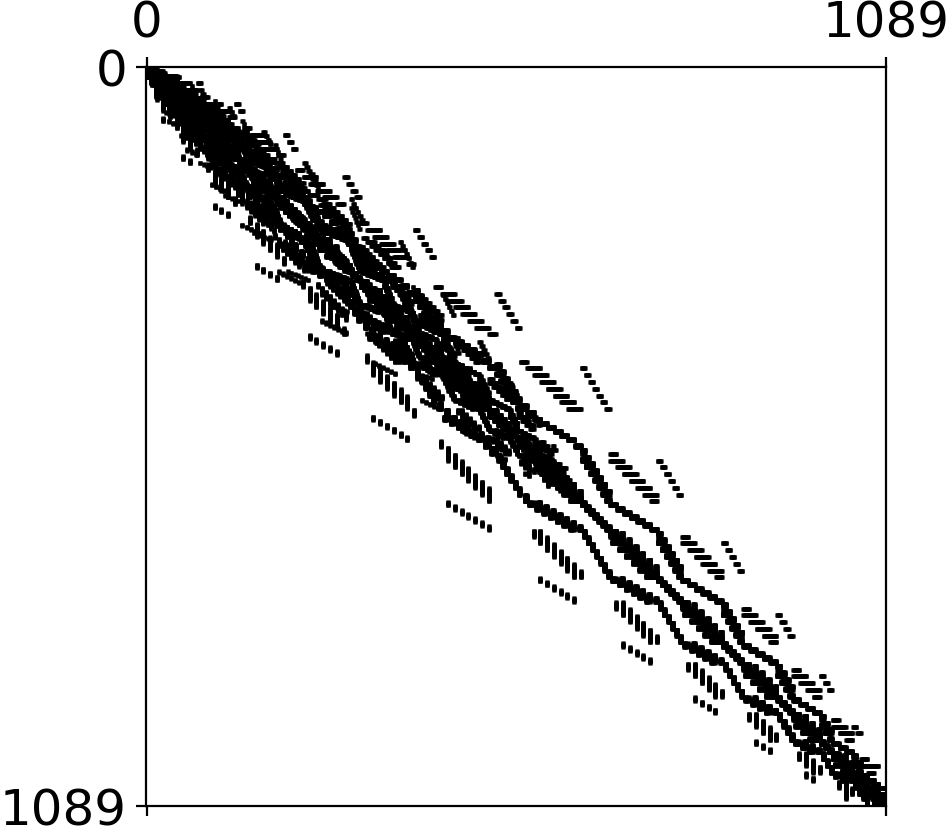}
    \caption{$P^4$}
  \end{subfigure}
  \hspace{0.002\textwidth}
  \begin{subfigure}[t]{0.15\textwidth}
    \centering
    \includegraphics[height=0.95\textwidth]{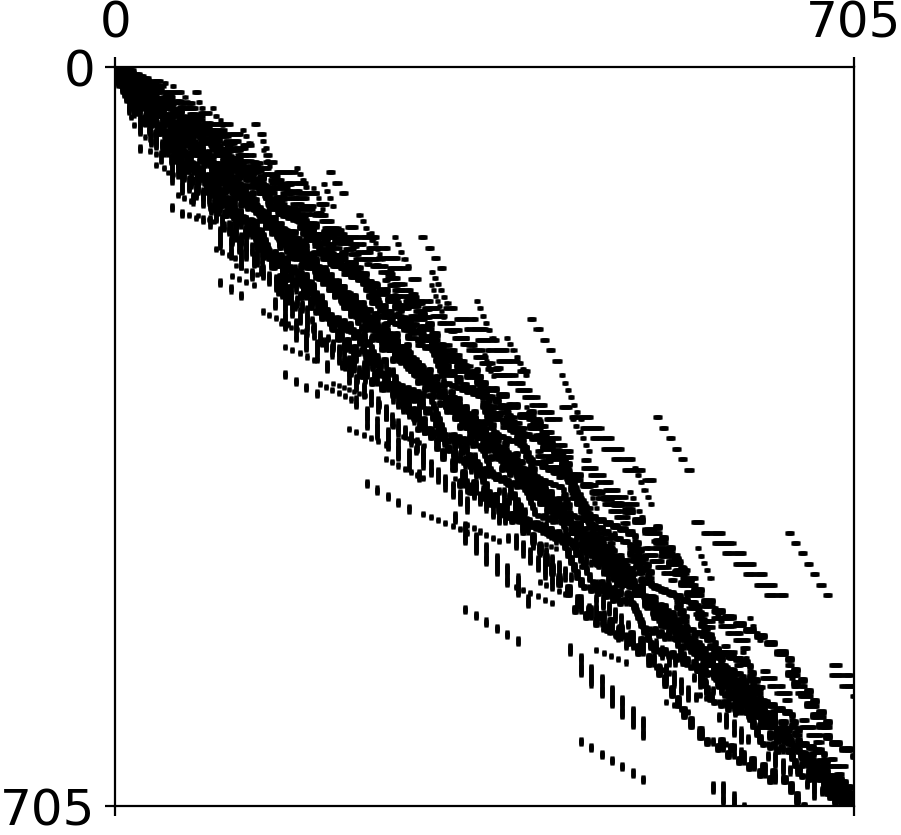}
    \caption{Condensed $P^4$}
  \end{subfigure}
  \hspace{0.02\textwidth}
  \begin{subfigure}[t]{0.15\textwidth}
    \centering
    \includegraphics[height=0.95\textwidth]{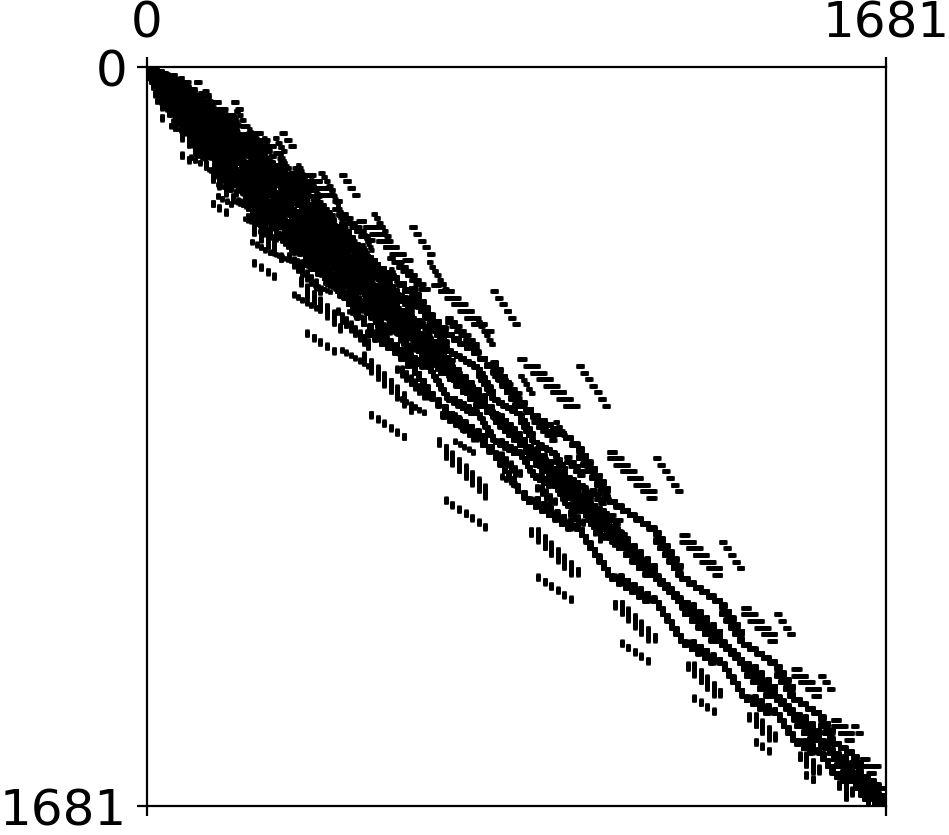}
    \caption{$P^5$}
  \end{subfigure}
  \hspace{0.002\textwidth}
  \begin{subfigure}[t]{0.15\textwidth}
    \centering
    \includegraphics[height=0.95\textwidth]{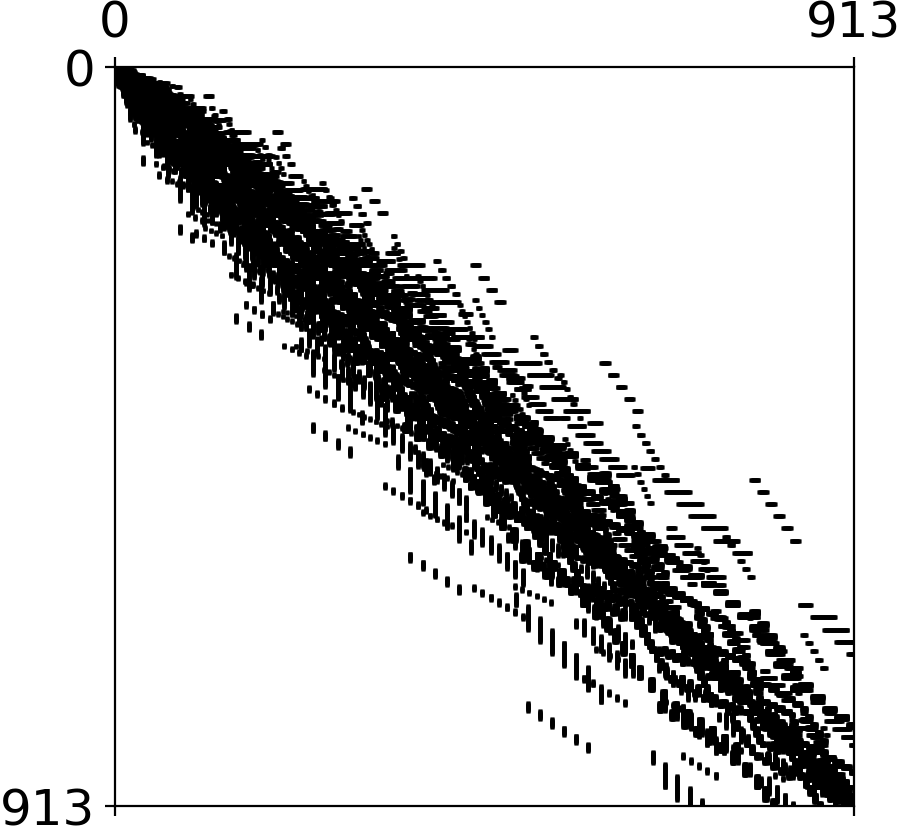}
    \caption{Condensed $P^5$}
  \end{subfigure} 
  \\[0.5\baselineskip]
  \begin{subfigure}[t]{0.3002\textwidth}
    \centering
    \includegraphics[height=0.475\textwidth]{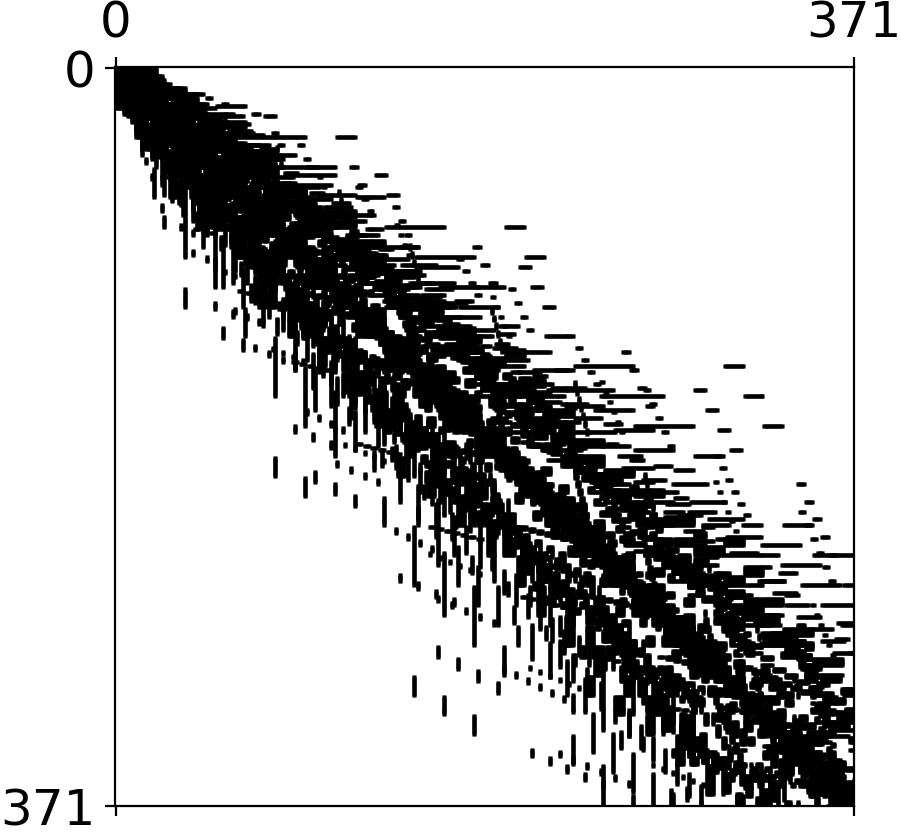}
    \caption{Hermite}
  \end{subfigure}
  \hspace{0.03\textwidth}
  \begin{subfigure}[t]{0.3002\textwidth}
    \centering
    \includegraphics[height=0.475\textwidth]{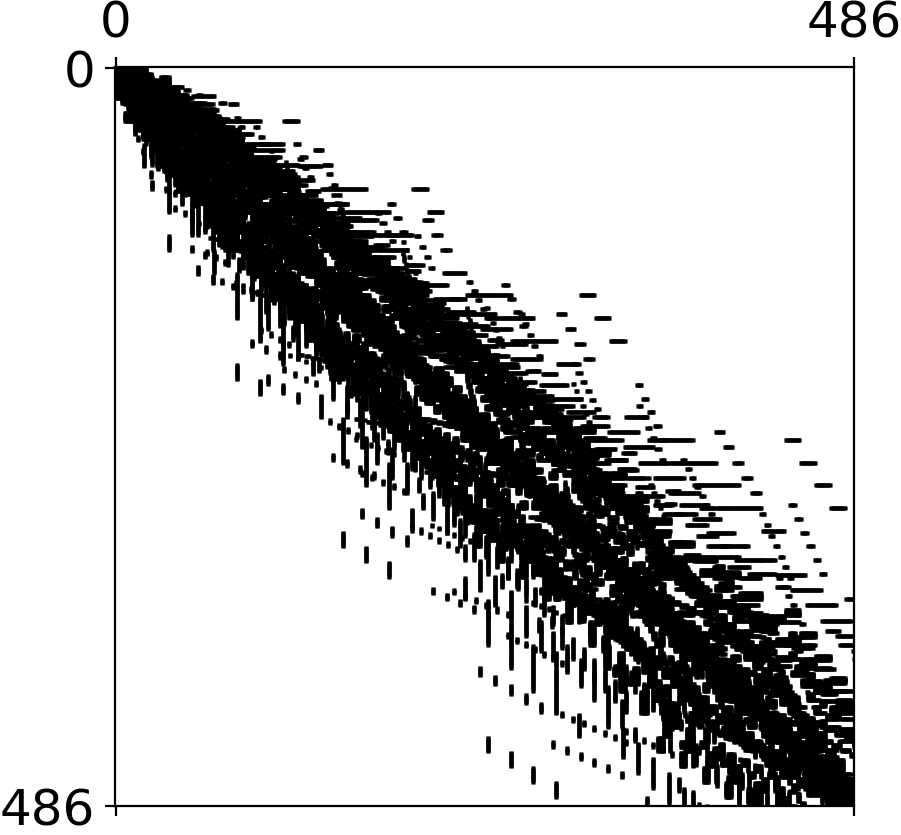}
    \caption{Bell}
  \end{subfigure}
  \hspace{0.03\textwidth}
  \begin{subfigure}[t]{0.3002\textwidth}
    \centering
    \includegraphics[height=0.475\textwidth]{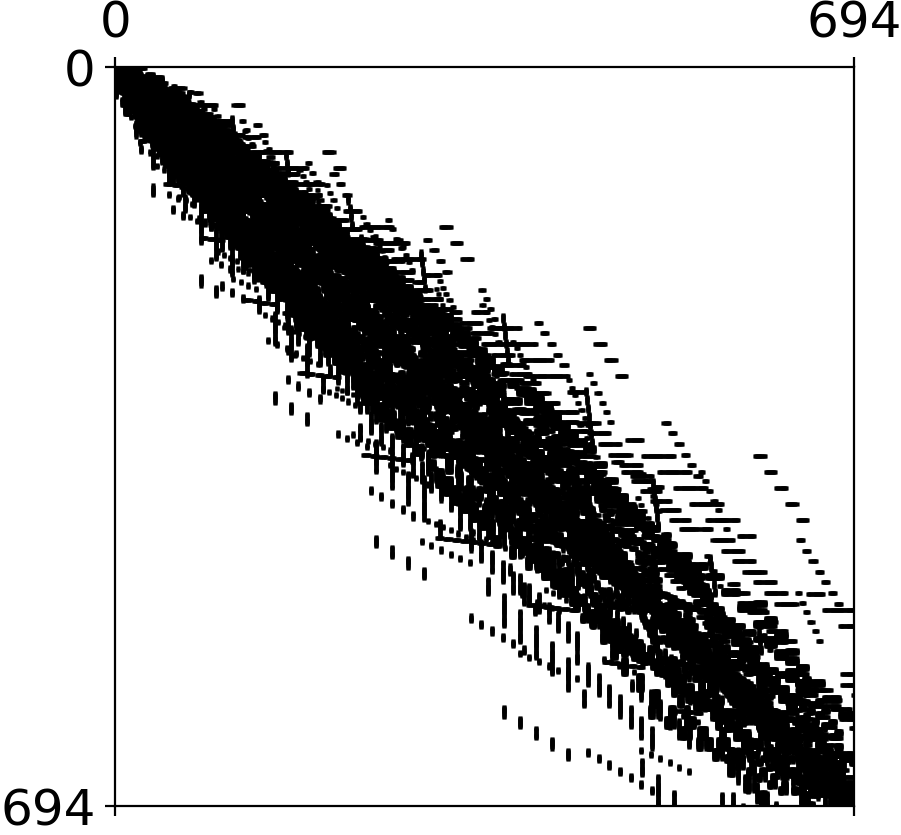}
    \caption{Argyris}
  \end{subfigure}
  \caption{Sparsity patterns of the discrete Laplacian on a regular $8
    \times 8$ mesh using $H^1$ elements (top row) and $H^2$
    elements (bottom row).}
  \label{fig:spyPoisson}
\end{figure}

\begin{figure}[htbp]
  \begin{subfigure}[t]{0.15\textwidth}
    \centering
    \includegraphics[height=0.95\textwidth]{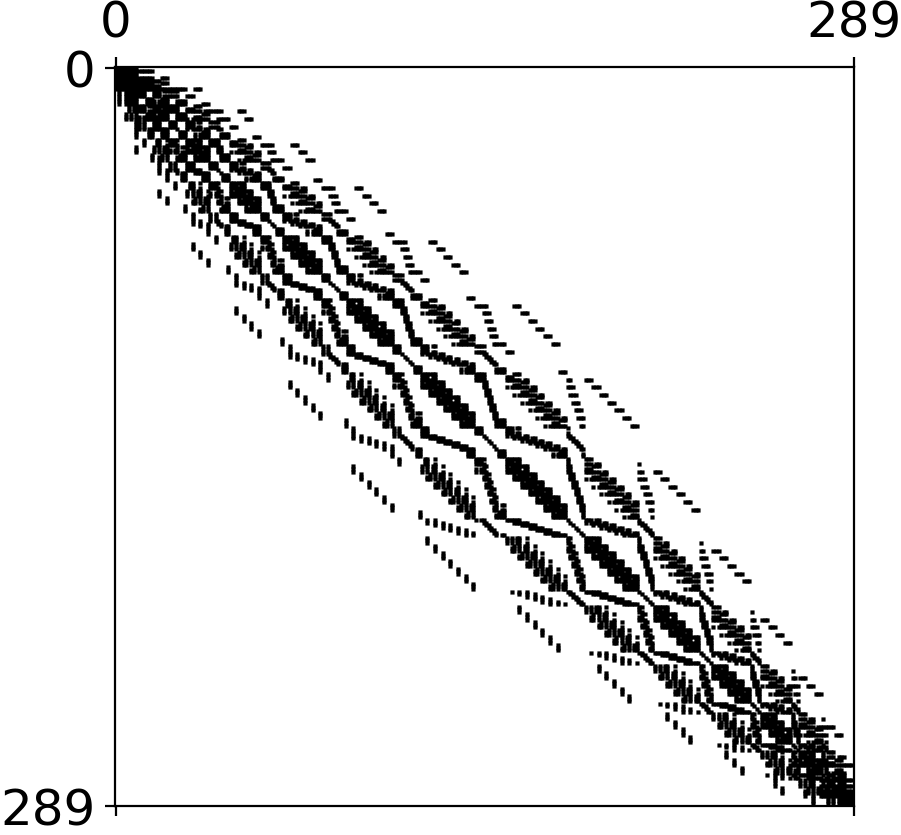}
    \caption{$P^2$}
  \end{subfigure}
  \hspace{0.002\textwidth}
  \begin{subfigure}[t]{0.15\textwidth}
    \centering
    \includegraphics[height=0.95\textwidth]{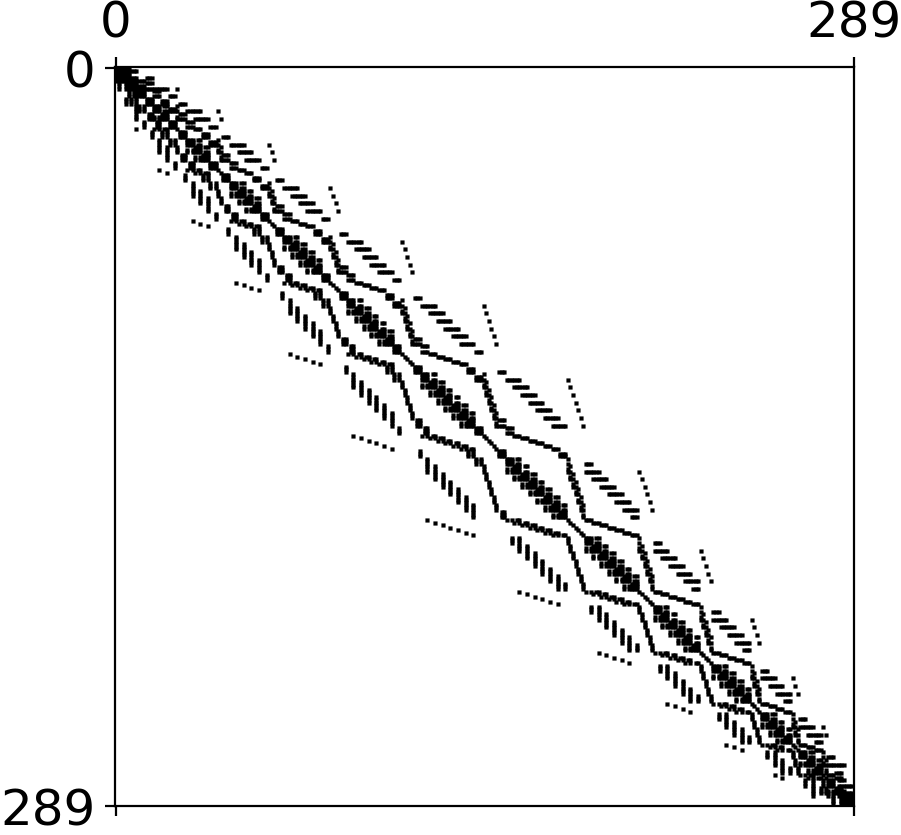}
    \caption{Morley}
  \end{subfigure}
  \hspace{0.02\textwidth}
  \begin{subfigure}[t]{0.15\textwidth}
    \centering
    \includegraphics[height=0.95\textwidth]{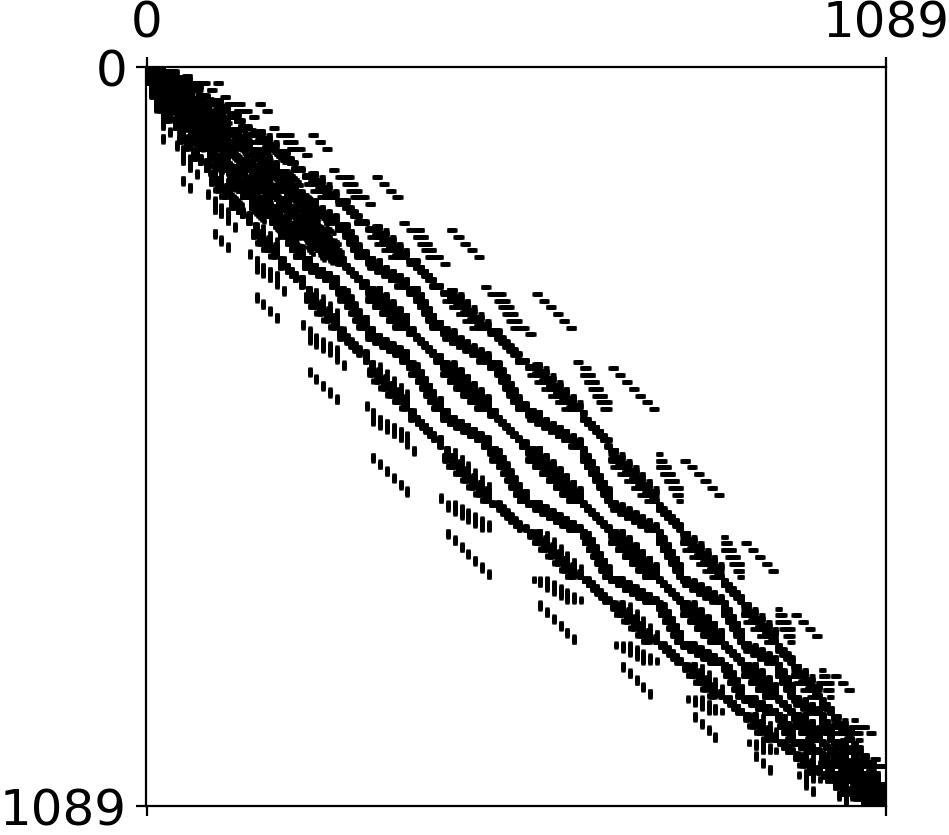}
    \caption{$P^4$}
  \end{subfigure}
  \hspace{0.002\textwidth}
  \begin{subfigure}[t]{0.15\textwidth}
    \centering
    \includegraphics[height=0.95\textwidth]{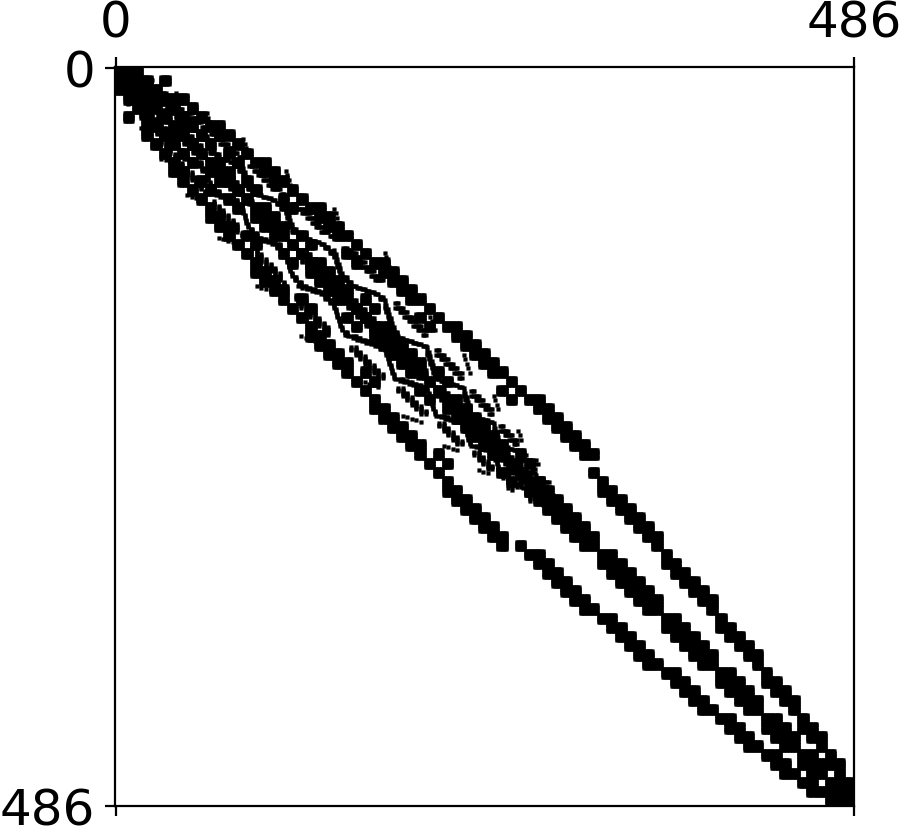}
    \caption{Bell}
  \end{subfigure}
  \hspace{0.02\textwidth}
  \begin{subfigure}[t]{0.15\textwidth}
    \centering
    \includegraphics[height=0.95\textwidth]{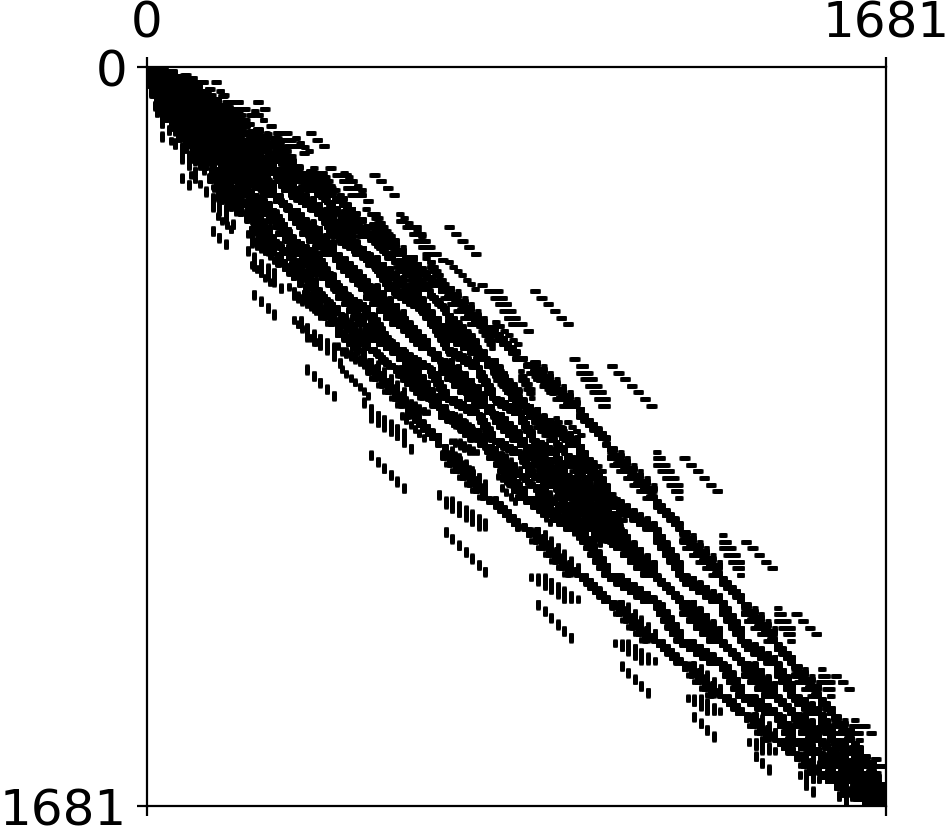}
    \caption{$P^5$}
  \end{subfigure}
  \hspace{0.002\textwidth}
  \begin{subfigure}[t]{0.15\textwidth}
    \centering
    \includegraphics[height=0.95\textwidth]{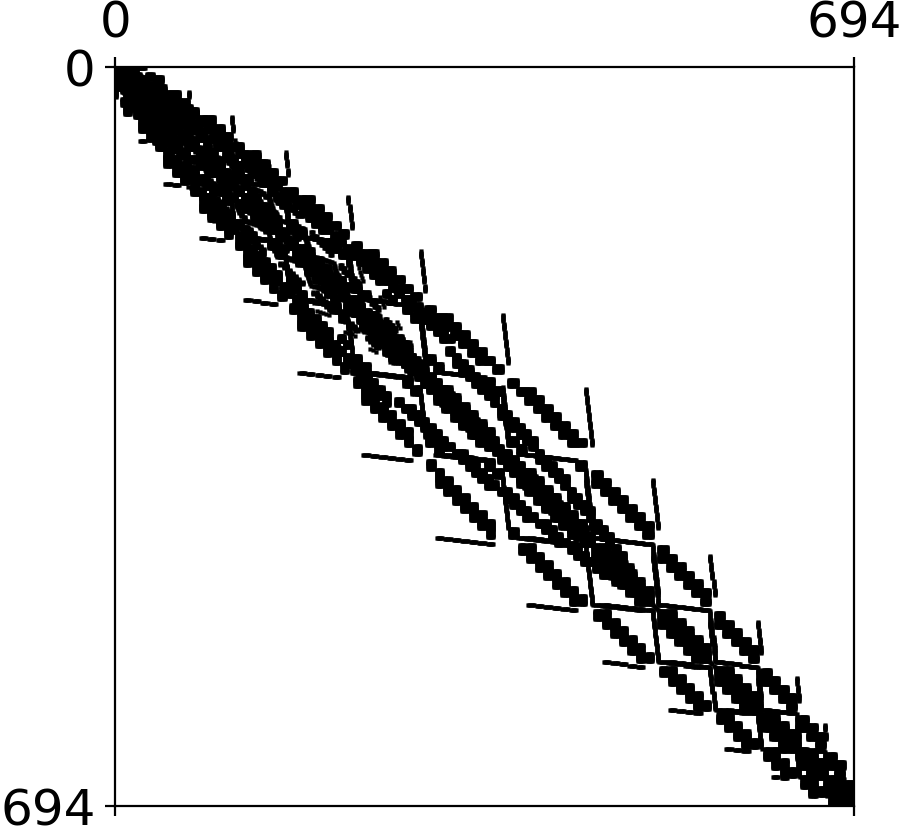}
    \caption{Argyris}
  \end{subfigure}
  \caption{Sparsity patterns of the discrete biharmonic operator on a
    regular $8 \times 8$ mesh for interior penalty discretizations
    (a, c, e) and using $H^2$ discretizations (b, d, f).}
  \label{fig:spyBiharm}
\end{figure}

As a further illustration, we present the global sparsity pattern on
an coarse $8 \times 8$ mesh for each of our discretizations for
Poisson and the biharmonic operator in
\cref{fig:spyPoisson,fig:spyBiharm}.  For Poisson, we compare $P^k$
elements with their smaller and slightly denser statically condensed counterparts in
the first row.  The $H^2$ element with
comparable accuracy appears in below each pair of $P^k$ patterns, demonstrating the
smaller, but denser system in each case. For the biharmonic operator,
edge terms in the interior penalty formulation prevent static
condensation for $P^k$ elements, and $H^2$ elements have few, if any,
interior degrees of freedom to eliminate, so we just consider the
global stiffness matrix.  We see that the $H^2$
elements in fact give smaller and \emph{sparser} systems than $P^k$
elements, owing to the absence of edge terms.  Since the sparsity
pattern is driven by the elements and mesh connectivity rather than
bilinear forms, we note that these trends in sparsity patterns are
rather generic for second- and fourth-order scalar problems,
respectively.

\begin{figure}[htbp]
  \begin{subfigure}[t]{0.475\textwidth}
    \begin{tikzpicture}[scale=0.88]
      \begin{loglogaxis}[xlabel={$N$}, ylabel={Time ($s$)},
        log basis x=2,
        ylabel near ticks,
        ymin=1e-3, ymax=2e1]
        \addplot[dotted,mark=*,mark options={solid,fill=gray}] table [x=N, y=KSPSolve, col sep=comma]
        {code/data/poisson.Lagrange.3.csv};
        % \addlegendentry{$P^3$}
        \addplot[dotted,mark=*,mark options={solid,fill}] table [x=N, y=KSPSolve, col sep=comma]
        {code/data/poisson.Lagrange.4.csv};
        % \addlegendentry{$P^4$}
        \addplot[dotted,mark=o, mark options={solid}] table [x=N, y=KSPSolve, col sep=comma]
        {code/data/poisson.Lagrange.5.csv};
        % \addlegendentry{$P^5$}
        \addplot[dashdotted,mark=diamond*,mark options={solid,fill=gray}] table [x=N, y=KSPSolve, col sep=comma]
        {code/data/poisson.Hermite.3.csv};
        % \addlegendentry{Hermite}
        \addplot[dashdotted,mark=diamond*,mark options={solid,fill}] table [x=N, y=KSPSolve, col sep=comma]
        {code/data/poisson.Bell.5.csv};
        % \addlegendentry{Bell}
        \addplot[dashdotted,mark=diamond, mark options={solid}] table [x=N, y=KSPSolve, col sep=comma]
        {code/data/poisson.Argyris.5.csv};
        % \addlegendentry{Argyris}
      \end{loglogaxis}
      \path[use as bounding box] (current bounding box.south west)
      rectangle ($(current bounding box.north east) + (0, 1.2cm)$);
    \end{tikzpicture}
    \caption{Time to factor and solve the sparse linear system for the
      discrete Laplacian on $N \times N$ mesh, including one iteration
      of iterative refinement.}
    \label{fig:poissonsolve}
  \end{subfigure}
  \hspace{0.04\textwidth}
  \begin{subfigure}[t]{0.475\textwidth}
    \begin{tikzpicture}[scale=0.88]
      \begin{loglogaxis}[xlabel={$N$}, ylabel={Time ($s$)},
        log basis x=2,
        ymin=1e-3, ymax=2e1,
        ylabel near ticks,
        legend cell align=left,
        legend style={overlay, at={(-0.16, 1.05)}, anchor=south},
        legend columns=6]
        \addplot[dotted,mark=*,mark options={solid,fill=gray}] table [x=N, y=SNESSolve, col sep=comma]
        {code/data/poisson.Lagrange.3.csv};
        \addlegendentry{$P^3$}
        \addplot[dotted,mark=*,mark options={solid,fill}] table [x=N, y=SNESSolve, col sep=comma]
        {code/data/poisson.Lagrange.4.csv};
        \addlegendentry{$P^4$}
        \addplot[dotted,mark=o, mark options={solid}] table [x=N, y=SNESSolve, col sep=comma]
        {code/data/poisson.Lagrange.5.csv};
        \addlegendentry{$P^5$}
        \addplot[dashed,mark=square*,mark options={solid,fill=gray}] table [x=N, y=SNESSolve, col sep=comma]
        {code/data/poisson.condensed.Lagrange.3.csv};
        \addlegendentry{Condensed $P^3$}
        \addplot[dashed,mark=square*,mark options={solid,fill}] table [x=N, y=SNESSolve, col sep=comma]
        {code/data/poisson.condensed.Lagrange.4.csv};
        \addlegendentry{Condensed $P^4$}
        \addplot[dashed,mark=square, mark options={solid}] table [x=N, y=SNESSolve, col sep=comma]
        {code/data/poisson.condensed.Lagrange.5.csv};
        \addlegendentry{Condensed $P^5$}
        \addplot[dashdotted,mark=diamond*,mark options={solid,fill=gray}] table [x=N, y=SNESSolve, col sep=comma]
        {code/data/poisson.Hermite.3.csv};
        \addlegendentry{Hermite}
        \addplot[dashdotted,mark=diamond*,mark options={solid,fill}] table [x=N, y=SNESSolve, col sep=comma]
        {code/data/poisson.Bell.5.csv};
        \addlegendentry{Bell}
        \addplot[dashdotted,mark=diamond, mark options={solid}] table [x=N, y=SNESSolve, col sep=comma]
        {code/data/poisson.Argyris.5.csv};
        \addlegendentry{Argyris}
      \end{loglogaxis}
      \path[use as bounding box] (current bounding box.south west)
      rectangle ($(current bounding box.north east) + (0, 1.2cm)$);
    \end{tikzpicture}
    \caption{Total solver time for Poisson's equation on $N \times N$
      mesh.  This includes element integration (and condensation),
      sparse matrix preallocation and assembly, and factorization.}
    \label{fig:poissonsolve-total}
  \end{subfigure}
  \caption{Time to solution for the Poisson equation using different
    discretisations.  We see that static condensation does not gain us
    anything for small problems, due to the overhead involved in
    forming the condensed system.}
  \label{fig:poissonsolve-combined}
\end{figure}
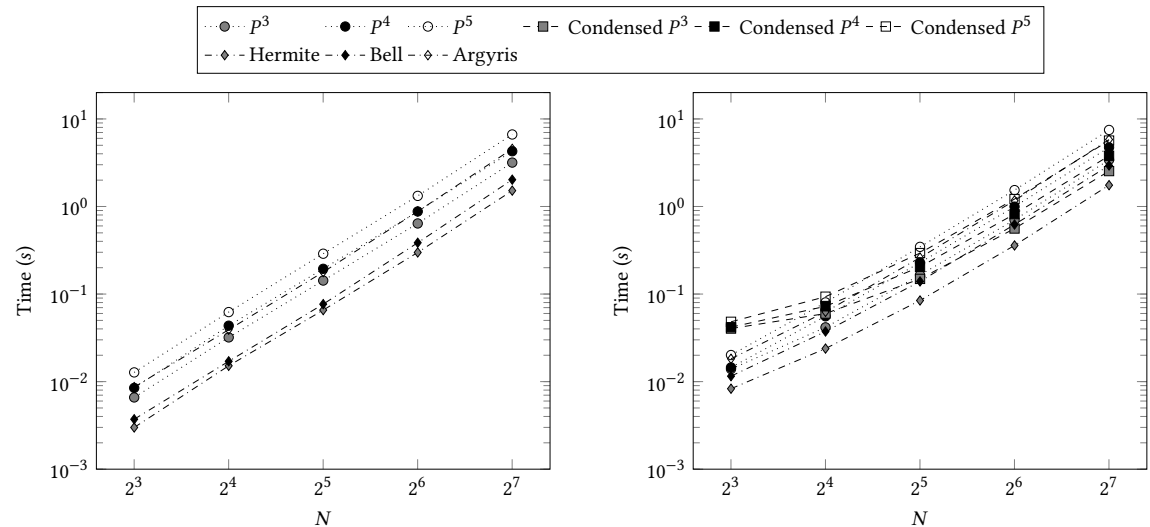

Having considered the cost of assembling the matrices and the
resulting sparsity patterns, we now present timings for solving the systems with
sparse LU factorization provided by PaStiX, plus iterative refinement.
These times are shown in \cref{fig:poissonsolve}.  For this
solver, it appears that Hermite and Bell systems are actually cheaper
to solve than any of the $P^k$ systems, and Argyris appears to have a
comparable cost to $P^4$ rather than $P^5$.
We note that these observations only on one sparse direct solver with
``out-of-the-box'' options and leave open the possibility that
reordering or other options for this or other sparse factorizations
could lead to different rankings among the solvers.
We also consider the effects of static condensation. In this case, the
full global stiffness matrix is never formed. In
\cref{fig:poissonsolve-total}, we compare the overall solver time for
statically condensed systems to the combined times of building and
solving the uncondensed systems. It appears that condensation improves
the total run-time for $P^k$ elements on the finer meshes, and we
would expect this trend to continue under additional mesh refinement.
In addition to the visualized sparsity patterns, we have also computed
the average number of nonzeros per row and condition number on a
coarse ($8 \times 8$) mesh for the Poisson operator, shown in
\cref{tab:poisson-sparsity-stats}. Notice how the conditioning of the
$H^2$ elements is significantly worse. In contrast, when applied to
the biharmonic operator, the picture is reversed
(\cref{tab:biharmonic-sparsity-stats}). The $H^2$ discretizations have
both better conditioning (Bell apart) and lower numbers of nonzeros
per row than an interior penalty discretization using $H^1$ elements.

\begin{table}[htbp]
  \begin{subfigure}[b]{0.475\textwidth}
    \centering
    \begin{filecontents*}{code/data/poisson-spy-cond.csv}
Name,Nrows,NNZ,NNZperRow,kappa
{$P^3$},625,10033,16.0528,448.7251001954773
{$P_c^3$},497,7601,15.293762575452716,277.4792752497565
Hermite,371,6905,18.61185983827493,3128.75043795687
{$P^4$},1089,24449,22.45087235996327,1290.6171244582542
{$P_c^4$},705,14081,19.97304964539007,635.9530173404564
Bell,486,17892,36.81481481481482,15571626.031014875
{$P^5$},1681,50161,29.839976204640095,3978.996672421671
{$P_c^5$},913,22513,24.65826944140197,2991.4294620175474
Argyris,694,28468,41.02017291066282,9583207.253466908
\end{filecontents*}
\pgfplotstabletypeset[col sep=comma,
    columns={Name,Nrows,NNZperRow,kappa},
    columns/Name/.style={string type, column name=Element, column type={r},
    },
    columns/Nrows/.style={column name=Total DoFs, column type={c},
    },
    columns/NNZperRow/.style={column name=Nonzeros/row, column type={c},
      precision=1, fixed zerofill,
    },
    columns/kappa/.style={column name={$\kappa$}, column type={c}, sci,precision=2,
    },
    every head row/.style={
      before row=\toprule,after row=\midrule},
    every nth row={3}{before row=\midrule[0.25pt]},
    every last row/.style={after row=\bottomrule}]{code/data/poisson-spy-cond.csv}

    \vspace{\baselineskip}
    \caption{Laplace operator, $P^k_c$ denotes statically condensed $P^k$.}
    \label{tab:poisson-sparsity-stats}
  \end{subfigure}
  \hspace{0.04\textwidth}
  \begin{subfigure}[b]{0.475\textwidth}
    \pgfplotstableset{col sep=comma}
    % Load data for both
    \pgfplotstablevertcat{\biharmonicstats}{code/data/biharmonic-h1-stats.csv}
    \pgfplotstablevertcat{\biharmonicstats}{code/data/biharmonic-h2-stats.csv}
    \centering
    \pgfplotstabletypeset[columns={Name,Nrows,NNZperRow,kappa},
    sort,sort key=sort order,
    % Reorder so that matching approximation degrees come next to each other.
    create on use/sort order/.style={
      create col/set list={1,3,5,2,4,6},
    },
    columns/Name/.style={string type, column name=Element, column type={r},
    },
    columns/Nrows/.style={column name=Total DoFs, column type={c},
    },
    columns/NNZperRow/.style={column name=Nonzeros/row, column type={c},
      precision=1, fixed zerofill,
    },
    columns/kappa/.style={column name={$\kappa$}, column type={c}, sci,precision=2,
    },
    every head row/.style={
      before row=\toprule,after row=\midrule},
    every nth row={2}{before row=\midrule[0.25pt]},
    every last row/.style={after row=\bottomrule}]\biharmonicstats

    \vspace{\baselineskip}
    \caption{Biharmonic operator.}
    \label{tab:biharmonic-sparsity-stats}
  \end{subfigure}
  \caption{Sparsity and conditioning information for $H^1$- and
    $H^2$-conforming discretizations of the Laplace and biharmonic
    operators on a coarse $8\times8$ mesh.}
  \label{tab:sparsity-stats}
\end{table}

\begin{figure}[htbp]
  \begin{subfigure}[t]{0.475\textwidth}
    \begin{tikzpicture}[scale=0.88]
      \begin{loglogaxis}[xlabel={$N$}, ylabel={Time ($s$)},
        log basis x=2,
        ylabel near ticks,
        ymin=1e-3, ymax=1e2]
        \addplot[dotted,mark=otimes,mark options={solid,fill=gray}] table [x=N, y=KSPSolve, col sep=comma]
        {code/data/biharmonicdg.Lagrange.2.csv};
        % \addlegendentry{$P^2$}
        \addplot[dotted,mark=*,mark options={solid,fill=gray}] table [x=N, y=KSPSolve, col sep=comma]
        {code/data/biharmonicdg.Lagrange.3.csv};
        % \addlegendentry{$P^3$}
        \addplot[dotted,mark=*,mark options={solid,fill}] table [x=N, y=KSPSolve, col sep=comma]
        {code/data/biharmonicdg.Lagrange.4.csv};
        % \addlegendentry{$P^4$}
        \addplot[dotted,mark=o, mark options={solid}] table [x=N, y=KSPSolve, col sep=comma]
        {code/data/biharmonicdg.Lagrange.5.csv};
        % \addlegendentry{$P^5$}
        \addplot[dashdotted,mark=diamond*,mark options={solid,fill=gray}] table [x=N, y=KSPSolve, col sep=comma]
        {code/data/biharmonic.Morley.2.csv};
        % \addlegendentry{Morley}
        \addplot[dashdotted,mark=diamond*,mark options={solid,fill}] table [x=N, y=KSPSolve, col sep=comma]
        {code/data/biharmonic.Bell.5.csv};
        % \addlegendentry{Bell}
        \addplot[dashdotted,mark=diamond, mark options={solid}] table [x=N, y=KSPSolve, col sep=comma]
        {code/data/biharmonic.Argyris.5.csv};
        % \addlegendentry{Argyris}
      \end{loglogaxis}
      \path[use as bounding box] (current bounding box.south west)
      rectangle ($(current bounding box.north east) + (0, 0.7cm)$);
    \end{tikzpicture}
    \caption{Time to factor and solve the sparse linear system for the
      discrete biharmonic operator on perturbed $N \times N$ mesh,
      including one iteration of iterative refinement.}
    \label{fig:biharmsolve}
  \end{subfigure}
  \hspace{0.04\textwidth}
  \begin{subfigure}[t]{0.475\textwidth}
    \begin{tikzpicture}[scale=0.88]
      \begin{loglogaxis}[xlabel={$N$}, ylabel={Time ($s$)},
        log basis x=2,
        ylabel near ticks,
        ymin=1e-3, ymax=1e2,
        legend cell align=left,
        legend style={overlay, at={(-0.16, 1.05)}, anchor=south},
        legend columns=7,]
        \addplot[dotted,mark=otimes,mark options={solid,fill=gray}] table [x=N, y=SNESSolve, col sep=comma]
        {code/data/biharmonicdg.Lagrange.2.csv};
        \addlegendentry{$P^2$}
        \addplot[dotted,mark=*,mark options={solid,fill=gray}] table [x=N, y=SNESSolve, col sep=comma]
        {code/data/biharmonicdg.Lagrange.3.csv};
        \addlegendentry{$P^3$}
        \addplot[dotted,mark=*,mark options={solid,fill}] table [x=N, y=SNESSolve, col sep=comma]
        {code/data/biharmonicdg.Lagrange.4.csv};
        \addlegendentry{$P^4$}
        \addplot[dotted,mark=o, mark options={solid}] table [x=N, y=SNESSolve, col sep=comma]
        {code/data/biharmonicdg.Lagrange.5.csv};
        \addlegendentry{$P^5$}
        \addplot[dashdotted,mark=diamond*,mark options={solid,fill=gray}] table [x=N, y=SNESSolve, col sep=comma]
        {code/data/biharmonic.Morley.2.csv};
        \addlegendentry{Morley}
        \addplot[dashdotted,mark=diamond*,mark options={solid,fill}] table [x=N, y=SNESSolve, col sep=comma]
        {code/data/biharmonic.Bell.5.csv};
        \addlegendentry{Bell}
        \addplot[dashdotted,mark=diamond, mark options={solid}] table [x=N, y=SNESSolve, col sep=comma]
        {code/data/biharmonic.Argyris.5.csv};
        \addlegendentry{Argyris}
      \end{loglogaxis}
      \path[use as bounding box] (current bounding box.south west)
      rectangle ($(current bounding box.north east) + (0, 0.7cm)$);
    \end{tikzpicture}
    \caption{Total solver time for the biharmonic equation on $N
      \times N$ mesh.  This includes element integration, sparse
      matrix preallocation and assembly, and factorization.}
    \label{fig:biharmsolve-total}
  \end{subfigure}
  \caption{Time to solution for the biharmonic equation.  We see that
    the smaller, sparser $H^2$ discretizations give a significant
    advantage over the equivalent $H^1$ interior penalty scheme.}
  \label{fig:biharmsolve-combined}
\end{figure}
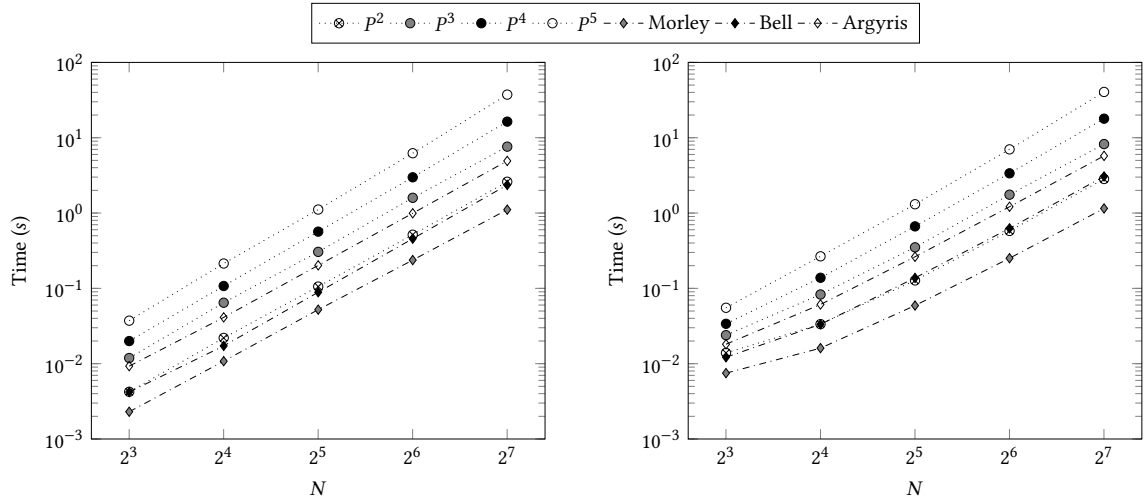

For the biharmonic operator, we have seen several factors
(lower local FLOP count, smaller system, lower bandwidth) that should
contribute to making sparse direct methods far more efficient for the
$H^2$ discretizations than for interior penalty methods.
\cref{fig:biharmsolve} confirms this hypothesis, with solution
of the Argyris element typically being cheaper than even the $P^3$
interior penalty method, not to mention $P^5$.

\subsection{Some advanced examples}
\subsubsection{The Cahn-Hilliard equation}
\label{sec:cahn-hilliard}
In this section, we demonstrate the use of these elements on further
problems.  First, we consider the two-dimensional Cahn-Hilliard
equation, a model of phase separation in binary fluids.  Here, the
chemical concentration $c$ satisfies the fourth-order time-dependent
equation
\begin{equation}
  \frac{\partial c}{\partial t} - \nabla \cdot M \left(\nabla\left(\frac{d f}{d c}
          - \lambda \nabla^{2}c\right)\right) = 0 \quad {\rm in}
          \ \Omega.
\end{equation}
$f$ is some typically non-convex function (in our case, we take $f(c)
= 100c^2(1-c)^2$, and $\lambda$ and $M$ are
scalar parameters controlling rates.  Although $M=M(c)$ (the so-called
degenerate mobility case) is possible, we have considered just
constant $M$ in our examples.
The system is closed with the boundary conditions
\begin{align}
M\left(\nabla\left(\frac{d f}{d c} - \lambda \nabla^{2}c\right)\right) \cdot n &= 0 \quad {\rm on} \ \partial\Omega, \\
M \lambda \nabla c \cdot n &= 0 \quad {\rm on} \ \partial\Omega.
\end{align}

The Cahn-Hilliard equation is often re-rewritten into a system of two
second-order equations~\citep{barrett1999finite}.  This eliminates the
need for a $C^1$ discretization, but it also increases the number of
unknowns and introduces a saddle point into the linear system.
Alternatively, one maintains a primal form via an interior penalty
method~\citep{wells2006discontinuous}, although we have seen above that
this leads to larger and less favorable linear systems for fourth-order operators.

\begin{figure}[htbp]
  \centering
\begin{lstlisting}
from firedrake import *
import numpy
mesh = UnitSquareMesh(16, 16)
# Set parameters
lmbda = 1e-2
delta_t = 5e-6
theta = 0.5
M = 1
beta = 250
# Pick function space
V = FunctionSpace(mesh, "Bell", 5)
c = Function(V)
c0 = Function(V)
# Initial conditions
c0.vector()[::6] = 0.63 + 0.2*(0.5 - numpy.random.random(c0.vector().local_size() // 6))
c.assign(c0)

c_theta = theta*c + (1 - theta)*c0
dfdc = 200*(c_theta*(1 - c_theta)**2 - c_theta**2*(1 - c_theta))
n = FacetNormal(mesh)
h = CellSize(mesh)
# Nonlinear residual
v = TestFunction(V)
F = (inner(c - c0, v)*dx +
     delta_t*(inner(M*grad(dfdc), grad(v))*dx +
              inner(M*lmbda*div(grad(c_theta)), div(grad(v)))*dx -
              inner(M*lmbda*div(grad(c_theta)), dot(grad(v), n))*ds -
              inner(M*lmbda*dot(grad(c_theta), n), div(grad(v)))*ds +
              inner((beta/h)*M*lmbda*dot(grad(c_theta), n), dot(grad(v), n))*ds))
problem = NonlinearVariationalProblem(F, c)
solver = NonlinearVariationalSolver(problem, solver_parameters={"ksp_type": "preonly",
                                                                "pc_type": "lu"})
t = 0
T = 0.0025
while t < T:
    solver.solve()
    c0.assign(c)
    t += delta_t
\end{lstlisting}
  \caption{Code for solving the Cahn-Hilliard equation with Bell elements.}
  \label{fig:cahn-hilliard-code}
\end{figure}
\begin{figure}[htbp]
  \begin{subfigure}[t]{0.4\textwidth}
    \centering
    \includegraphics[height=4cm]{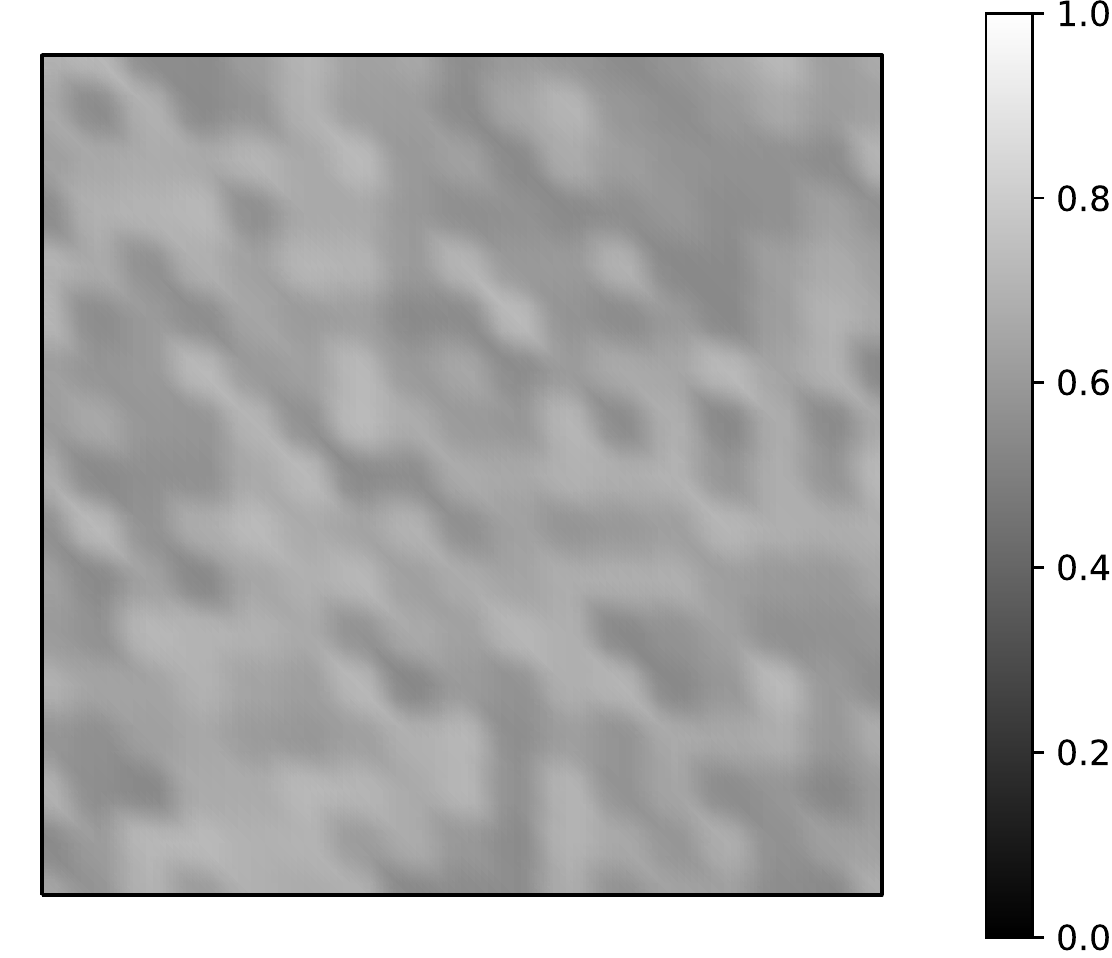}
  \caption{Initial condition, consisting of small perturbations of the concentration around $c=0.63$.}
  \label{fig:chic}
  \end{subfigure}
  \hspace{1em}
  \begin{subfigure}[t]{0.4\textwidth}
    \centering
    \includegraphics[height=4cm]{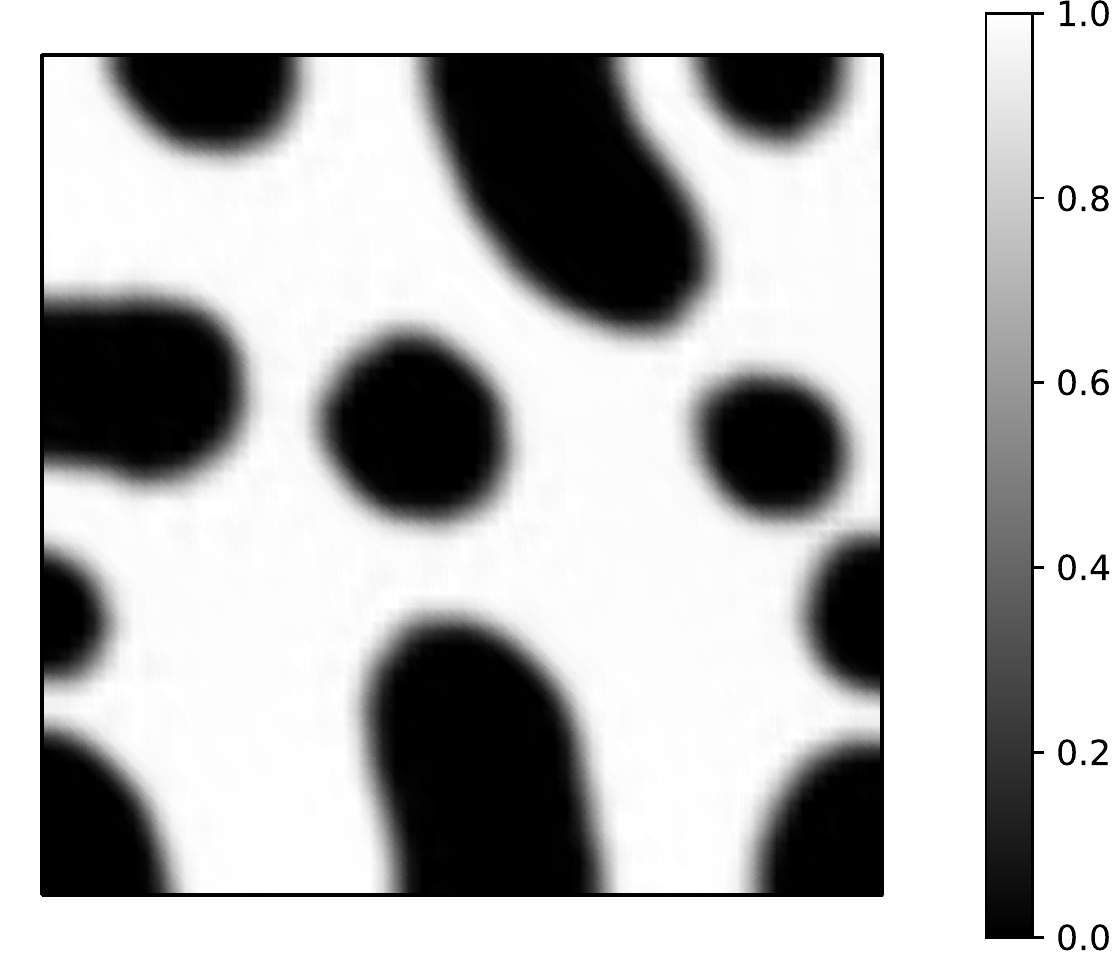}
    \caption{Final values of the concentration at time $t=0.25$, demonstrating phase separation.}
  \label{fig:chfinal}
  \end{subfigure}
  \label{fig:ch}
  \caption{Snapshots of the concentration field for the Cahn-Hilliard simulation, using the code of \cref{fig:cahn-hilliard-code}.}
\end{figure}

In our simulations, we use a $16 \times 16$ mesh, but with
higher-order elements, this leads to reasonable resolution.  In our
example, we use Bell elements, but it is simple to change to Argyris
or the lower-order nonconforming Morley element -- one merely changes
the requested element in the \lstinline[basicstyle=\normalsize\ttfamily]{FunctionSpace} constructor.  The
complete code for simulating this problem, with the exception of
output routines, is shown in \cref{fig:cahn-hilliard-code}.
Using Crank-Nicolson time stepping with $\Delta t=5 \times 10^{-6}$,
we obtain the final state depicted in \cref{fig:chfinal}.  We remark
that visualization for higher-order polynomials is performed by
interpolating the solution onto piecewise linears on a refined mesh.

\subsubsection{Chladni plates}
\label{sec:chladni-plates}
As a final example, we consider the Chladni plate problem, which dates
back to the eighteenth century.  Chladni, a musician and
physicist, discovered beautiful patterns appearing when a metal plate
covered with dust or sand was excited with a violin bow.  These
patterns, now called \emph{Chladni figures}, changed dramatically as a
function of pitch.  After work by
\citet{germain1821recherches,germain1826remarques},
\citet{kirchoff1850uber} showed Chaldni's patterns were in fact
eigenpairs of the biharmonic operator under free boundary conditions
and was able to give a solution on circular plates where symmetries
offer simplification.  Realizing this connection and effectively
computing the figures in other geometries were historically different
matters, and it was not until 1909 when \citet{ritz1909theorie} was
able to give the first computation of Chladni figures for square
plates.  For many more details, a beautiful historical overview of the
computation of Chladni figures is given by \citet{gander2012ritz}.

\begin{figure}[htbp]
  \centering
  \begin{subfigure}[t]{0.12\textwidth}
    \hfill
  \end{subfigure}
  \begin{subfigure}[t]{0.12\textwidth}
    \hfill
  \end{subfigure}
  \begin{subfigure}[t]{0.12\textwidth}
    \includegraphics[width=\textwidth]{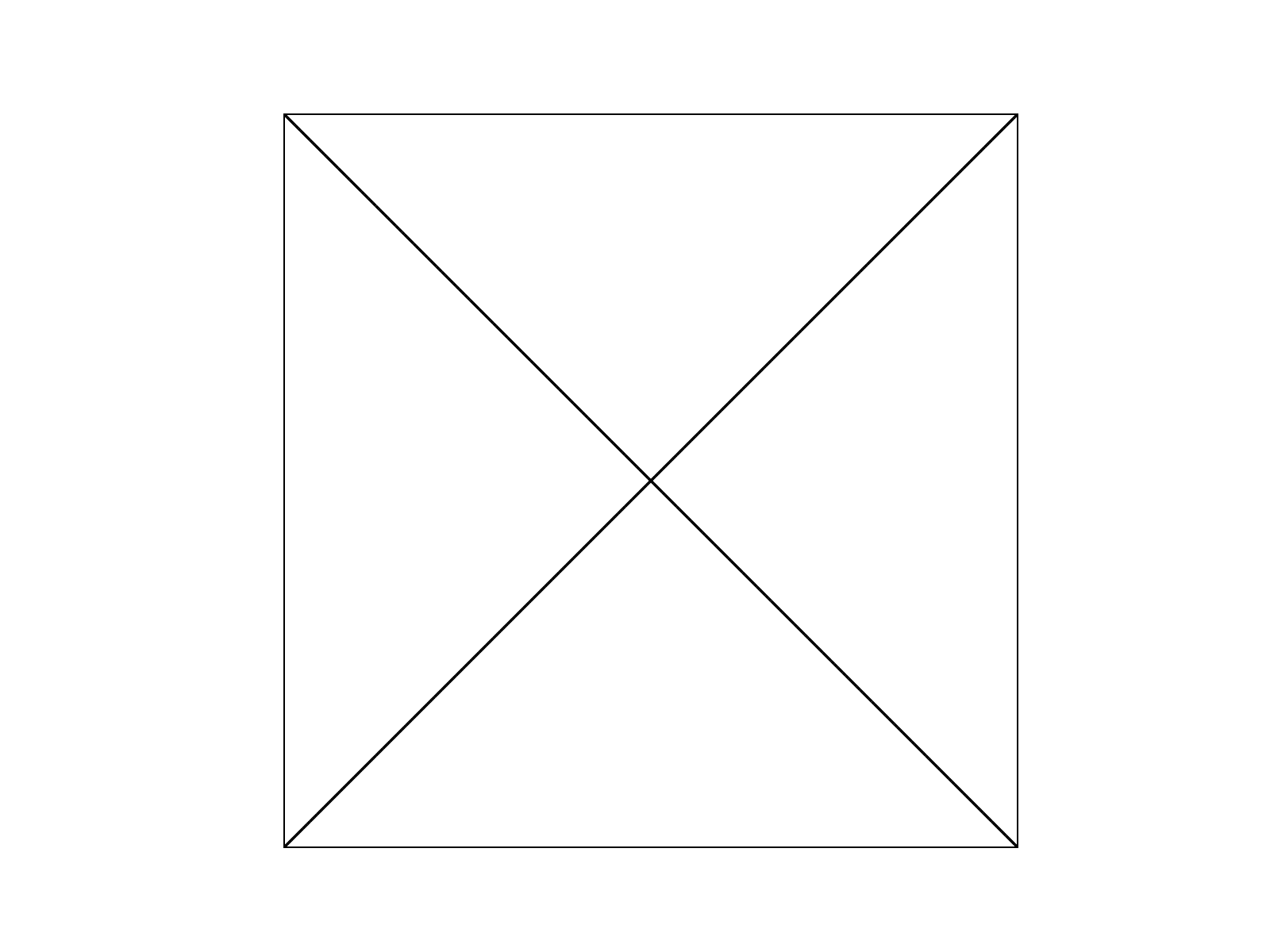}
  \end{subfigure}
  \begin{subfigure}[t]{0.12\textwidth}
    \includegraphics[width=\textwidth]{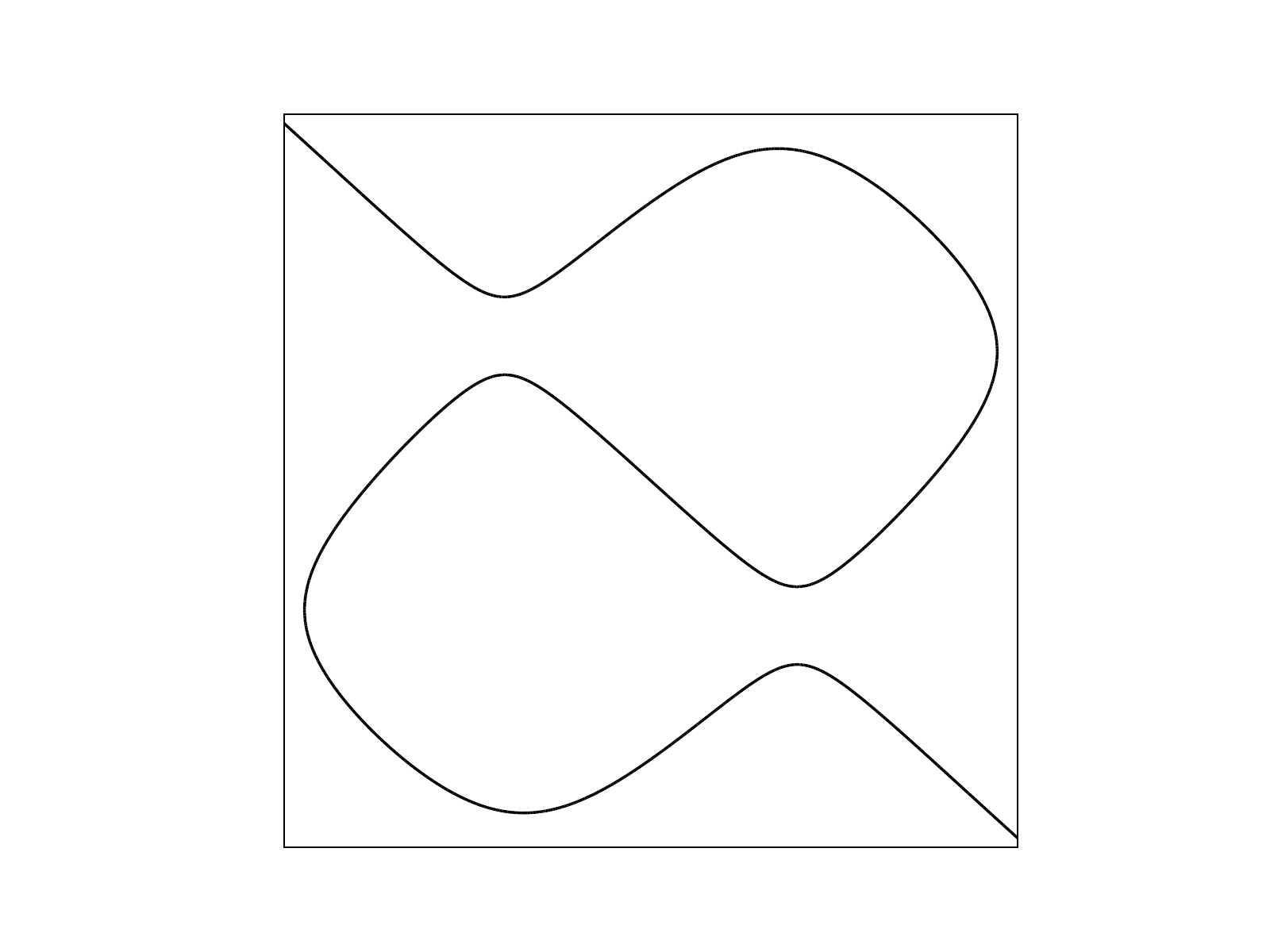}
  \end{subfigure}
  \begin{subfigure}[t]{0.12\textwidth}
    \includegraphics[width=\textwidth]{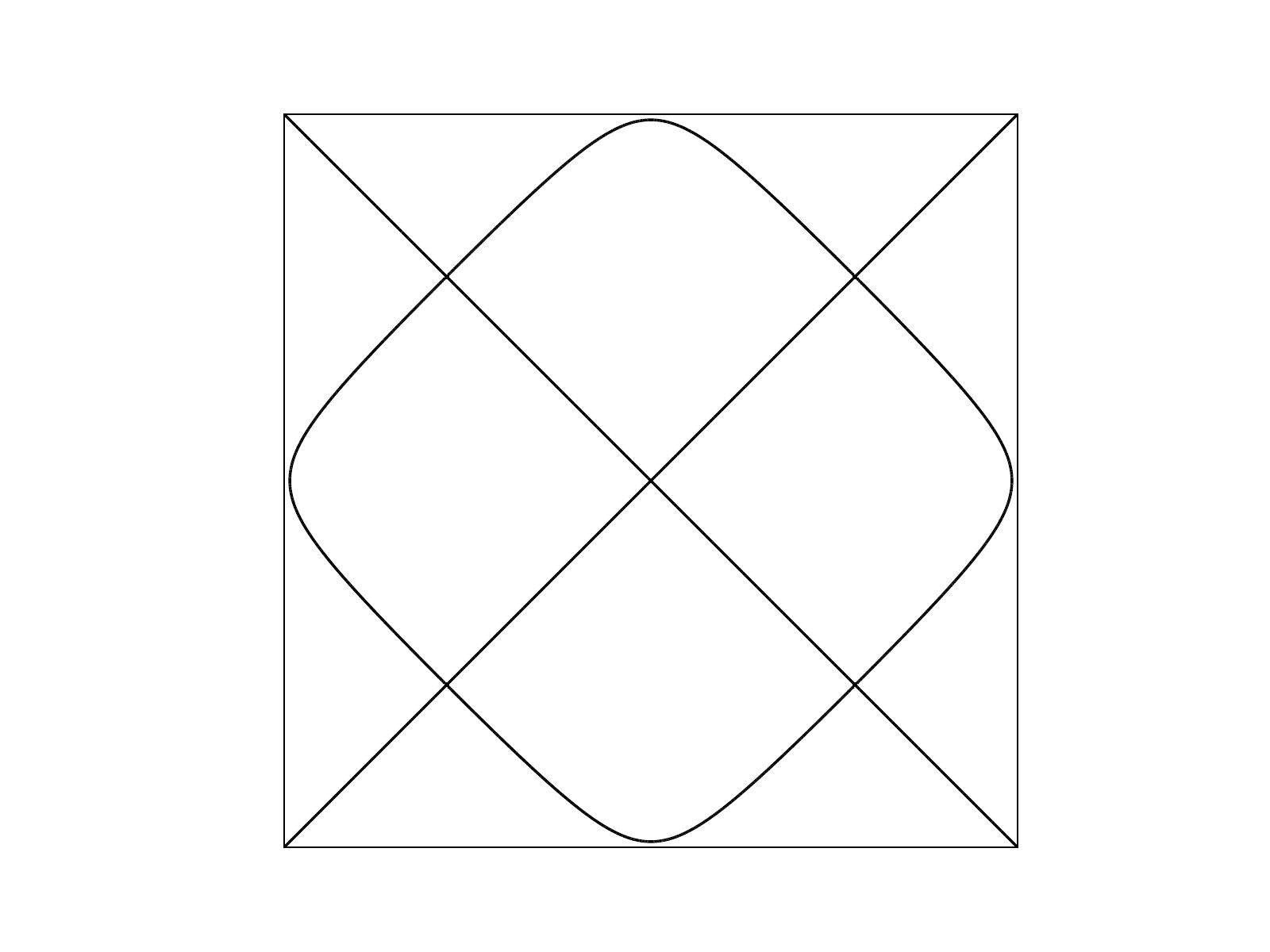}
  \end{subfigure}
  \\
  \begin{subfigure}[t]{0.12\textwidth}
    \hfill
  \end{subfigure}
  \begin{subfigure}[t]{0.12\textwidth}
    \includegraphics[width=\textwidth]{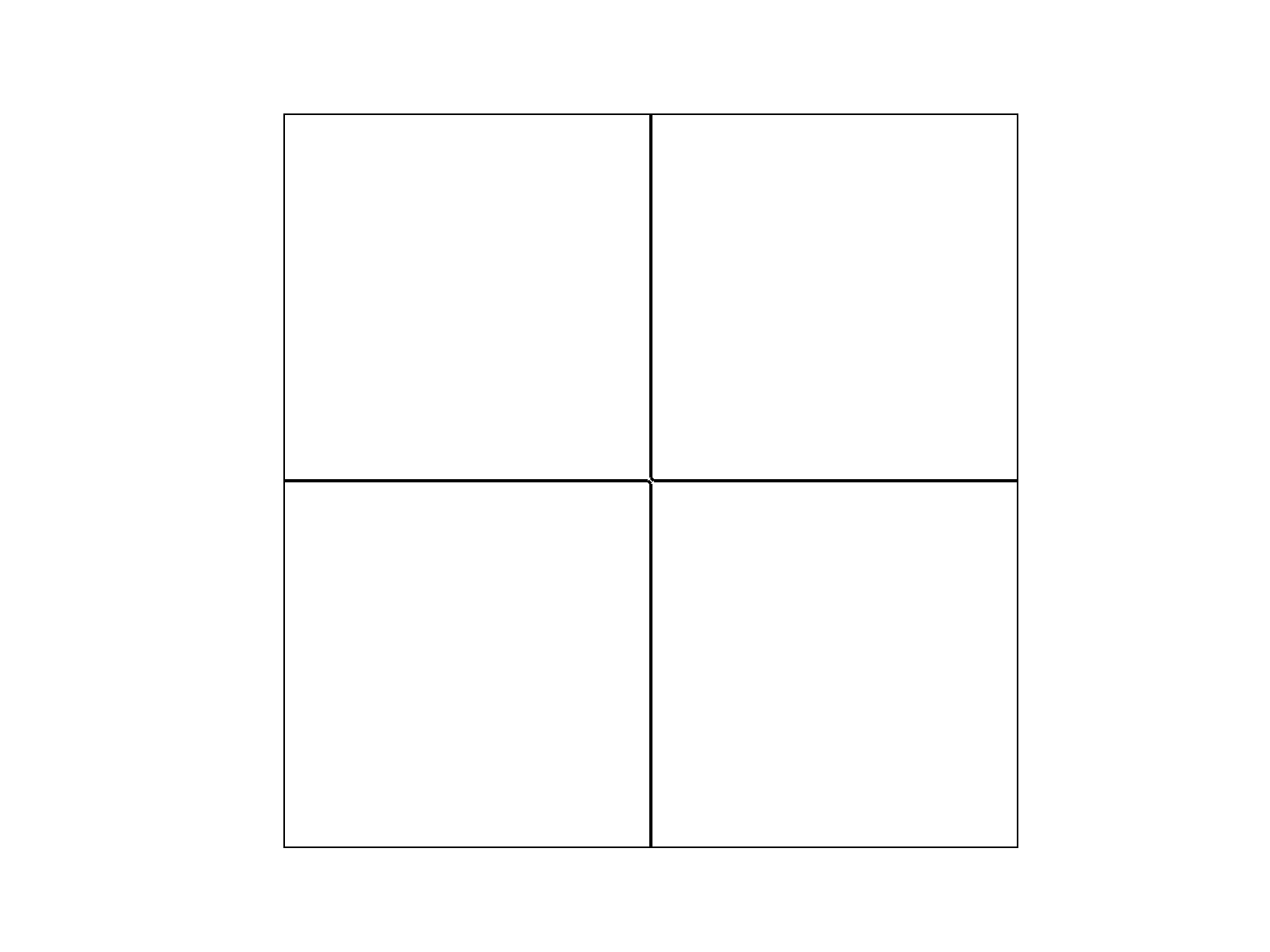}
  \end{subfigure}
  \begin{subfigure}[t]{0.12\textwidth}
    \includegraphics[width=\textwidth]{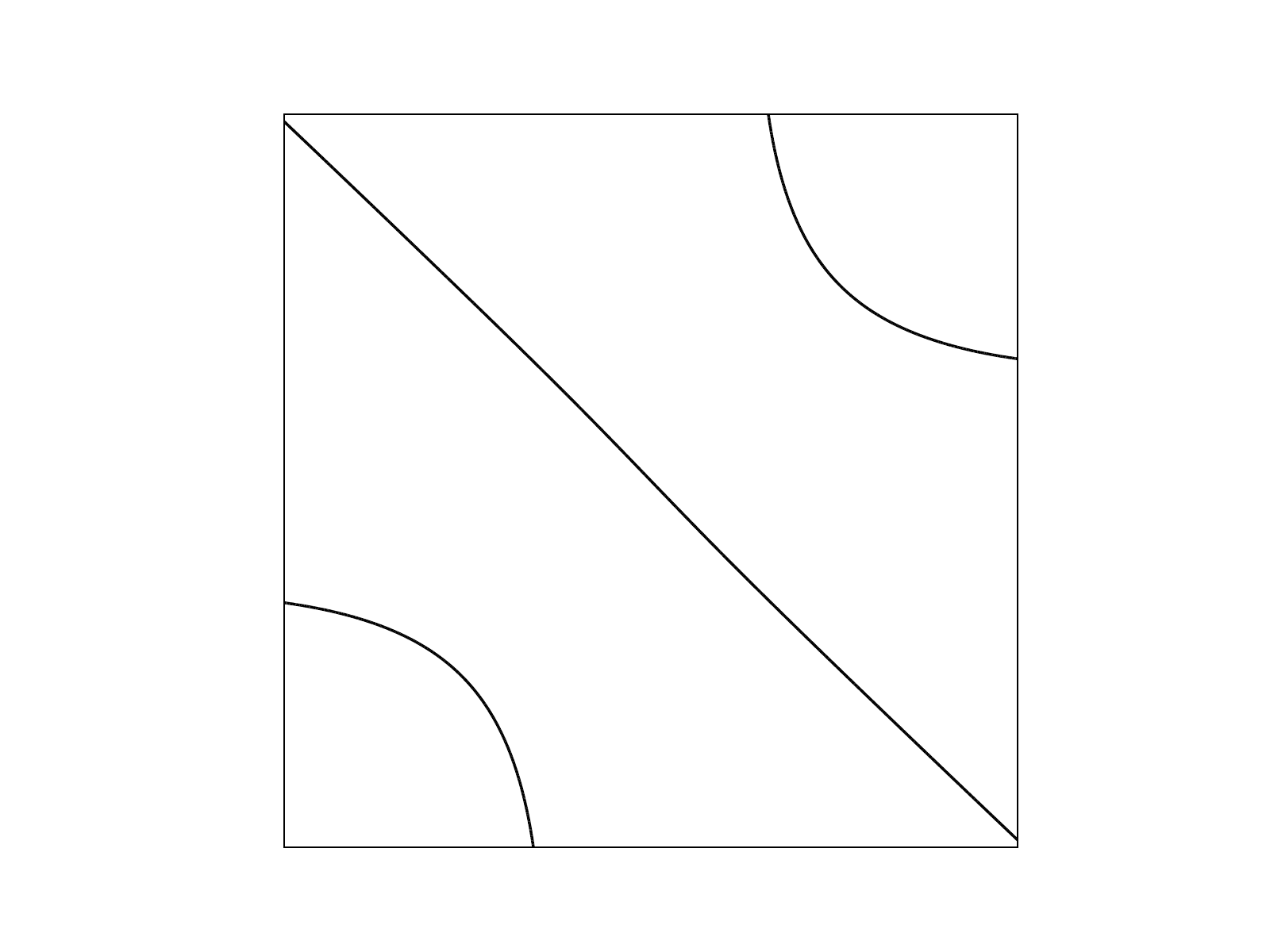}
  \end{subfigure}
  \begin{subfigure}[t]{0.12\textwidth}
    \includegraphics[width=\textwidth]{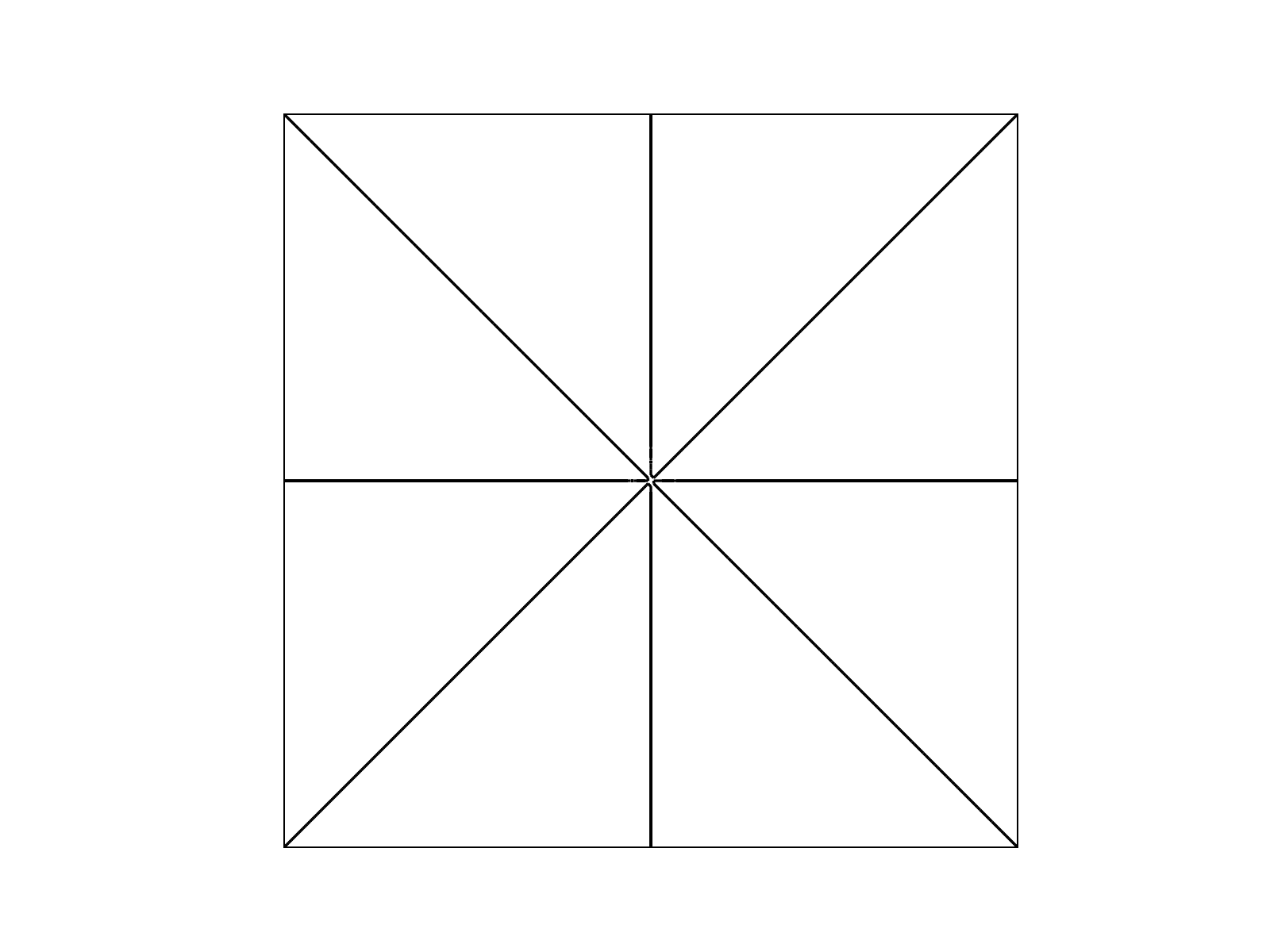}
  \end{subfigure}
  \begin{subfigure}[t]{0.12\textwidth}
    \includegraphics[width=\textwidth]{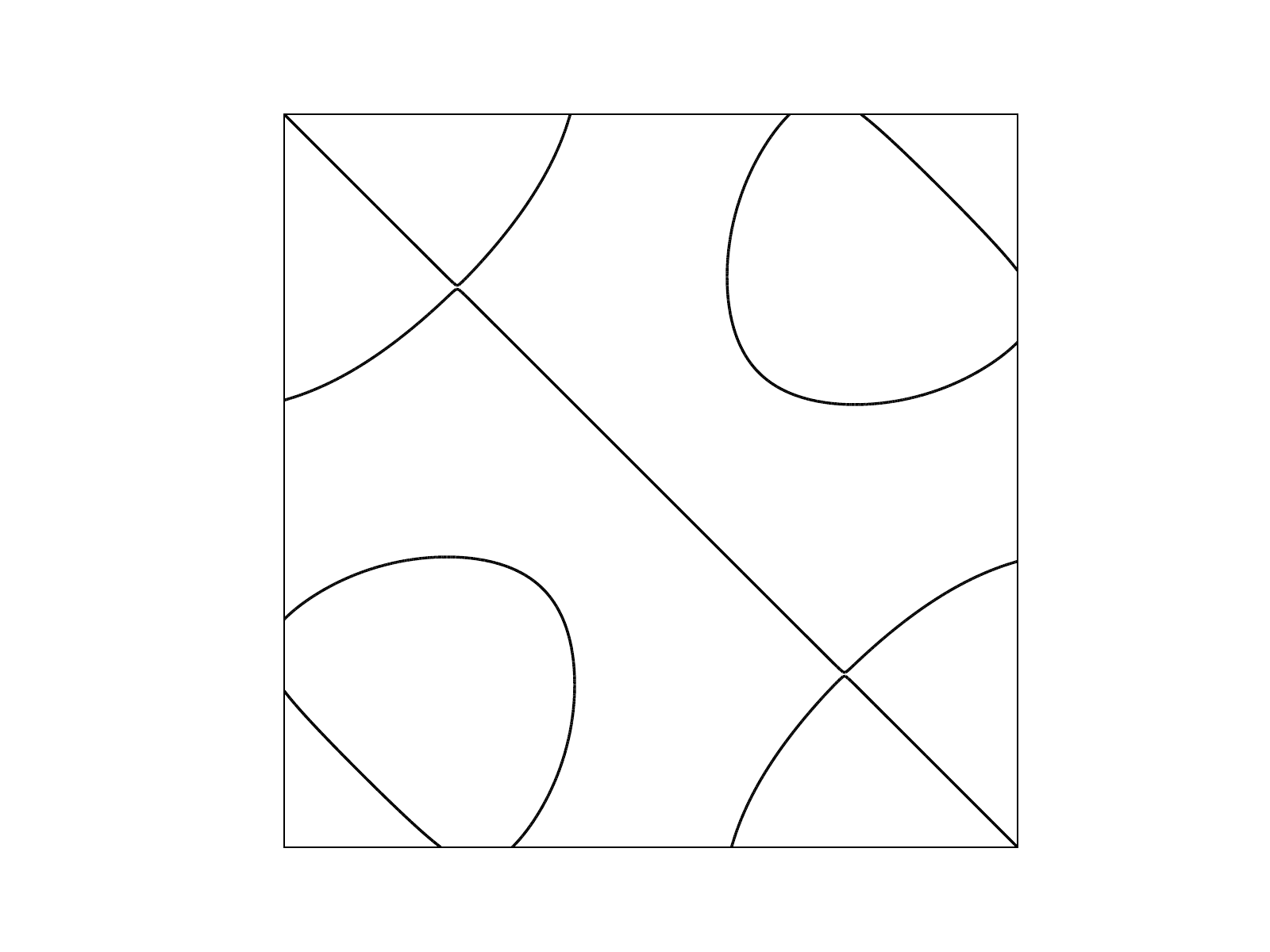}
  \end{subfigure}
  \\
  \begin{subfigure}[t]{0.12\textwidth}
    \includegraphics[width=\textwidth]{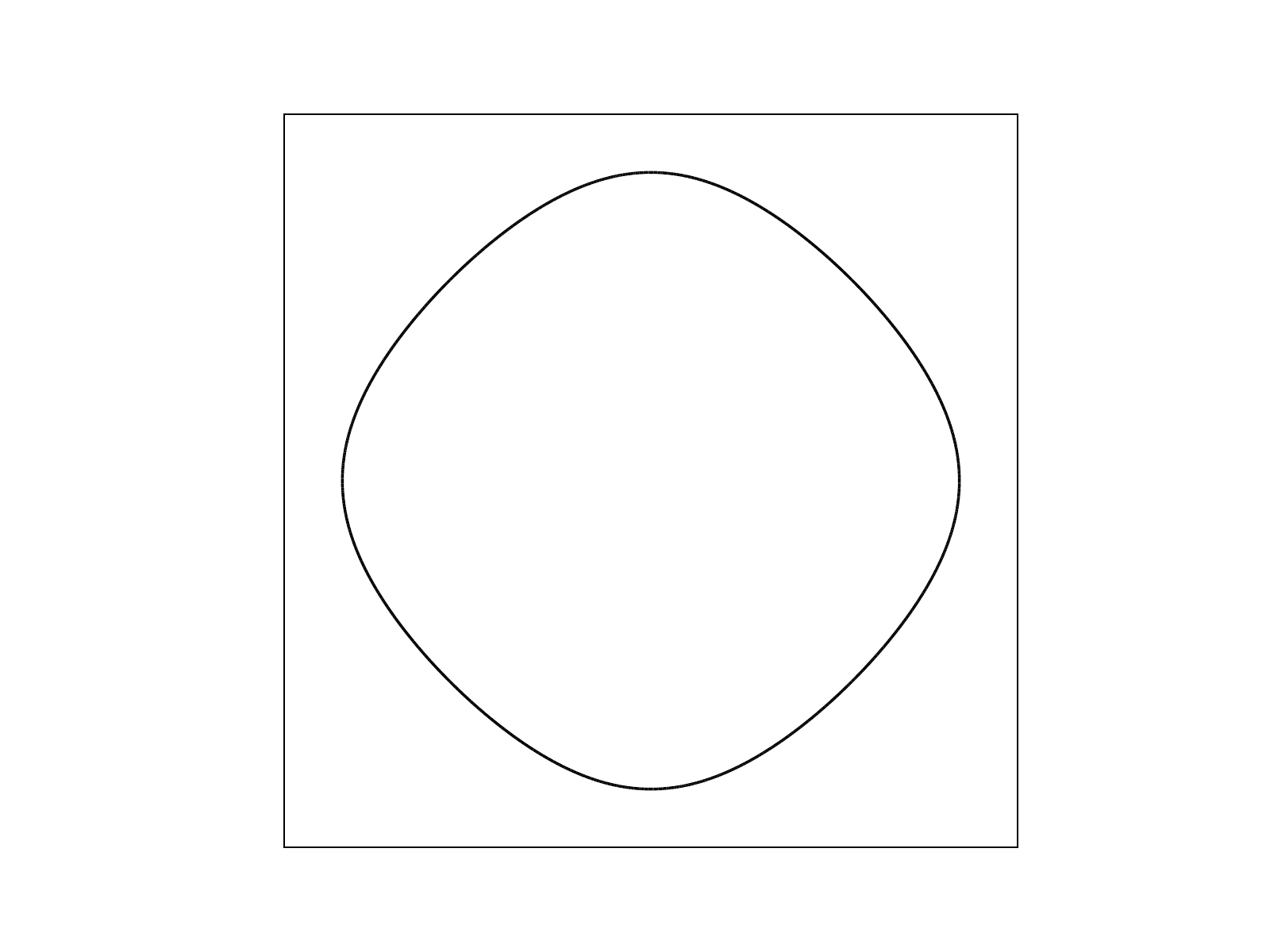}
  \end{subfigure}
  \begin{subfigure}[t]{0.12\textwidth}
    \includegraphics[width=\textwidth]{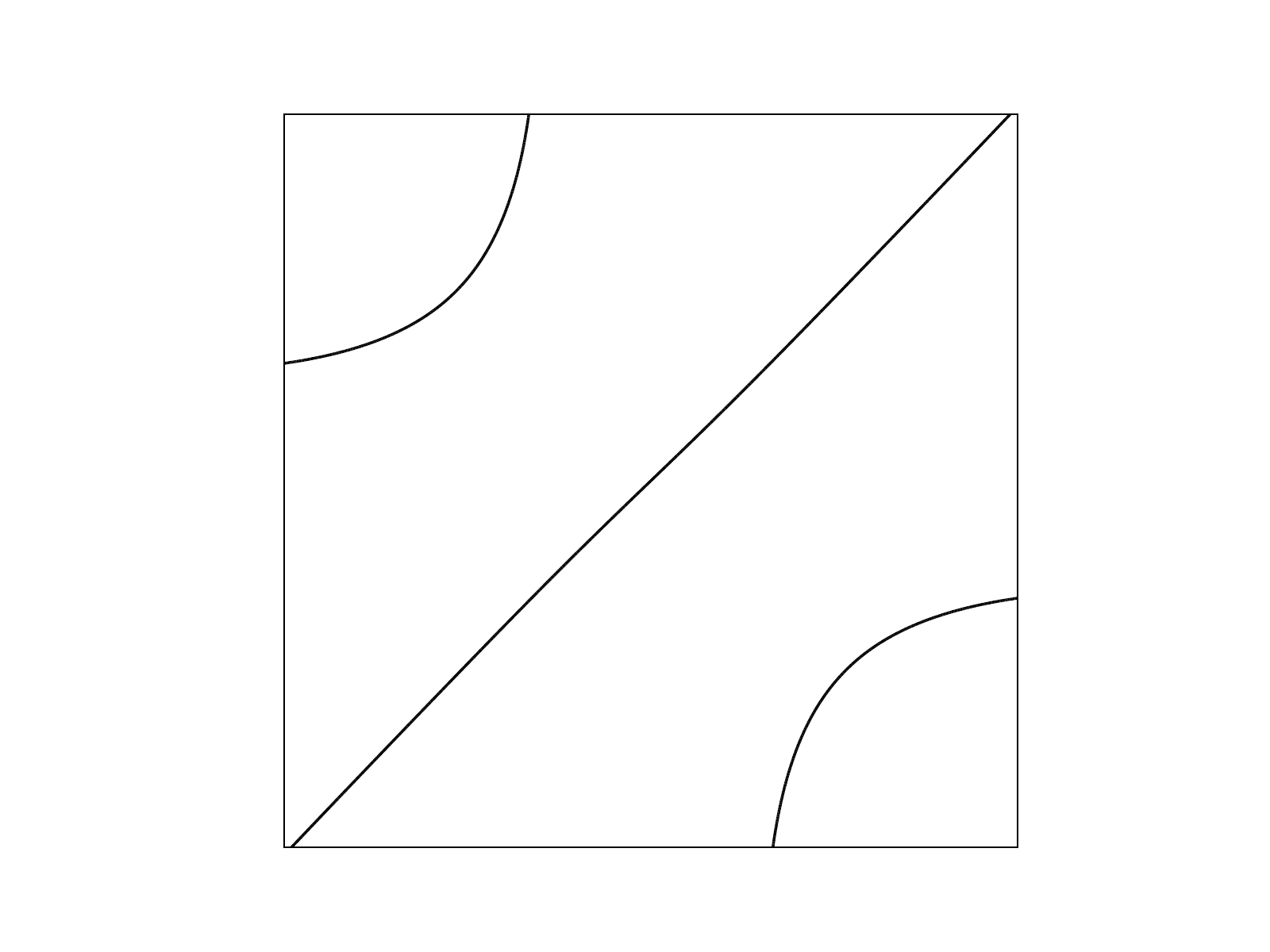}
  \end{subfigure}
  \begin{subfigure}[t]{0.12\textwidth}
    \includegraphics[width=\textwidth]{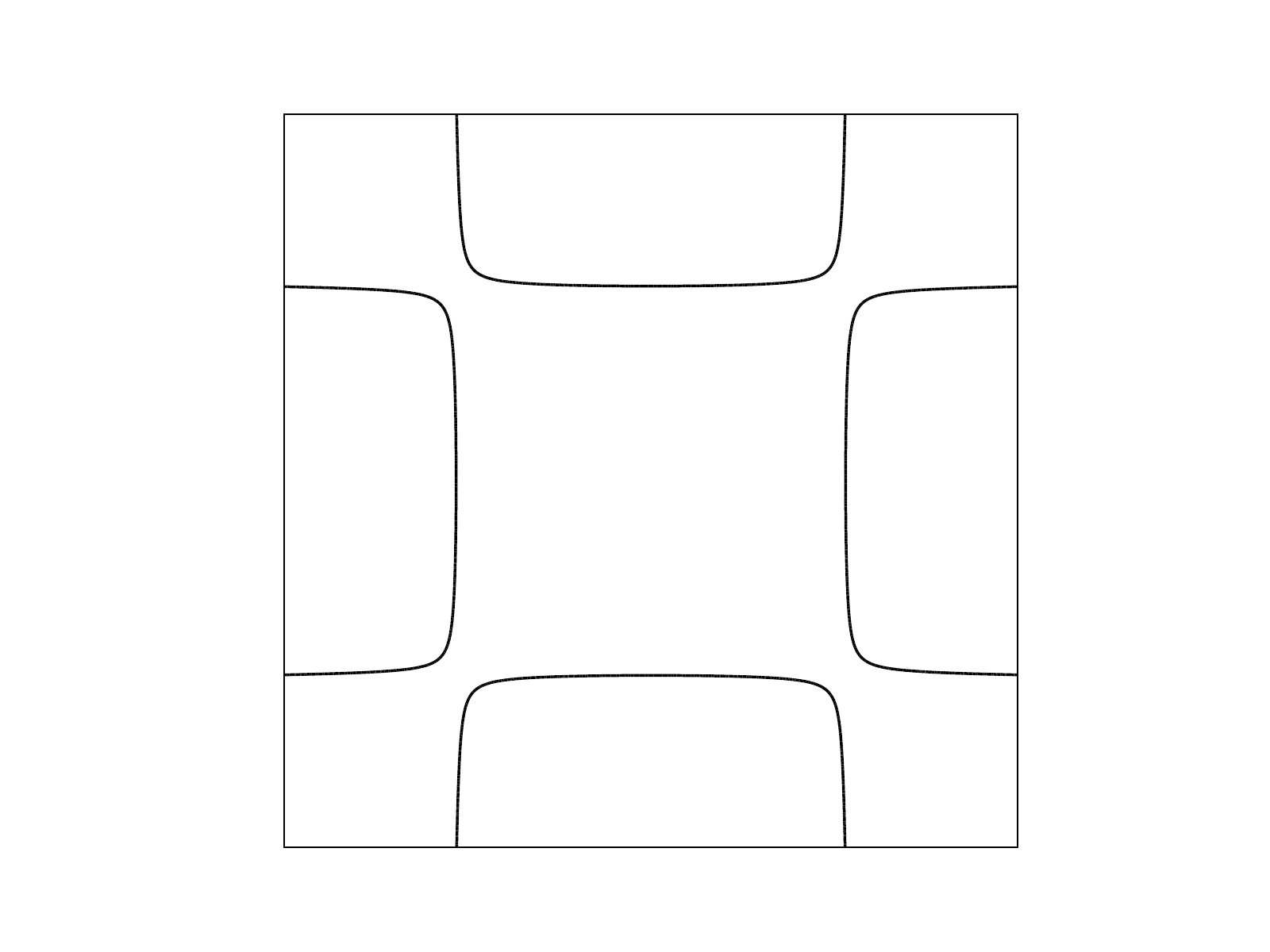}
  \end{subfigure}
  \begin{subfigure}[t]{0.12\textwidth}
    \includegraphics[width=\textwidth]{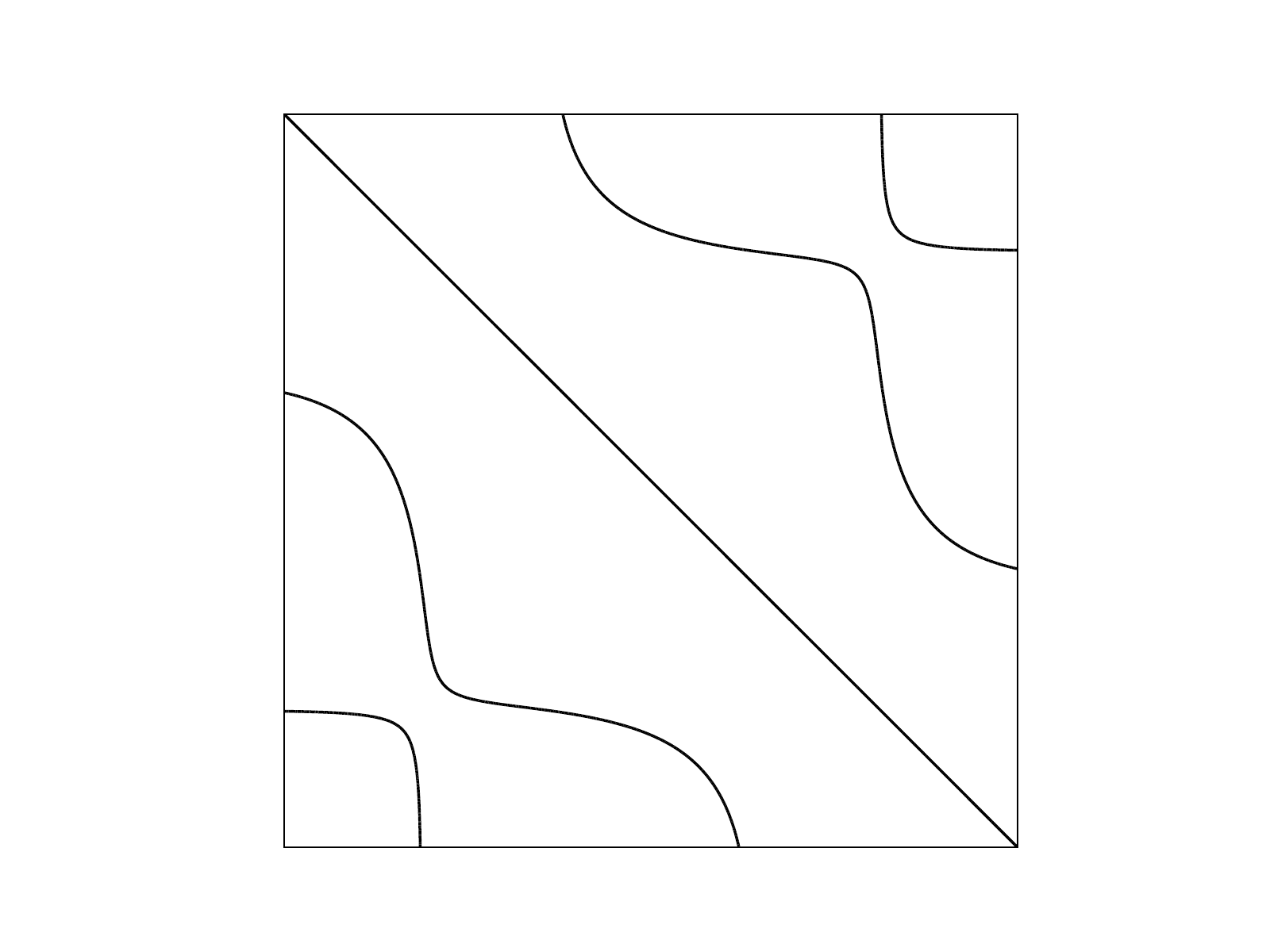}
  \end{subfigure}
  \begin{subfigure}[t]{0.12\textwidth}
    \includegraphics[width=\textwidth]{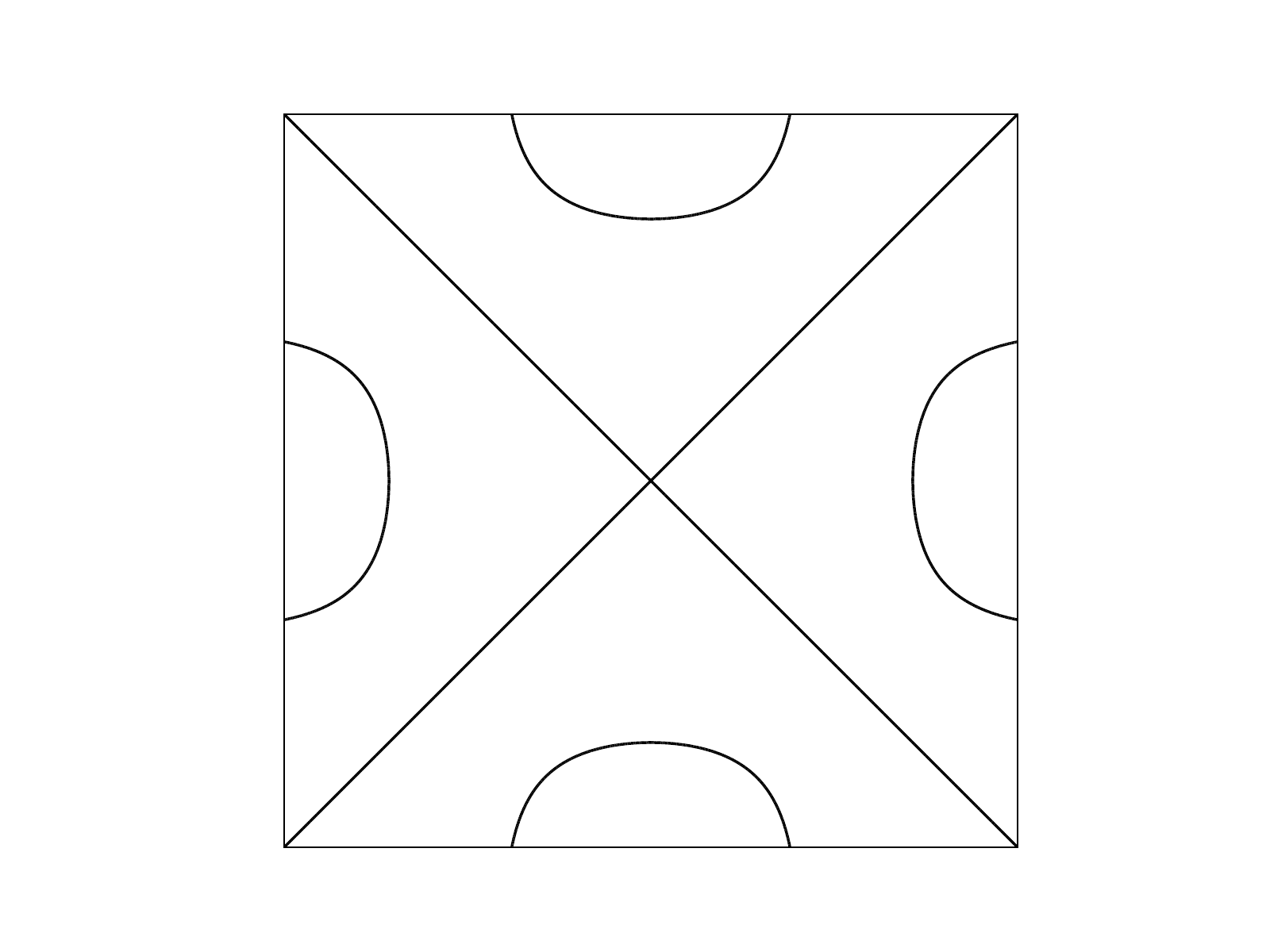}
  \end{subfigure}
  \\
  \begin{subfigure}[t]{0.12\textwidth}
    \includegraphics[width=\textwidth]{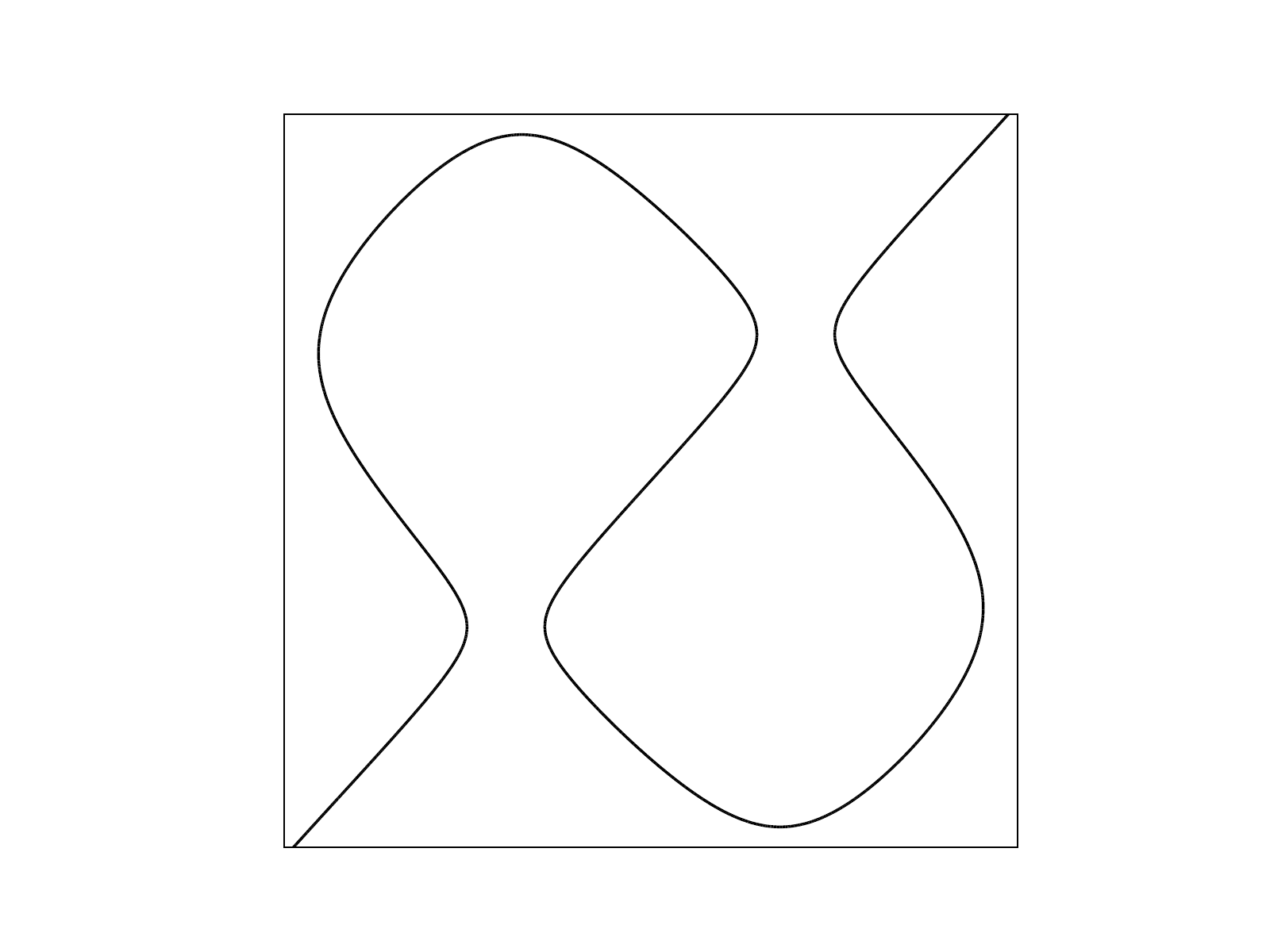}
  \end{subfigure}
  \begin{subfigure}[t]{0.12\textwidth}
    \includegraphics[width=\textwidth]{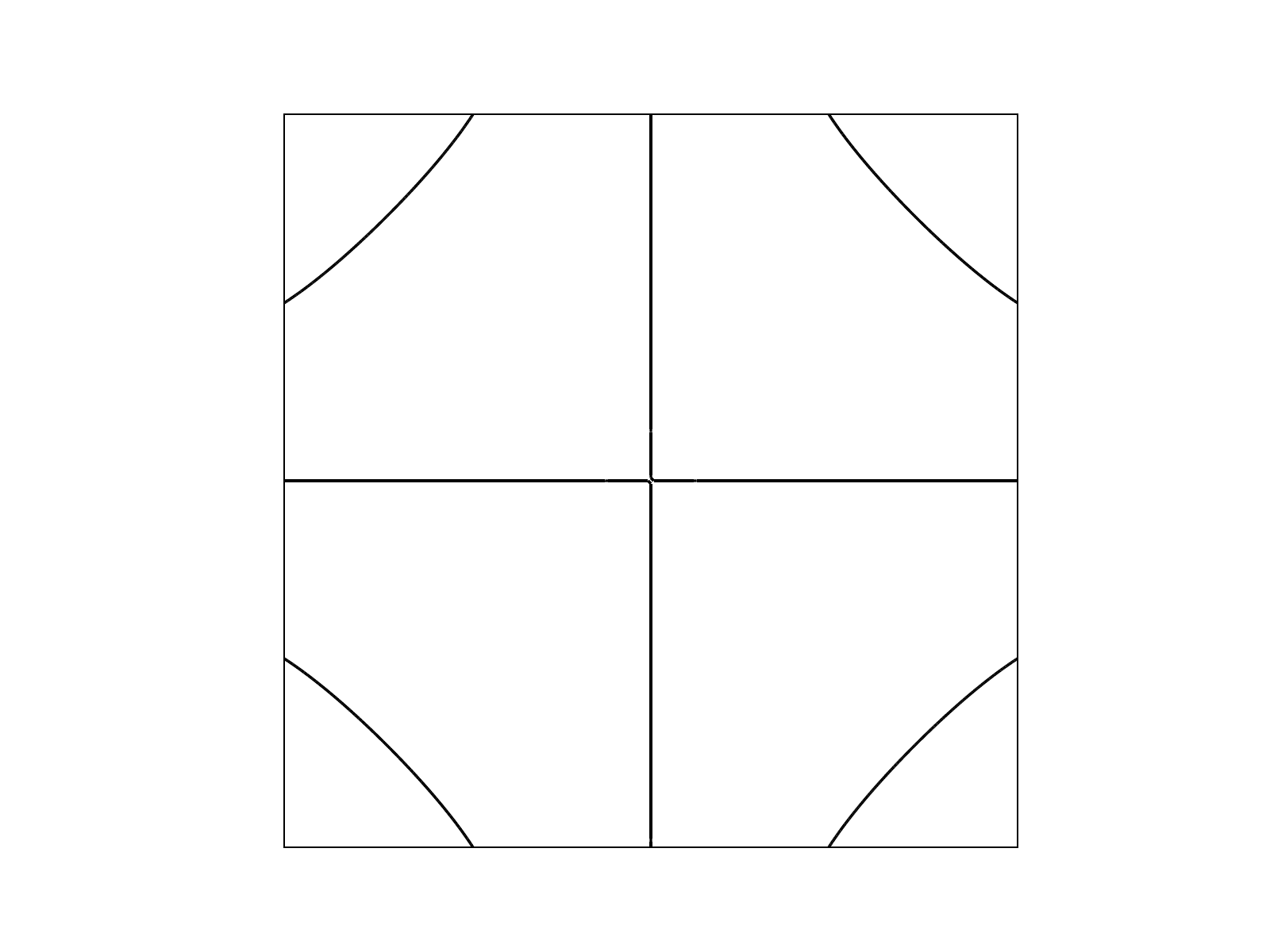}
  \end{subfigure}
  \begin{subfigure}[t]{0.12\textwidth}
    \includegraphics[width=\textwidth]{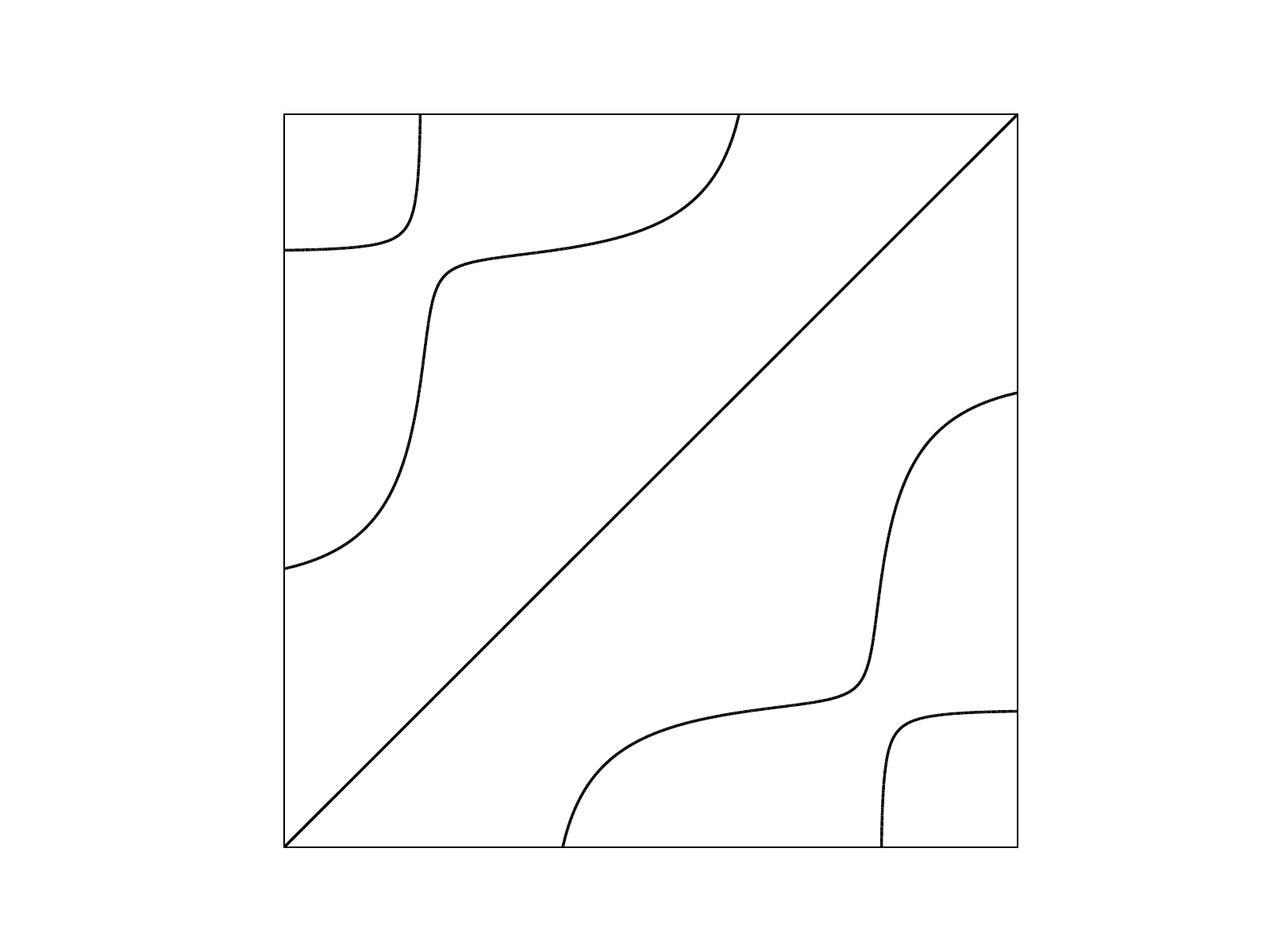}
  \end{subfigure}
  \begin{subfigure}[t]{0.12\textwidth}
    \includegraphics[width=\textwidth]{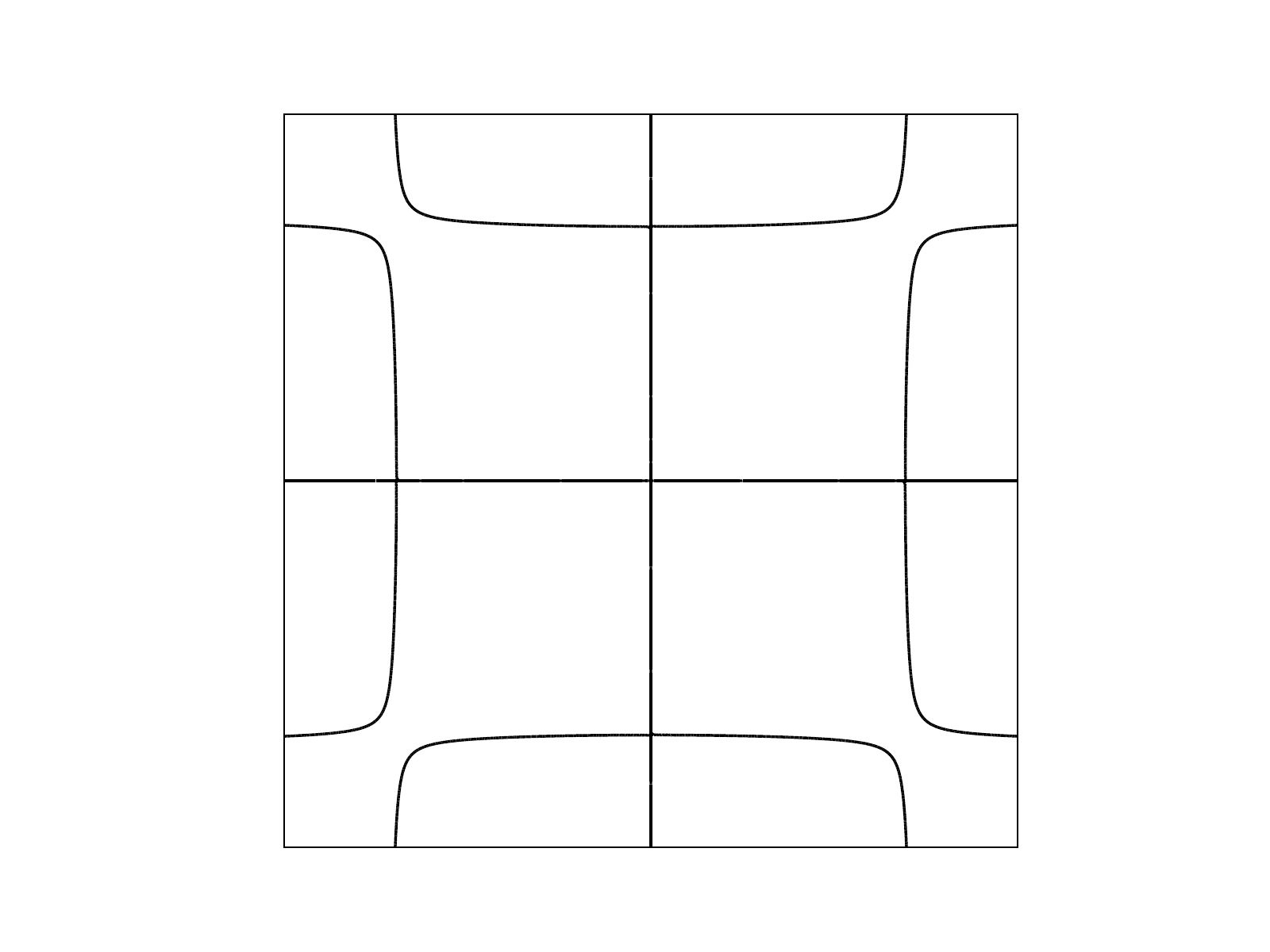}
  \end{subfigure}
  \begin{subfigure}[t]{0.12\textwidth}
    \includegraphics[width=\textwidth]{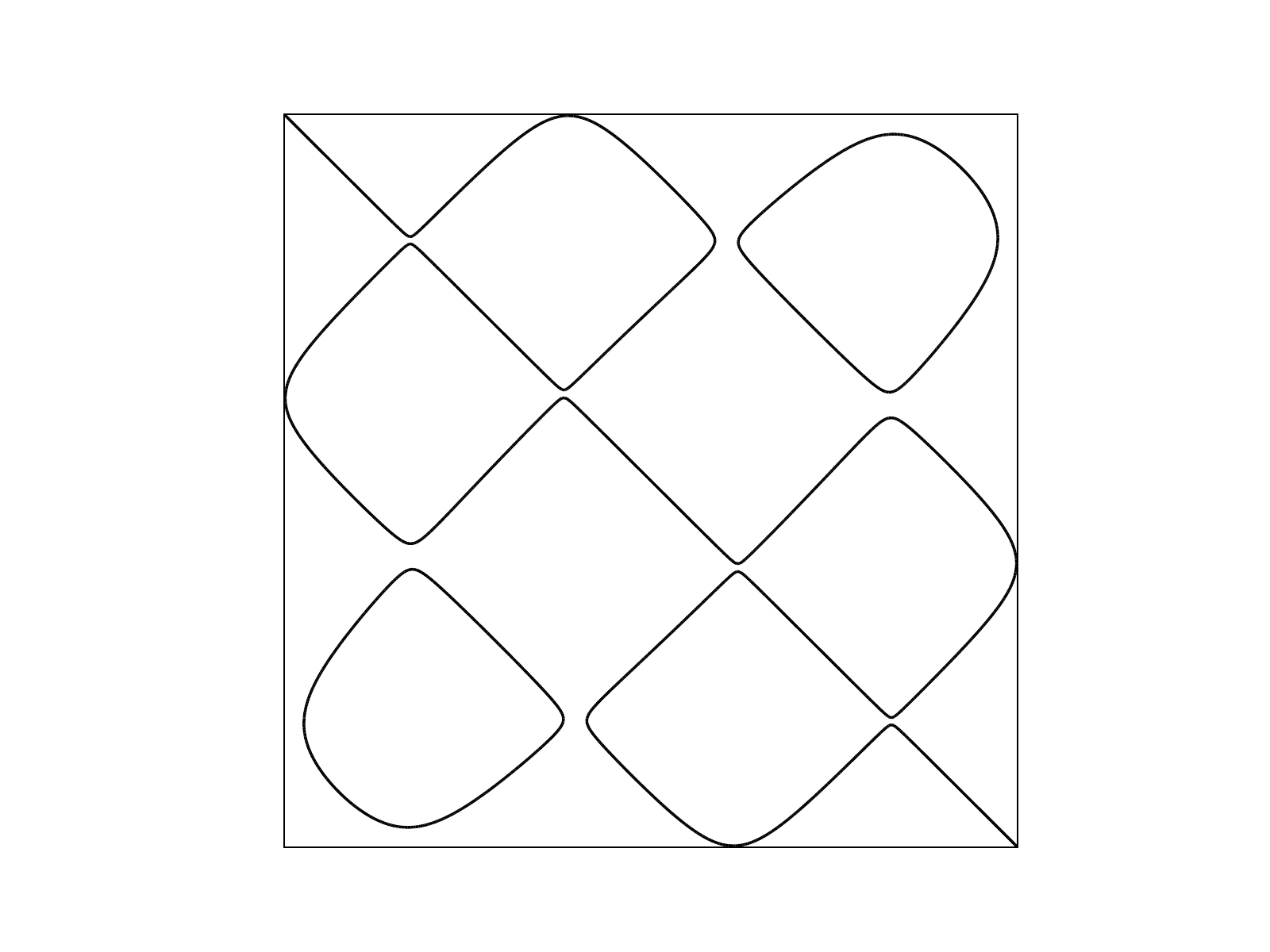}
  \end{subfigure}
  \\
  \begin{subfigure}[t]{0.12\textwidth}
    \includegraphics[width=\textwidth]{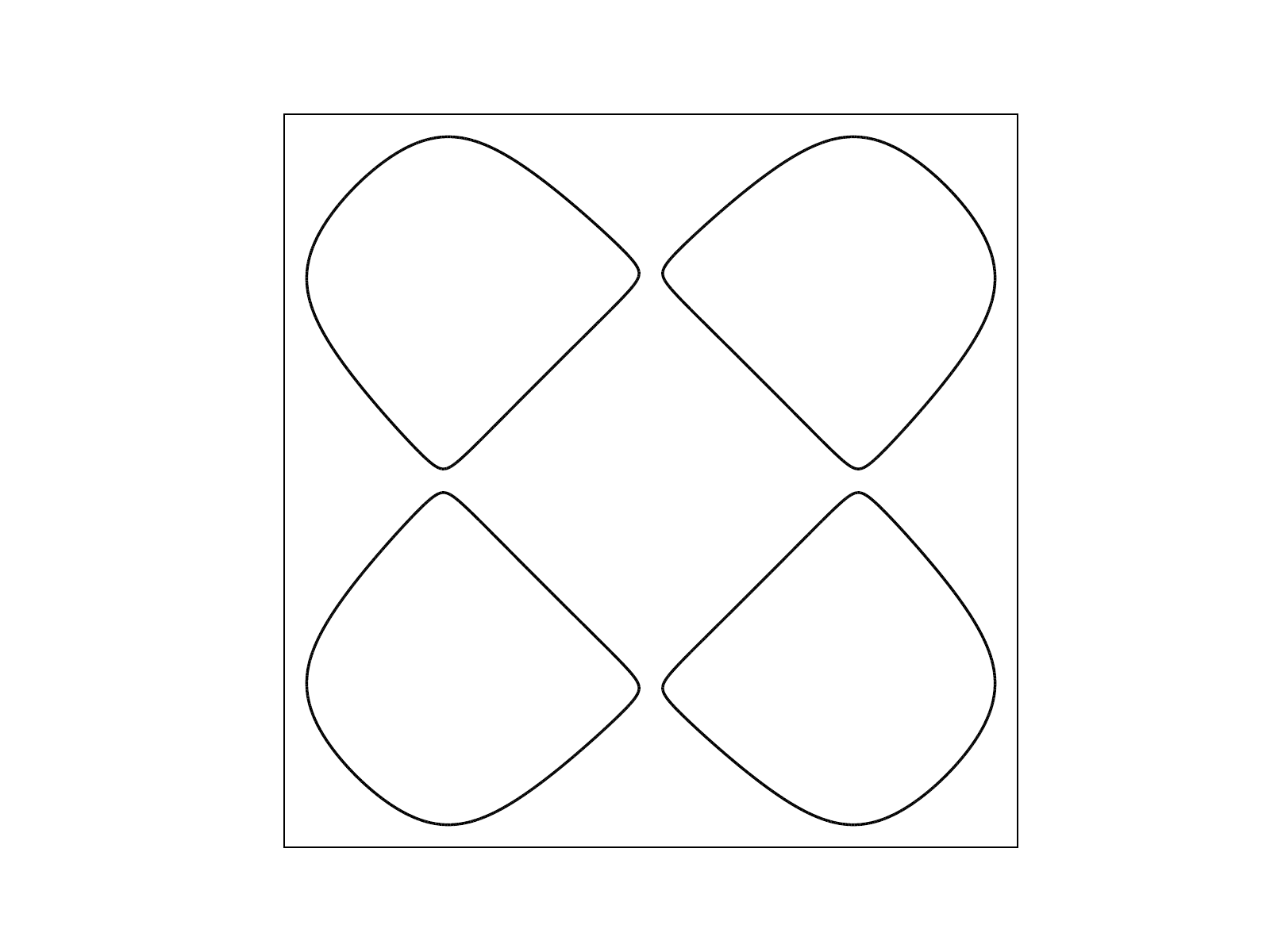}
  \end{subfigure}
  \begin{subfigure}[t]{0.12\textwidth}
    \includegraphics[width=\textwidth]{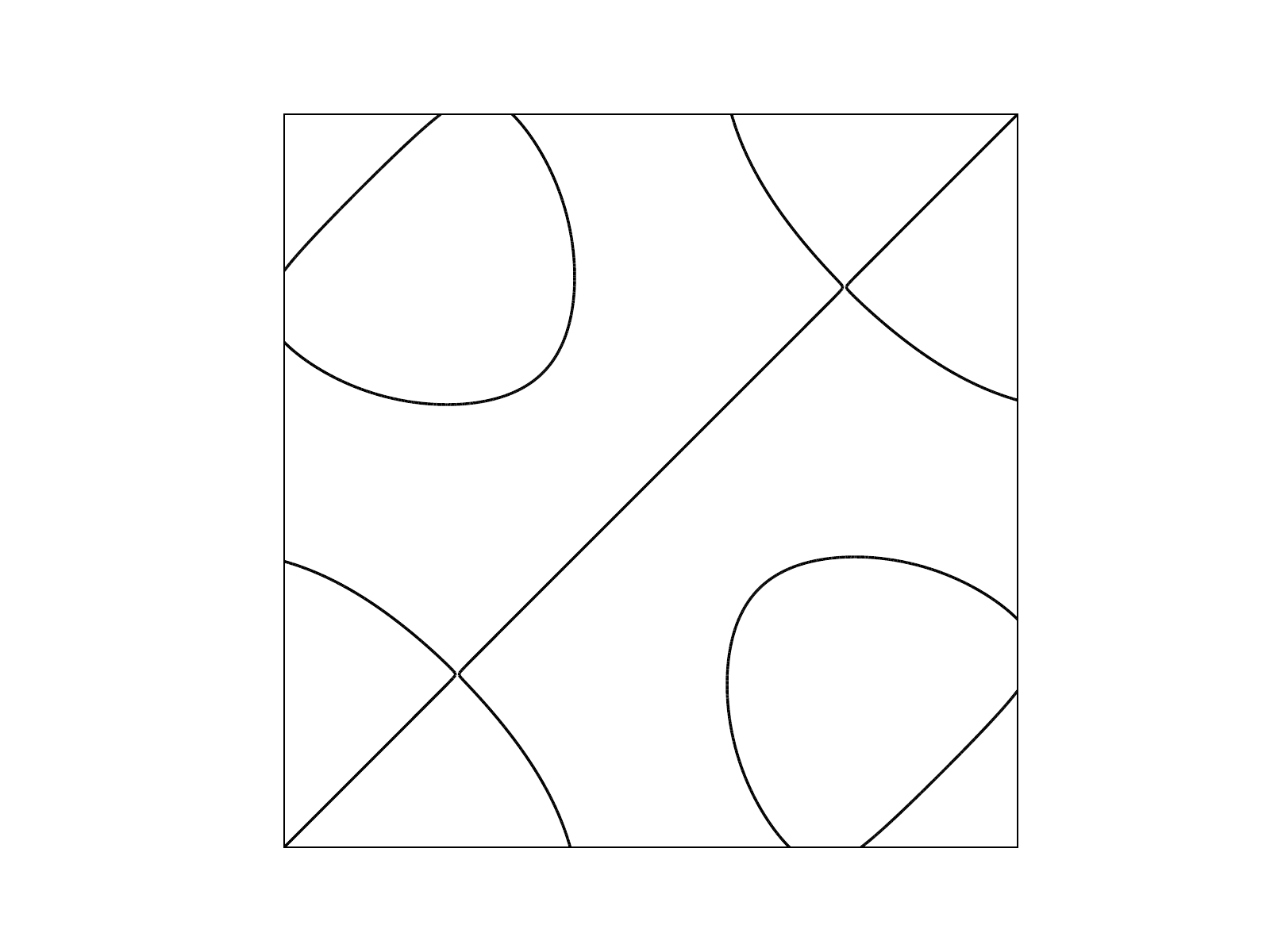}
  \end{subfigure}
  \begin{subfigure}[t]{0.12\textwidth}
        \includegraphics[width=\textwidth]{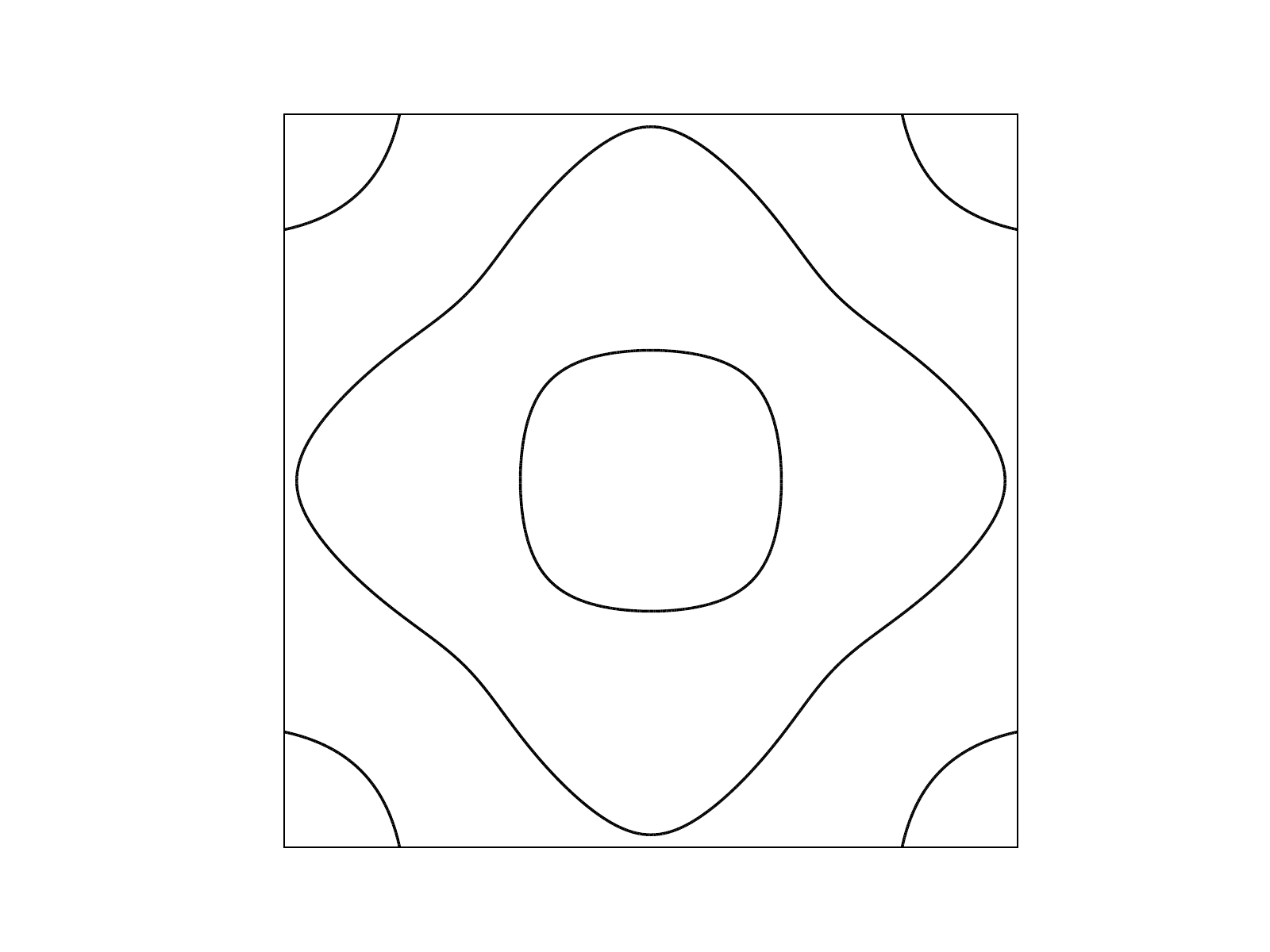}
  \end{subfigure}
  \begin{subfigure}[t]{0.12\textwidth}
        \includegraphics[width=\textwidth]{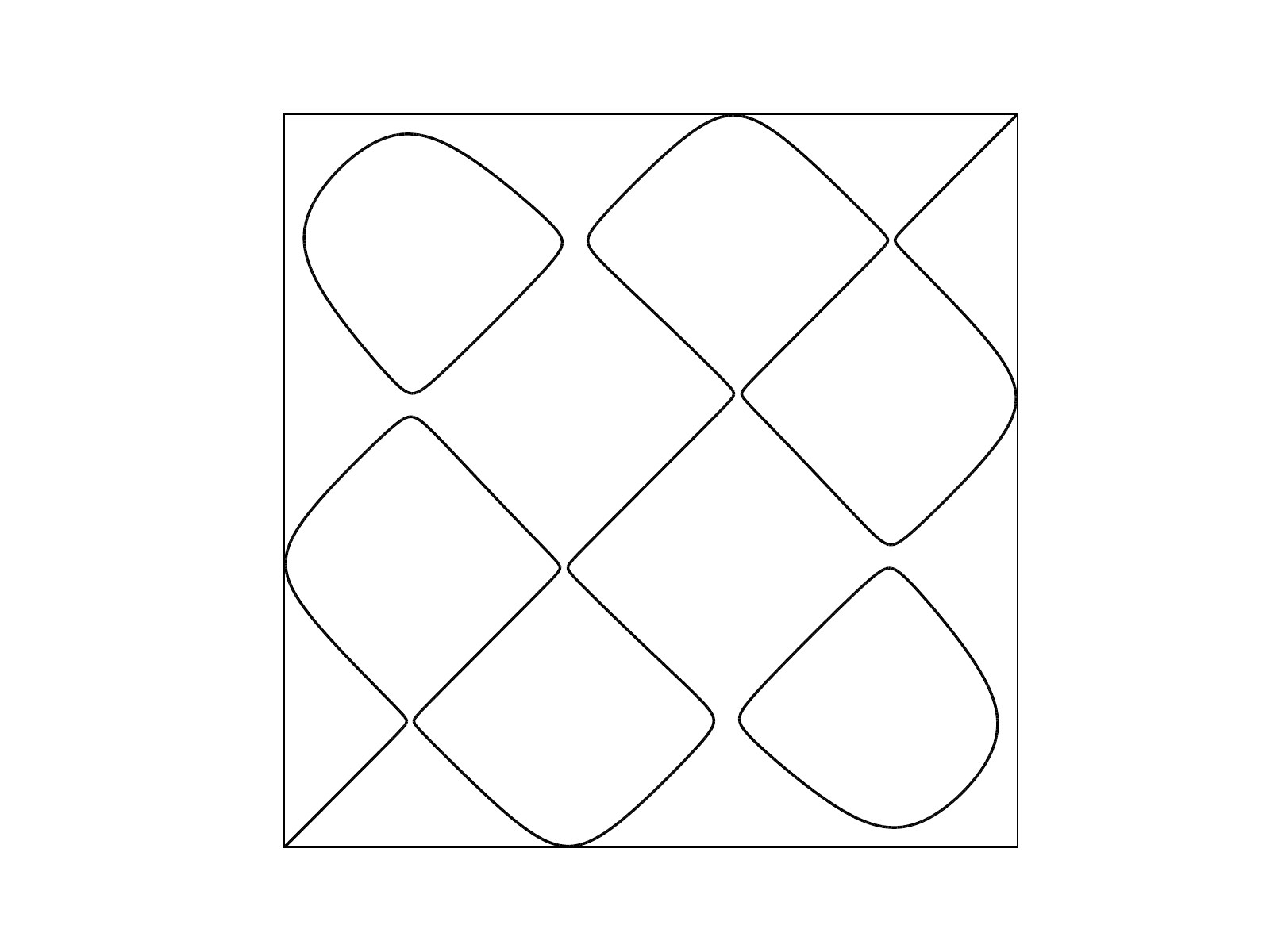}
  \end{subfigure}
  \begin{subfigure}[t]{0.12\textwidth}
    \includegraphics[width=\textwidth]{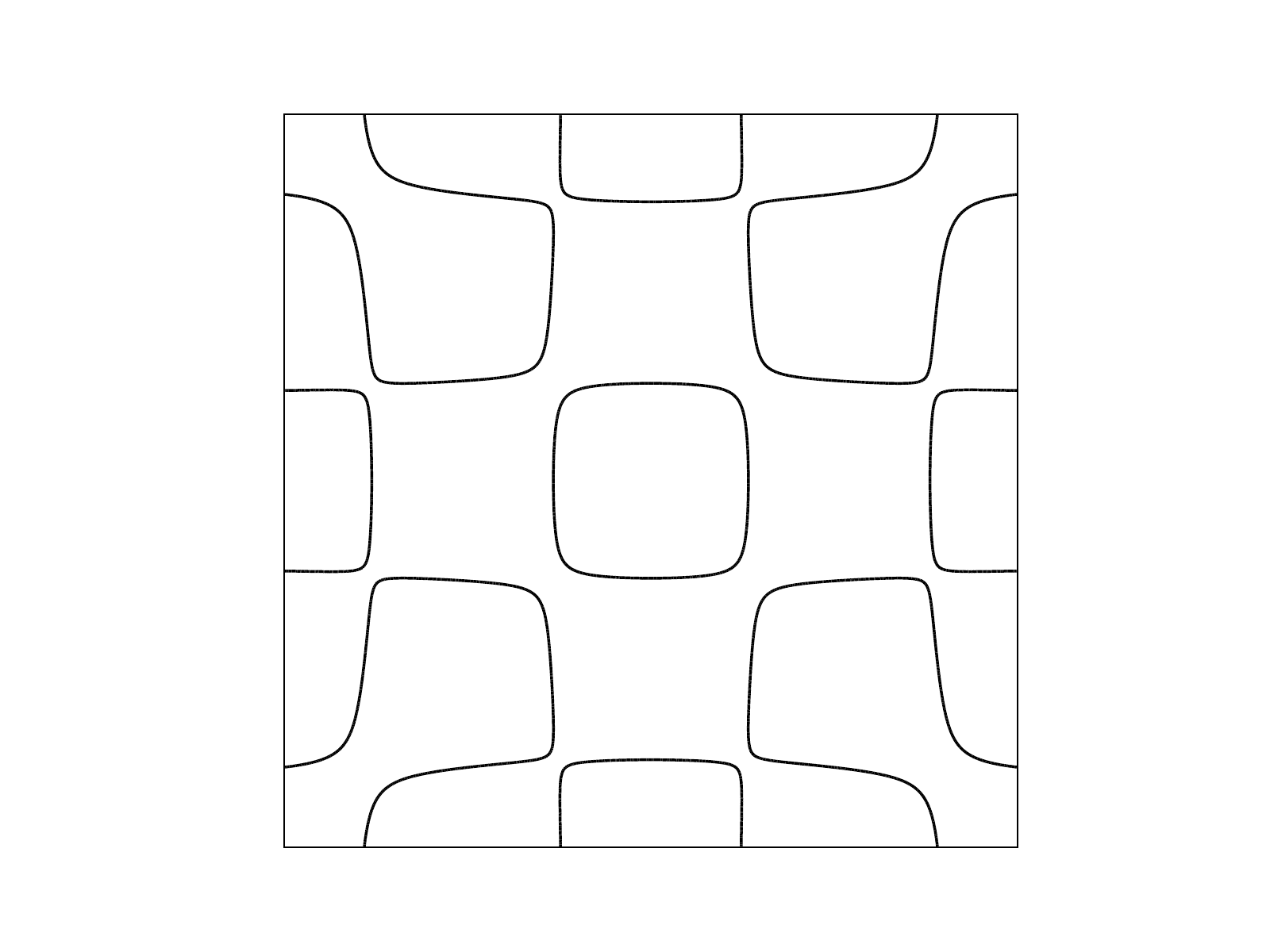}
  \end{subfigure}
  \caption{Several Chladni figures computed with Argyris elements on a
    $64\times 64$ mesh.  These are a subset of those computed
    in~\citet{gander2012ritz}.  The empty top-left figures correspond
    to the null modes of the operator under free boundary conditions.
    Figures for repeated eigenvalues may differ from those presented
    in other works since there is not a unique orthonormal basis for
    the eigenspace.}
  \label{fig:chladni-plates}
\end{figure}

To compute the Chladni figures, we use Firedrake to assemble the
discrete biharmonic \cref{eq:aplate} and mass matrices using Argyris elements on a
$64\times 64$ mesh of the square domain $\Omega = [-1, 1]^2$.  These are then fed into
SLEPc~\citep{hernandez2005slepc}.  The eigenpairs of the generalized eigenproblem are then constructed via
a shift-and-invert strategy (the options are
\lstinline[basicstyle=\normalsize\ttfamily]{-eps_nev 30 -eps_tol 1e-11 -eps_target 10 -st_type sinvert}).
We plot the zero contours of a number of eigenmodes in \cref{fig:chladni-plates}.

\begin{figure}[htbp]
  \begin{subfigure}[t]{0.12\textwidth}
    \includegraphics[width=\textwidth]{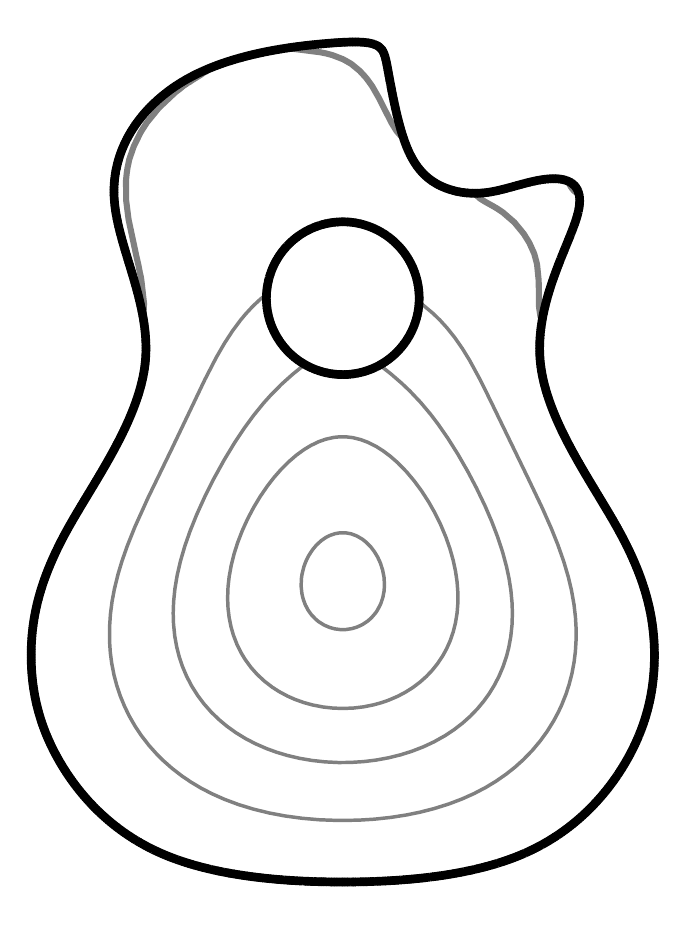}
  \end{subfigure}
  \begin{subfigure}[t]{0.12\textwidth}
    \includegraphics[width=\textwidth]{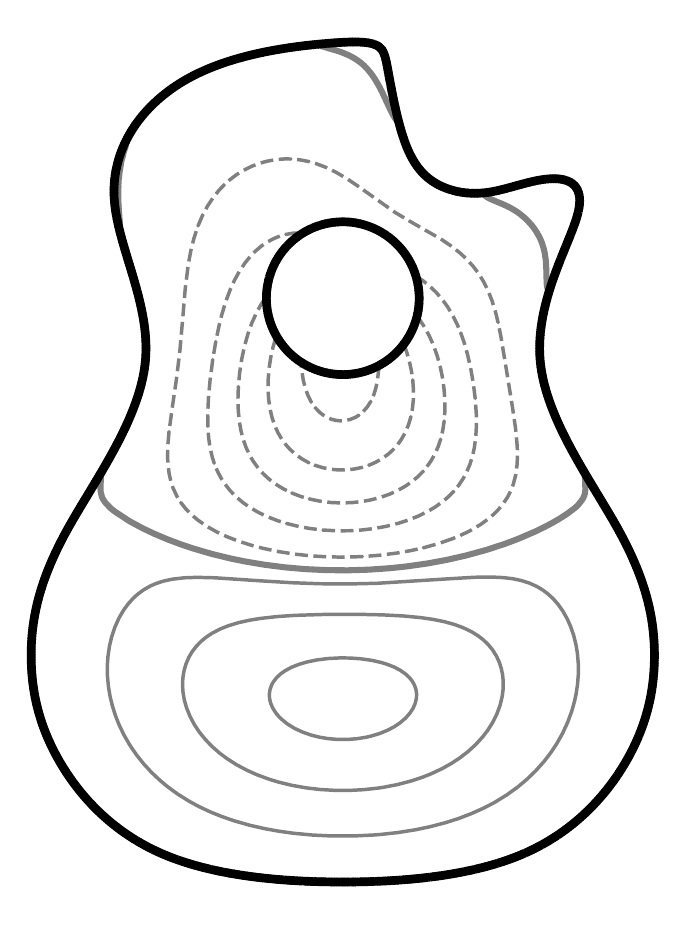}
  \end{subfigure}
  \begin{subfigure}[t]{0.12\textwidth}
    \includegraphics[width=\textwidth]{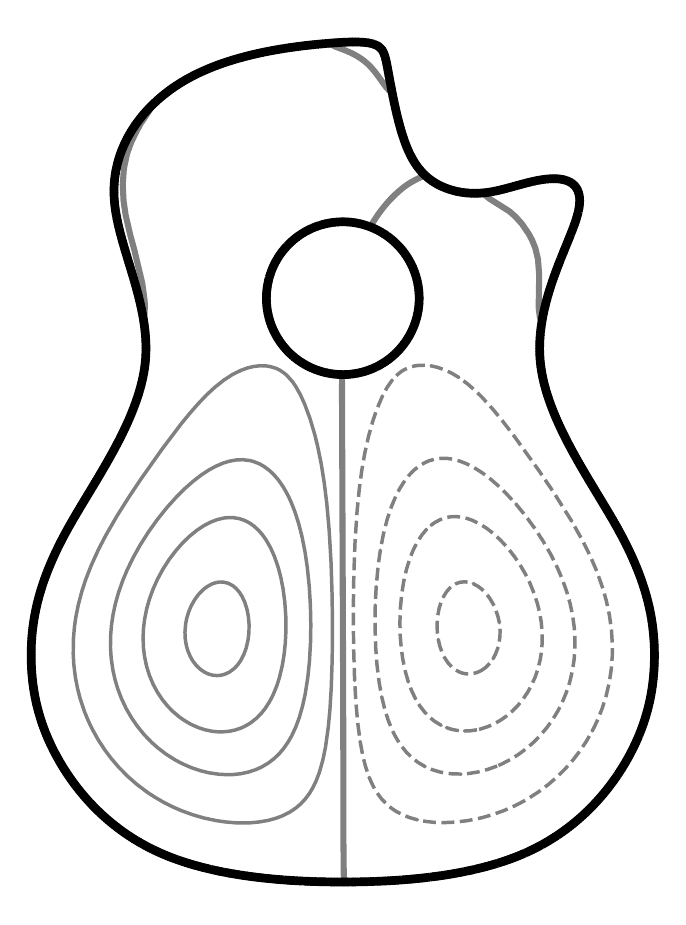}
  \end{subfigure}
  \begin{subfigure}[t]{0.12\textwidth}
    \includegraphics[width=\textwidth]{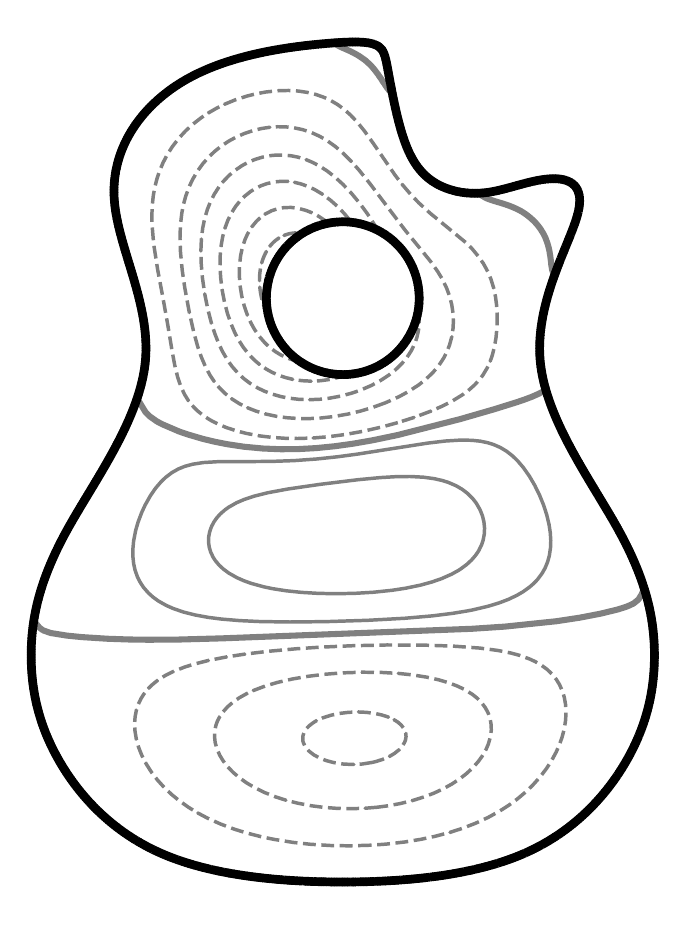}
  \end{subfigure}
  \begin{subfigure}[t]{0.12\textwidth}
    \includegraphics[width=\textwidth]{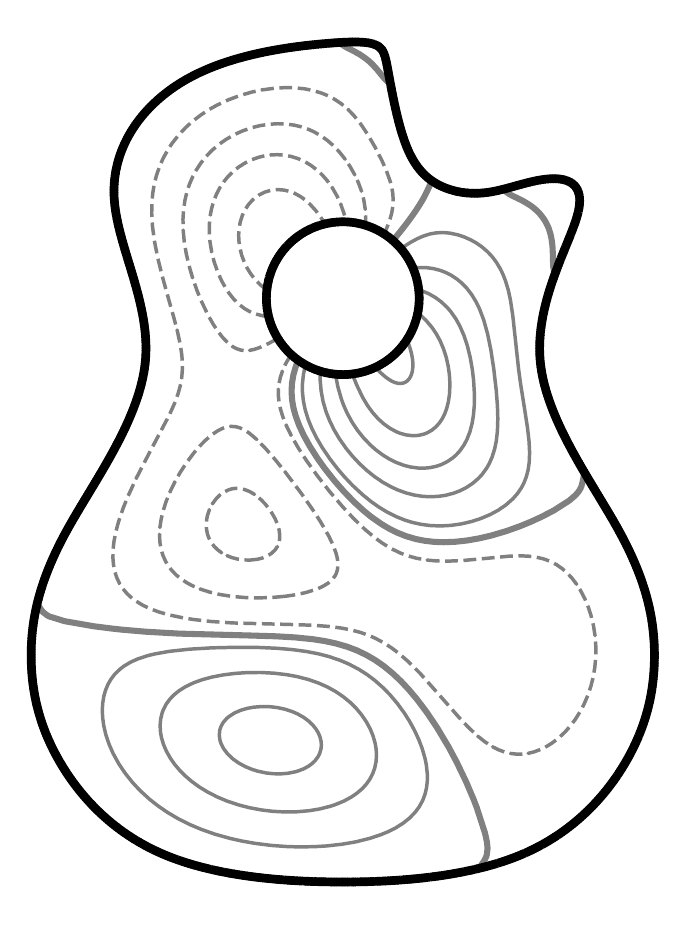}
  \end{subfigure}
  \begin{subfigure}[t]{0.12\textwidth}
    \includegraphics[width=\textwidth]{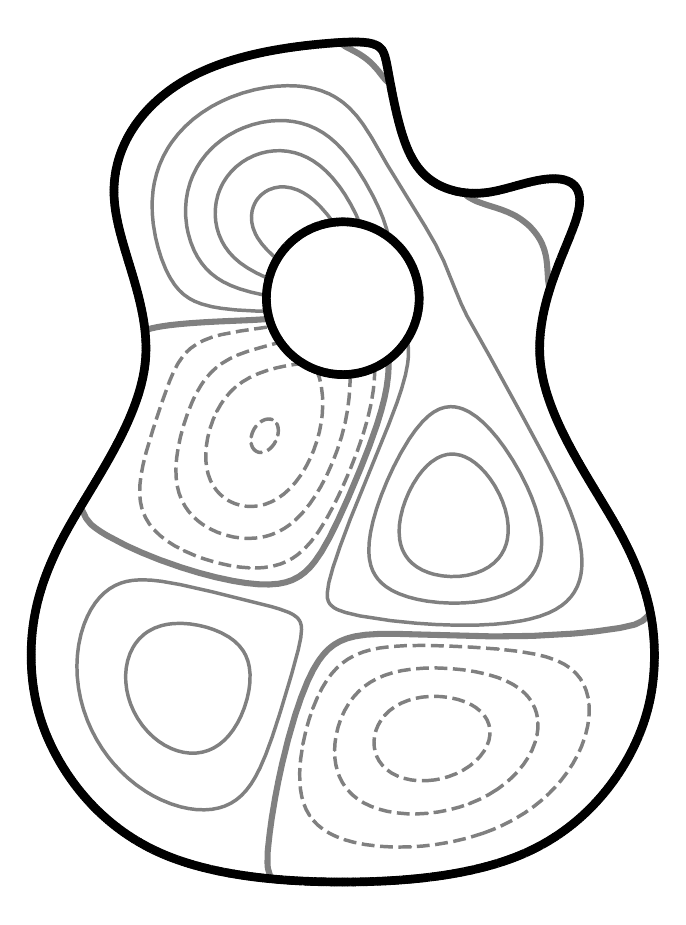}
  \end{subfigure}
  \begin{subfigure}[t]{0.12\textwidth}
    \includegraphics[width=\textwidth]{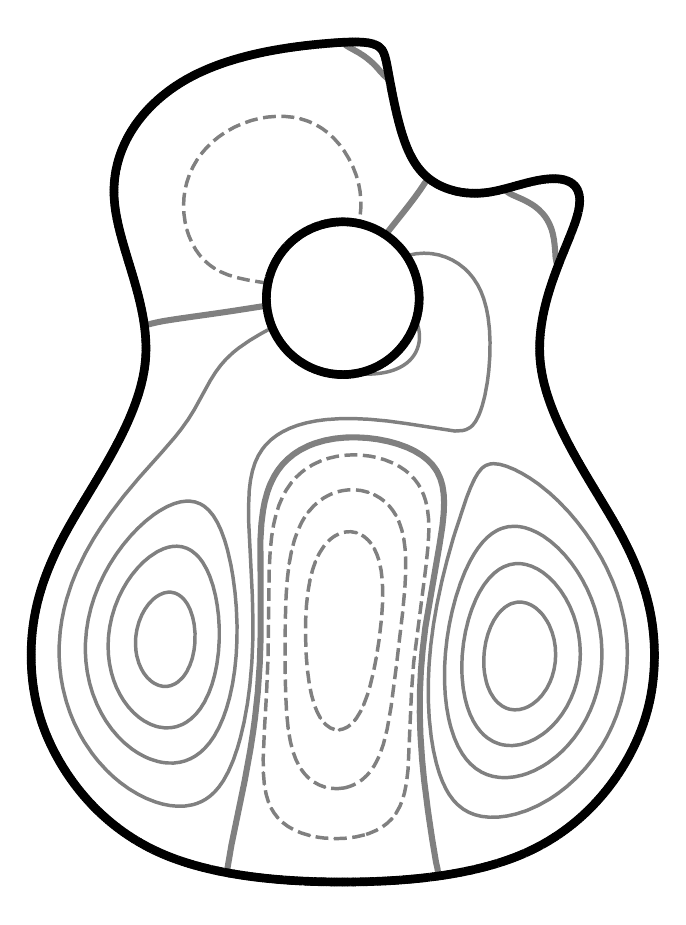}
  \end{subfigure}
  \\
  \begin{subfigure}[t]{0.12\textwidth}
    \includegraphics[width=\textwidth]{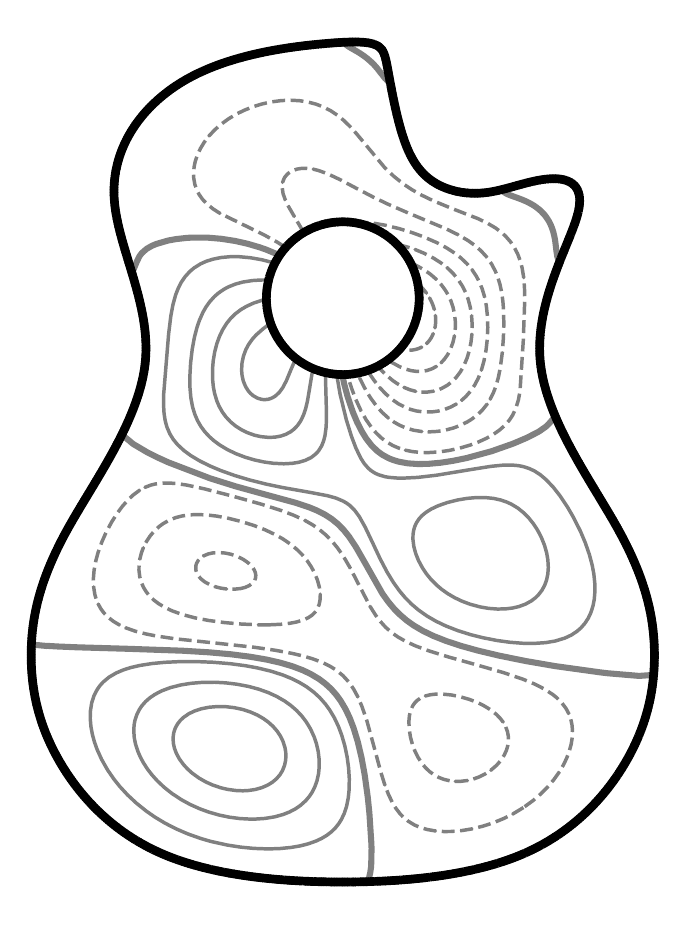}
  \end{subfigure}
  \begin{subfigure}[t]{0.12\textwidth}
    \includegraphics[width=\textwidth]{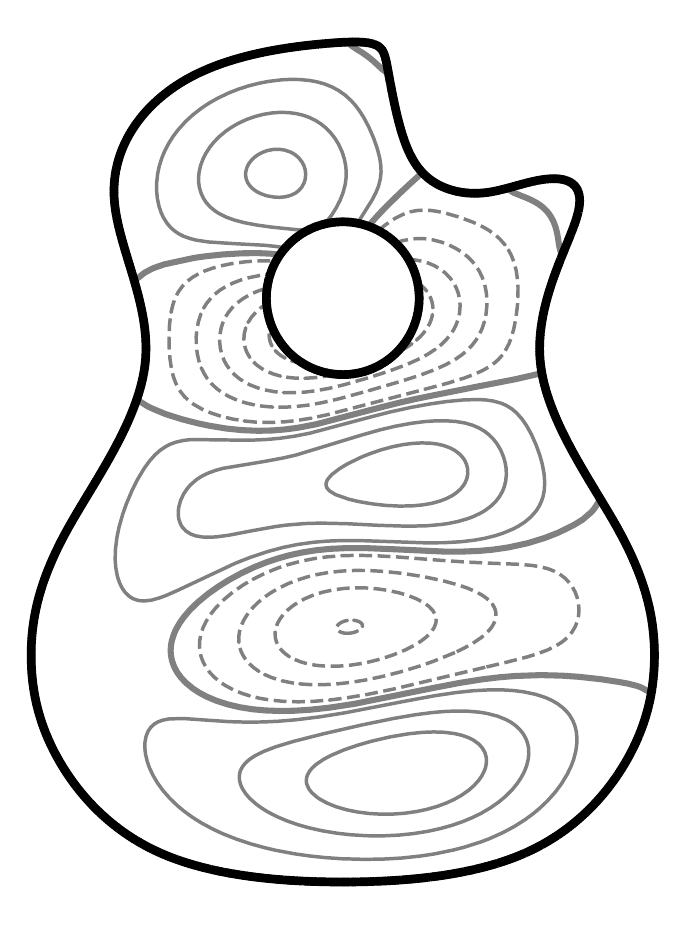}
  \end{subfigure}
  \begin{subfigure}[t]{0.12\textwidth}
    \includegraphics[width=\textwidth]{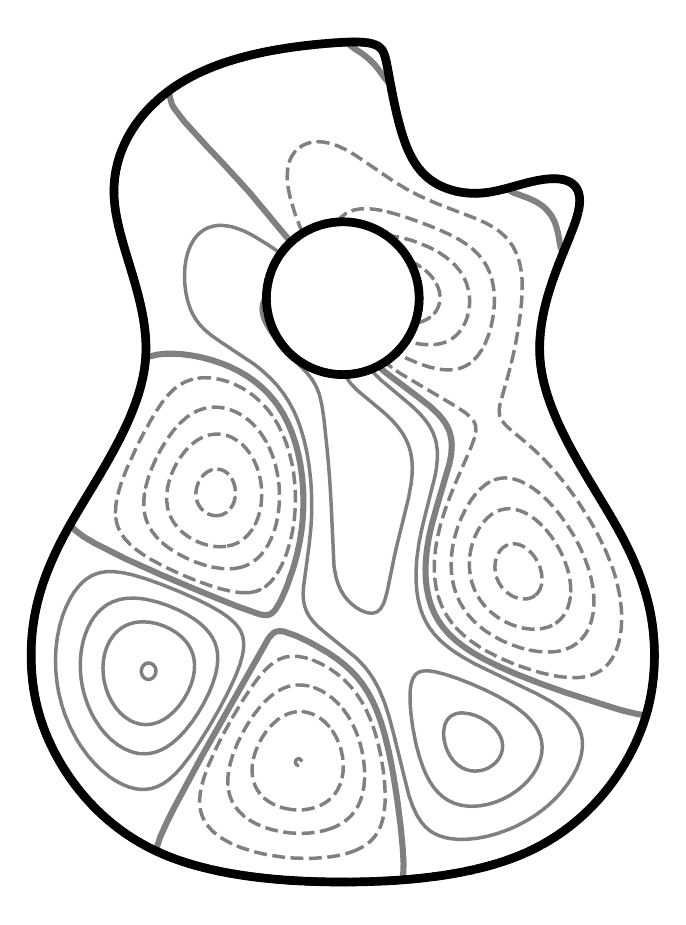}
  \end{subfigure}
  \begin{subfigure}[t]{0.12\textwidth}
    \includegraphics[width=\textwidth]{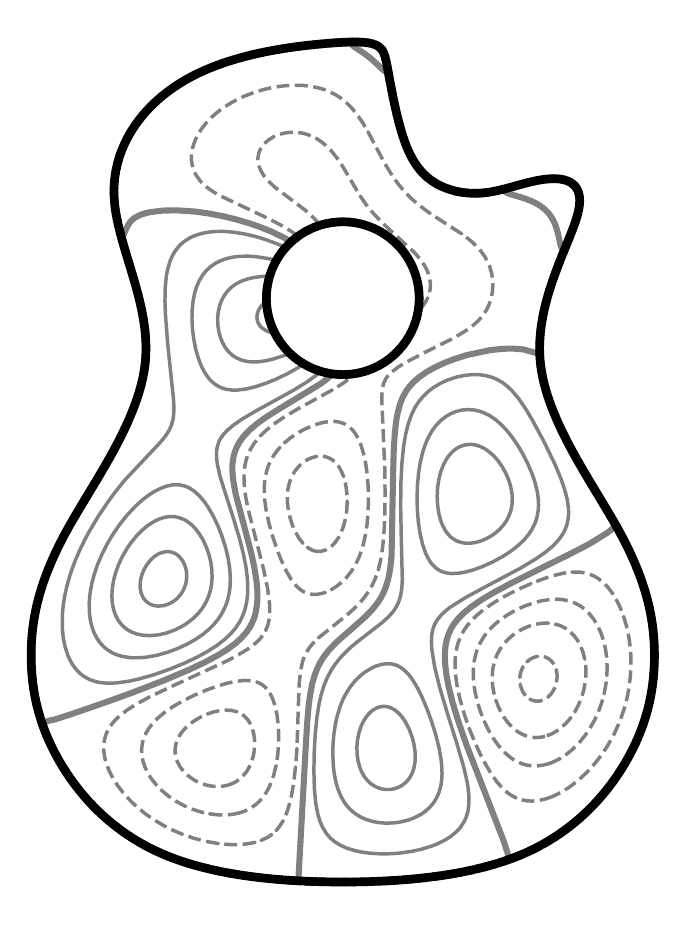}
  \end{subfigure}
  \begin{subfigure}[t]{0.12\textwidth}
    \includegraphics[width=\textwidth]{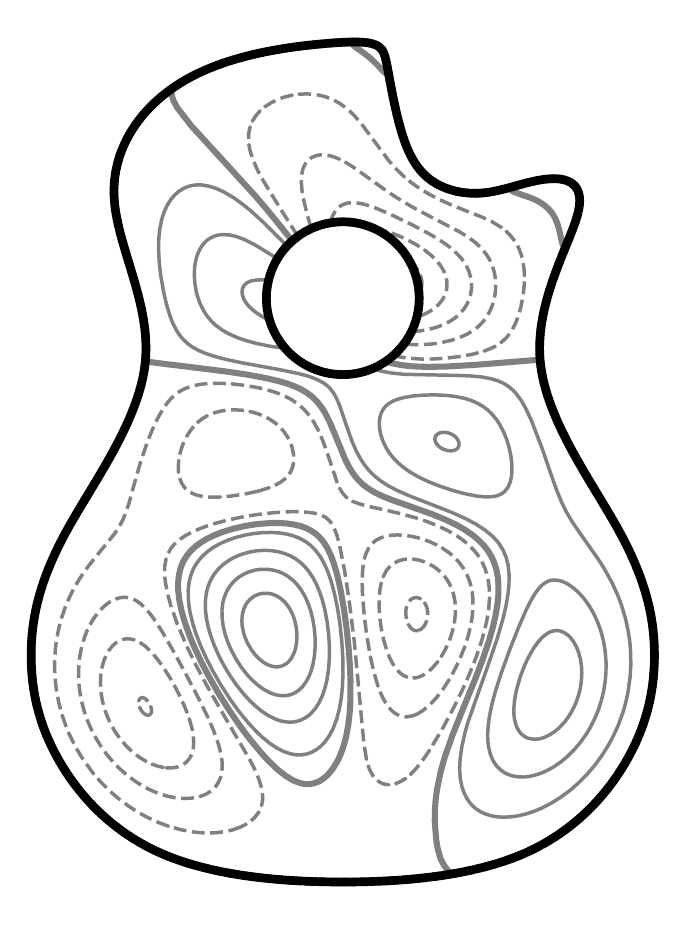}
  \end{subfigure}
  \begin{subfigure}[t]{0.12\textwidth}
    \includegraphics[width=\textwidth]{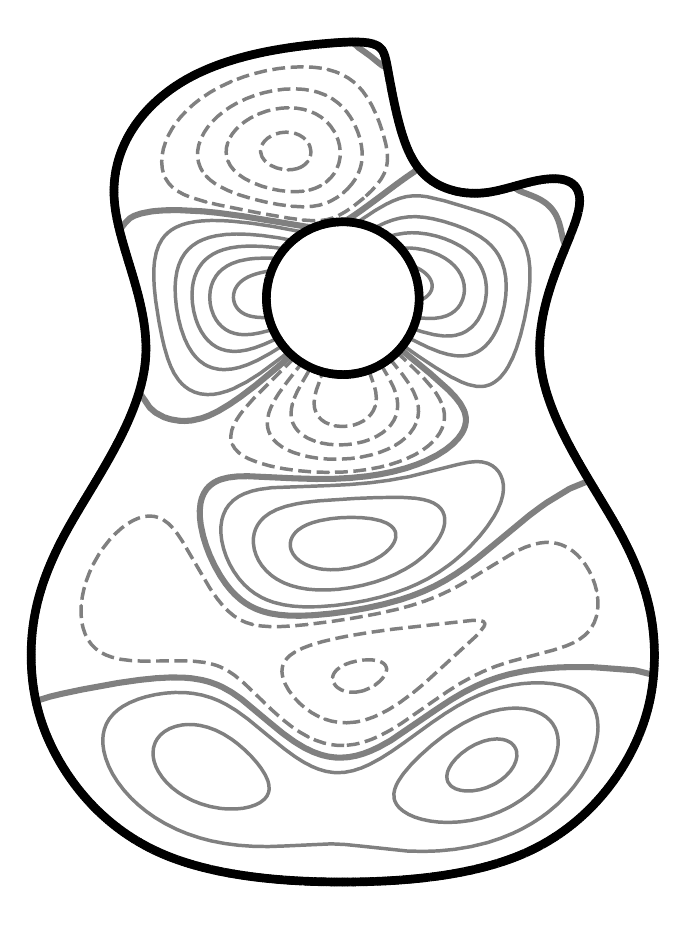}
  \end{subfigure}
  \begin{subfigure}[t]{0.12\textwidth}
    \includegraphics[width=\textwidth]{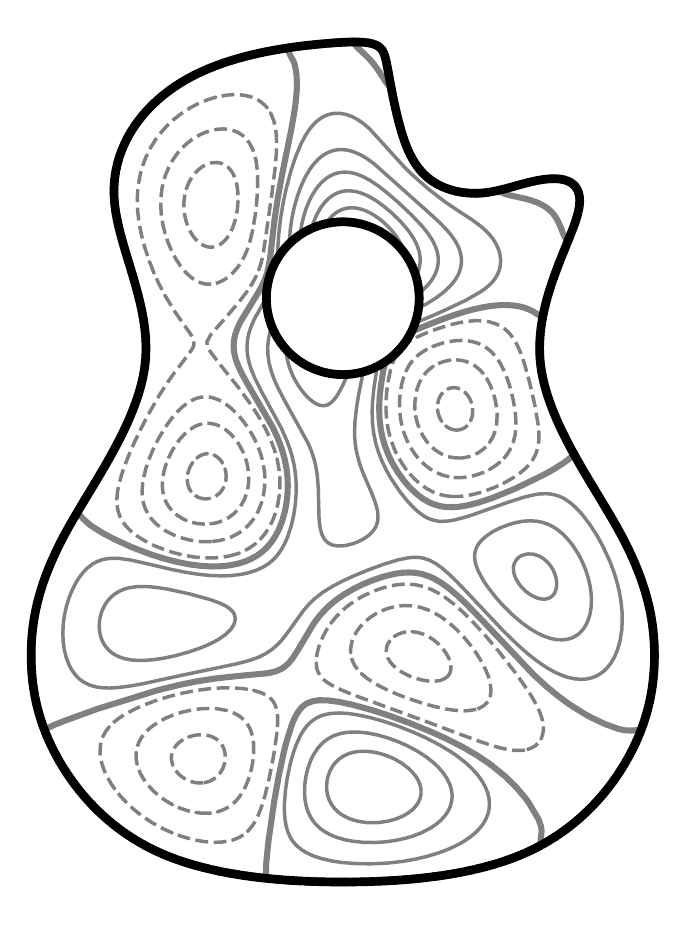}
  \end{subfigure}
  \\
  \begin{subfigure}[t]{0.12\textwidth}
    \includegraphics[width=\textwidth]{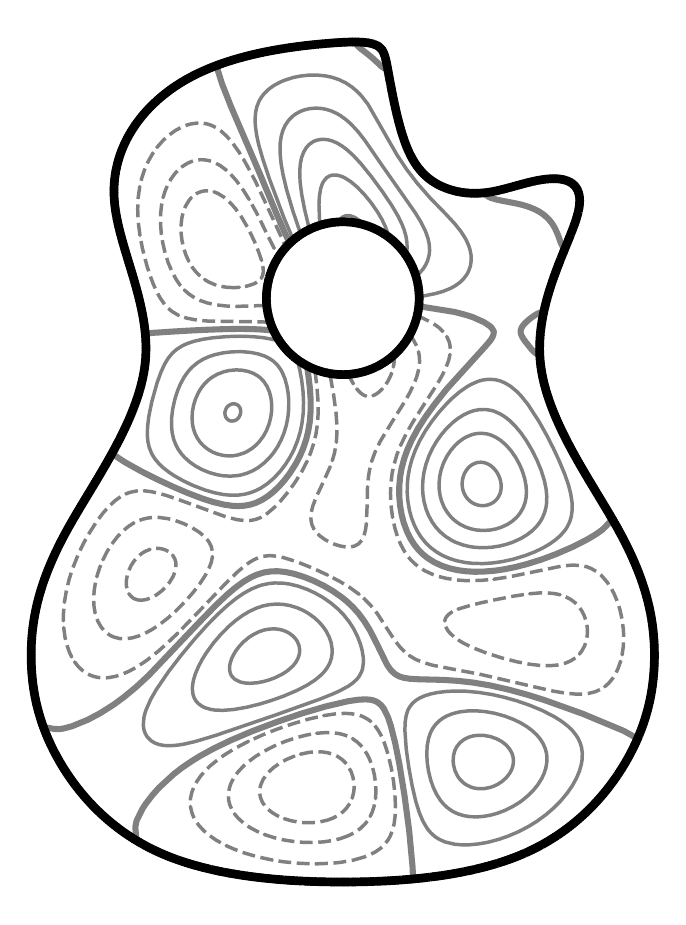}
  \end{subfigure}
  \begin{subfigure}[t]{0.12\textwidth}
    \includegraphics[width=\textwidth]{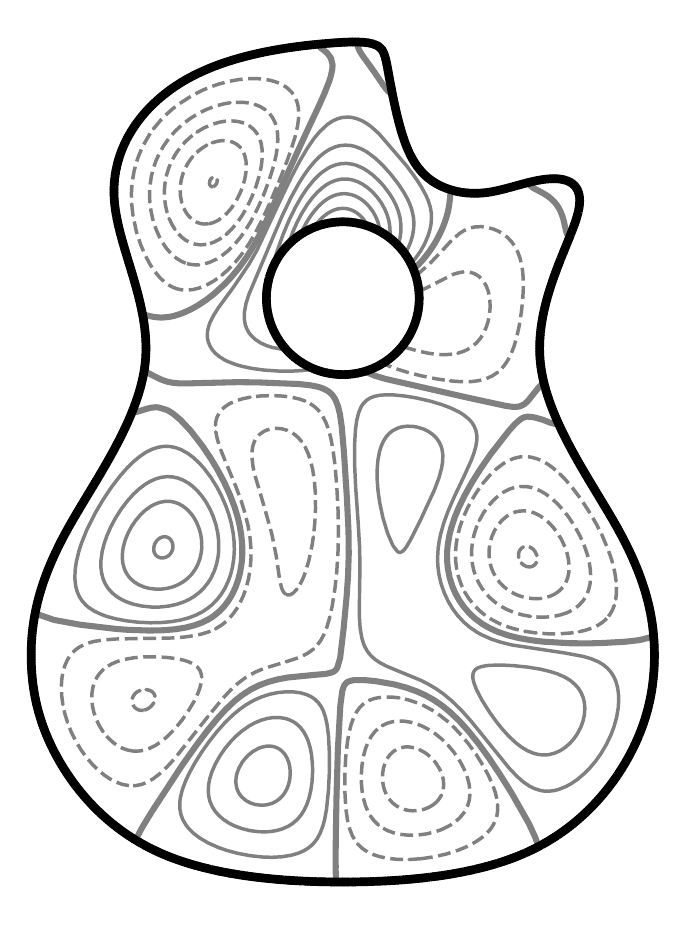}
  \end{subfigure}
  \begin{subfigure}[t]{0.12\textwidth}
    \includegraphics[width=\textwidth]{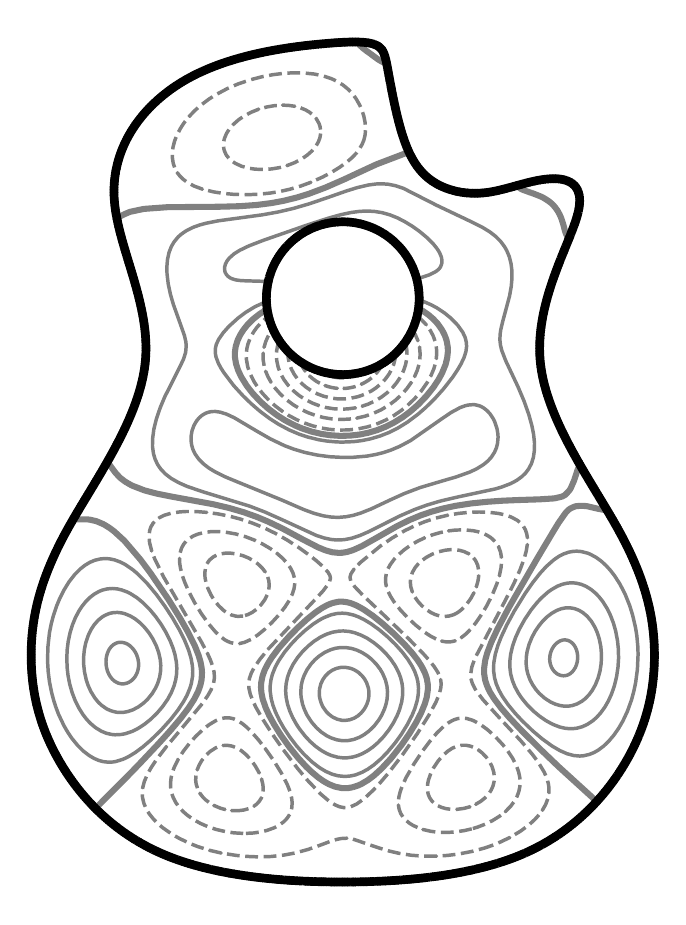}
  \end{subfigure}
  \begin{subfigure}[t]{0.12\textwidth}
    \includegraphics[width=\textwidth]{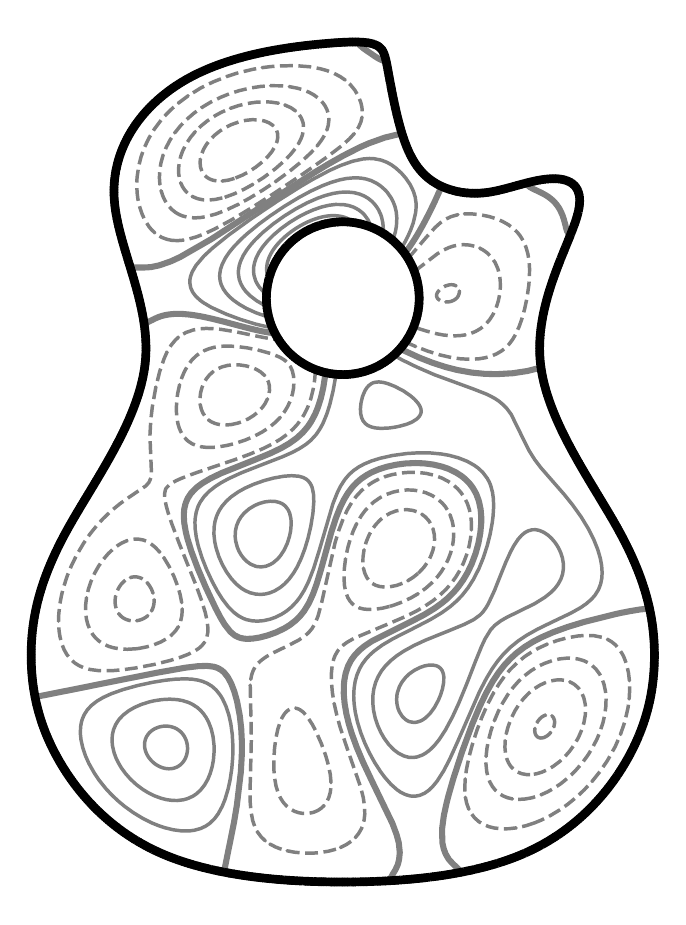}
  \end{subfigure}
  \begin{subfigure}[t]{0.12\textwidth}
    \includegraphics[width=\textwidth]{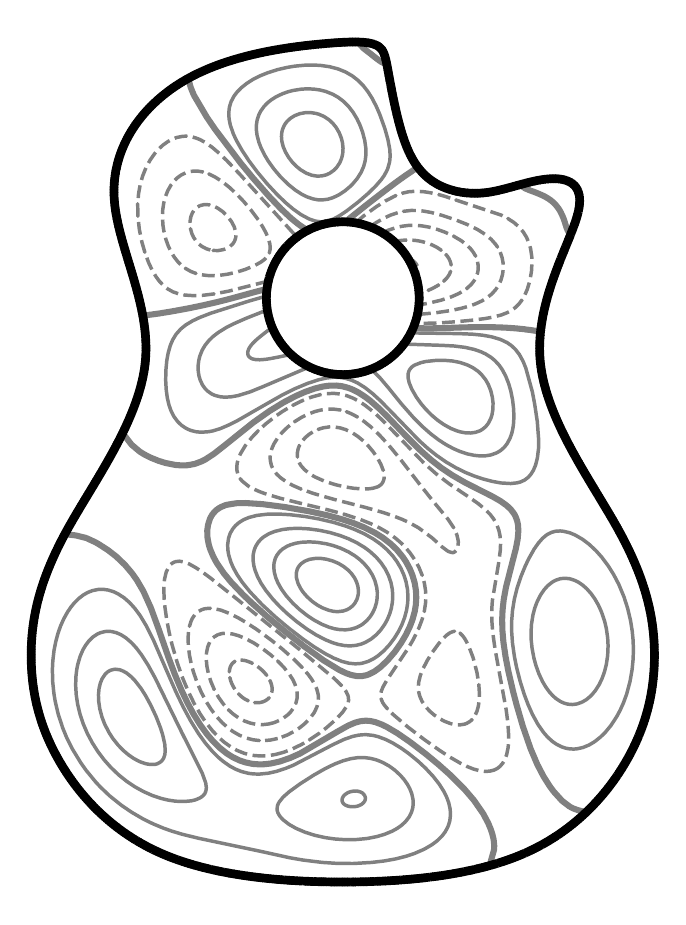}
  \end{subfigure}
  \begin{subfigure}[t]{0.12\textwidth}
    \includegraphics[width=\textwidth]{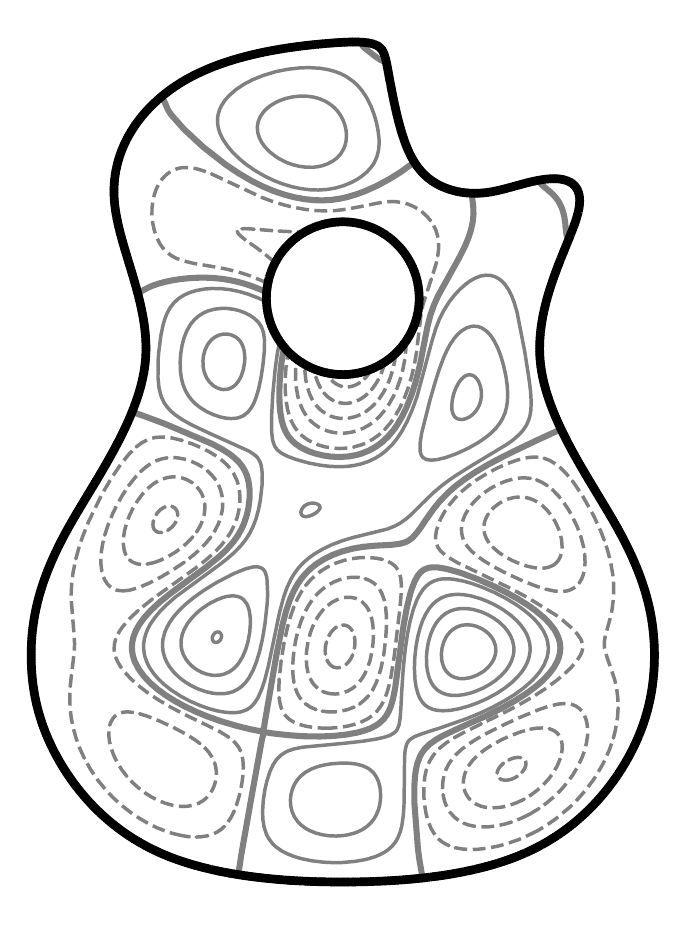}
  \end{subfigure}
  \begin{subfigure}[t]{0.12\textwidth}
    \includegraphics[width=\textwidth]{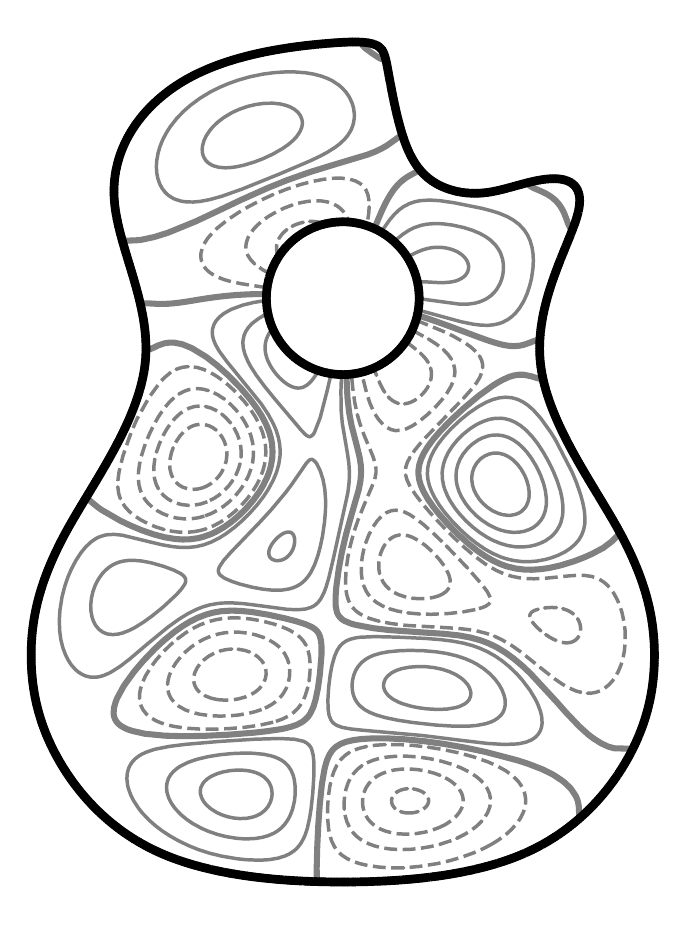}
  \end{subfigure}
  \\
  \begin{subfigure}[t]{0.12\textwidth}
    \includegraphics[width=\textwidth]{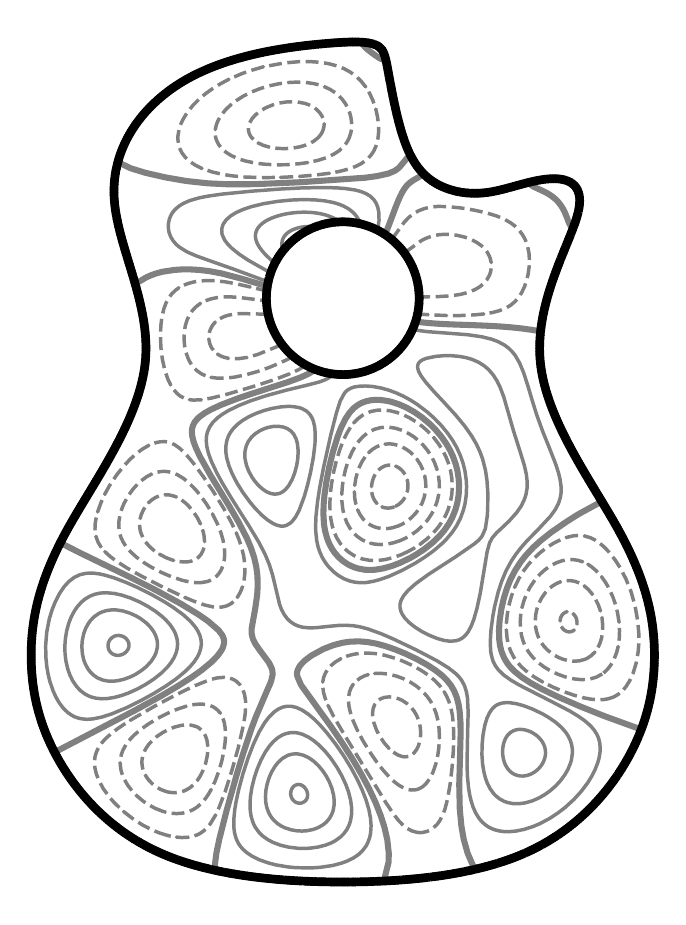}
  \end{subfigure}
  \begin{subfigure}[t]{0.12\textwidth}
    \includegraphics[width=\textwidth]{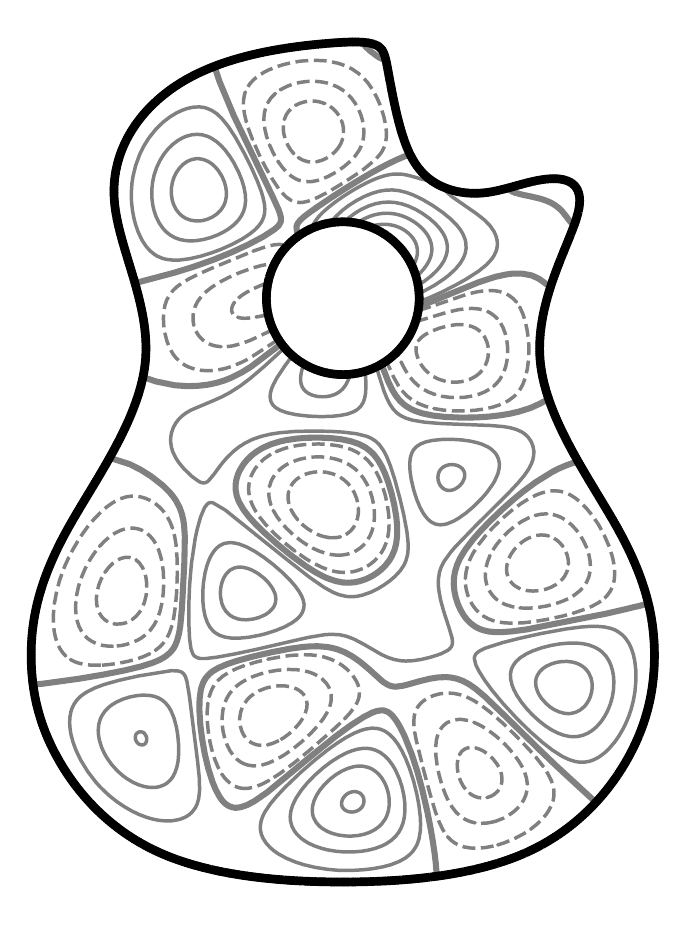}
  \end{subfigure}
  \begin{subfigure}[t]{0.12\textwidth}
    \includegraphics[width=\textwidth]{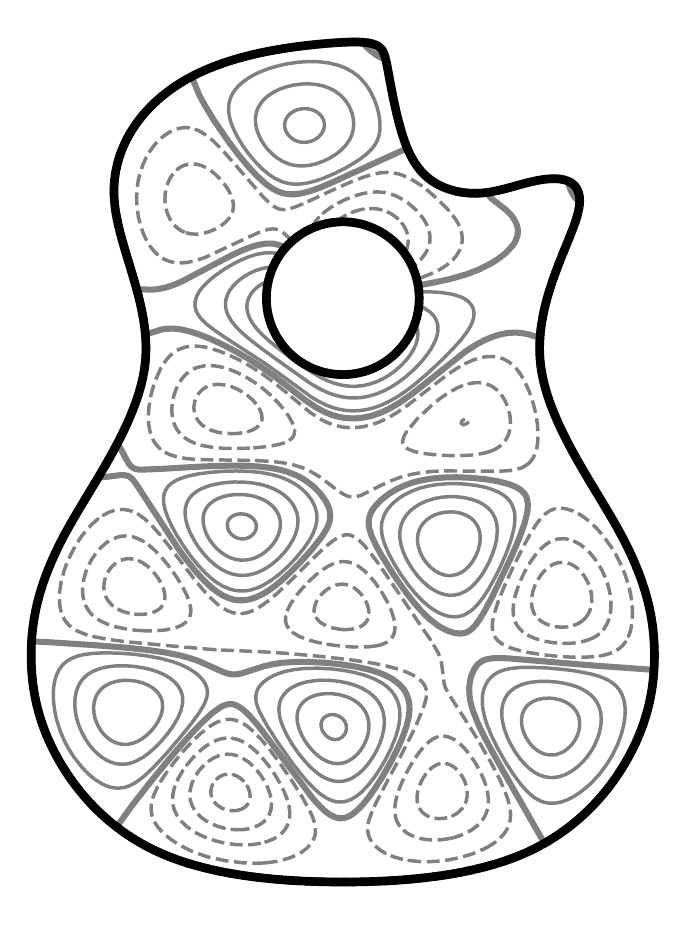}
  \end{subfigure}
  \begin{subfigure}[t]{0.12\textwidth}
    \includegraphics[width=\textwidth]{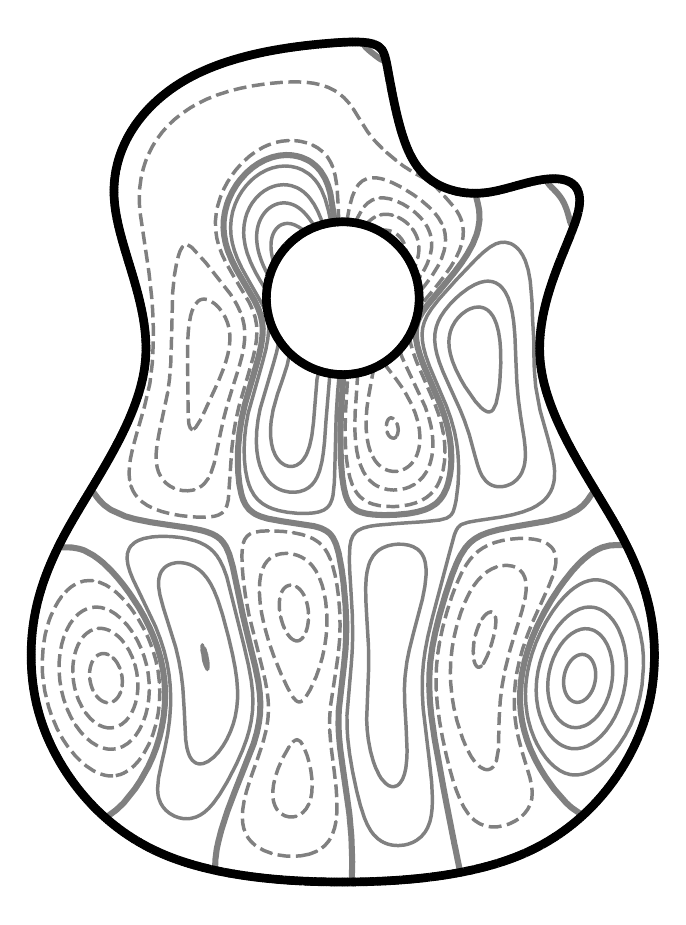}
  \end{subfigure}
  \begin{subfigure}[t]{0.12\textwidth}
    \includegraphics[width=\textwidth]{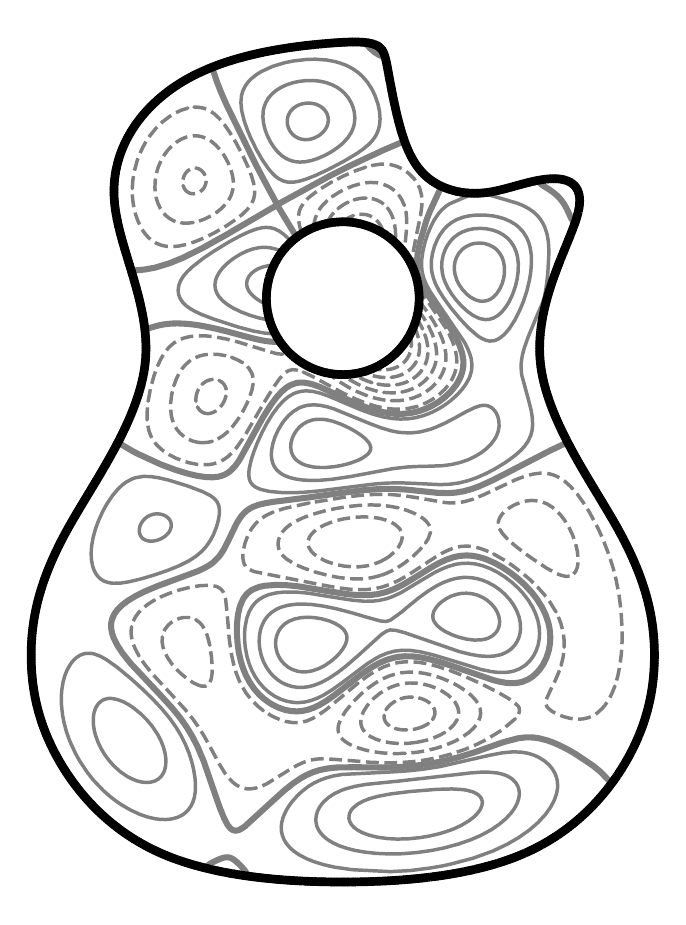}
  \end{subfigure}
  \begin{subfigure}[t]{0.12\textwidth}
    \includegraphics[width=\textwidth]{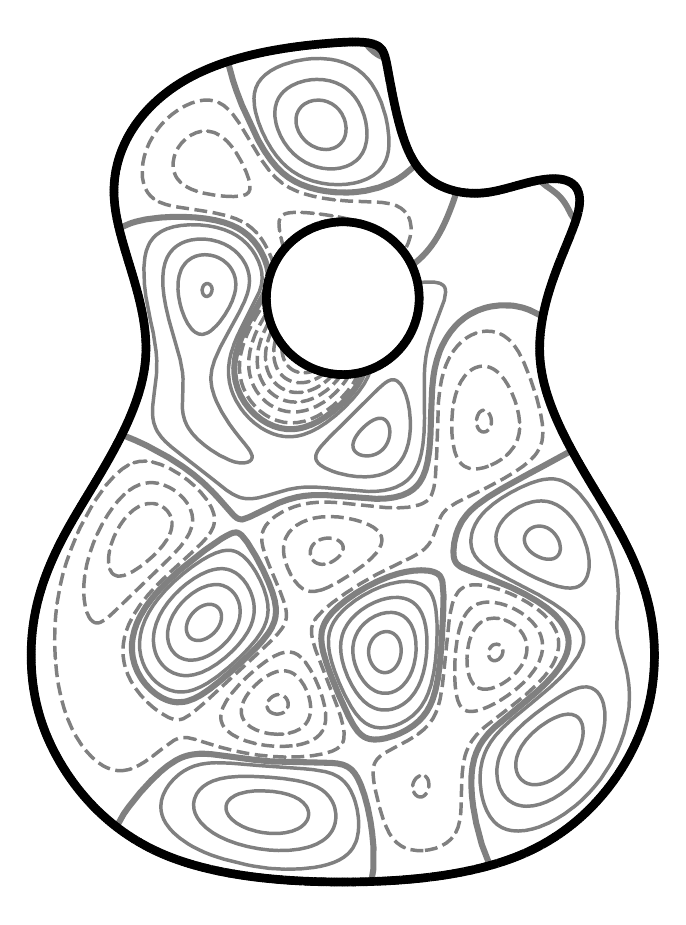}
  \end{subfigure}
  \begin{subfigure}[t]{0.12\textwidth}
    \includegraphics[width=\textwidth]{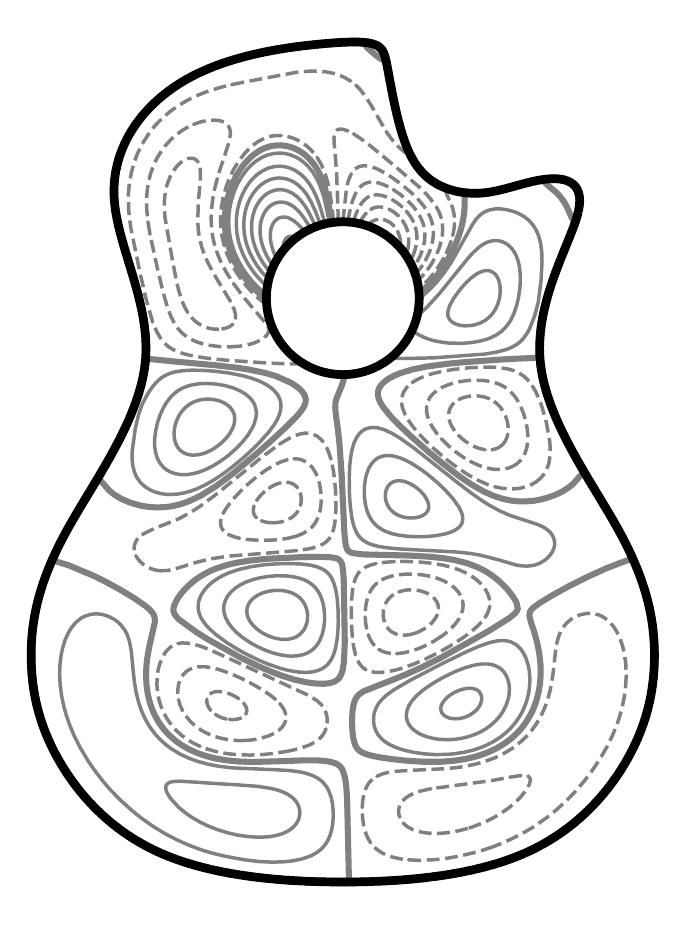}
  \end{subfigure}
  \caption{The first twenty-eight eigenmodes, arranged in order of increasing
    frequency, of the biharmonic operator \cref{eq:aplate-clamped}
    with clamped boundary conditions on the exterior boundary of the
    domain, and free conditions on the sound hole.  Contour lines are
    equally spaced, with negative contours dashed and the zero-contour
    slightly thicker.}
  \label{fig:guitar-eigenmodes}
\end{figure}

While our extensions now enable succinct solution of a challenging
classical problem, Firedrake also enables us to
perform similar analysis in more complex geometric settings
beyond the scope of analytic techniques.  For example, consider the
guitar-shaped domain illustrated in
\cref{fig:guitar-eigenmodes}.  We again 
compute eigenmodes of the plate-bending operator \cref{eq:aplate} on
this domain using Morley triangles on a mesh of 17277 vertices and
34536 elements, generated with \texttt{gmsh}~\citep{geuzaine2009gmsh}.
In this case, we only leave the boundary free on the sound hole, and
use clamped conditions on the exterior.  We remark that studying
eigenvalue computation with Morley elements is
well-established~\citep{rannacher1979nonconforming} but still receiving
research attention~\citep{gallistl2014morley, zhang2017multi}.
The additional
Nitsche terms are more involved than in the case of
\cref{eq:apoisson-nitsche}, and the full bilinear form is
\begin{equation}
  \begin{split}
    a_h(u, v) = &\int_{\Omega} \Delta u \Delta v - \left(1-\nu \right) \left( 2 u_{xx} v_{yy} + 2 u_{yy} v_{xx} - 4 u_{xy} v_{xy} \right) \dx \\
    &+ \frac{\beta_1}{h^2} \int_{\partial \Omega} u v \ds + \frac{\beta_2}{h} \int_{\partial \Omega} \Delta u \Delta v \ds\\ 
    &+ \int_{\partial \Omega} \left((\Delta u)_n - 2(1 - \nu)u_{ntt}\right) v \ds +
    \int_{\partial \Omega} u \left((\Delta v)_n - 2(1 - \nu)v_{ntt}\right) \ds\\
    &+ \int_{\partial \Omega} \left(\Delta u - 2(1 - \nu)u_{tt}\right) v_n \ds +
    \int_{\partial \Omega} u_n \left(\Delta v - 2(1 - \nu)v_{tt}\right) \ds,
  \end{split}
  \label{eq:aplate-clamped}
\end{equation}
where $h$ is again a suitable measure of the mesh cell size, $\beta_1$
and $\beta_2$ are positive constants, and $\bullet_n :=
(\nabla \bullet)\cdot n$, where $n$ is the outward pointing unit normal on a
facet, similarly for $\bullet_t$, only using the tangent to the
facet, obtained by a counter-clockwise rotation of $n$.
The first twenty-eight eigenmodes of the operator \cref{eq:aplate-clamped}
are shown \cref{fig:guitar-eigenmodes}.
Of course, this is a very crude approximation of the actual eigenmodes
supported by the top plate of a guitar.  To guide the design of an
instrument we would, at the very least, need to incorporate the
anisotropic properties of wood into the model, as well as thinking
about structural supports such as bass bars and ribbing.  Some of these modelling
issues are discussed, in the context of violins, in
\citet{Gough:2007}, and a simple multiphysics guitar model in
\citet{tapia2002simulating}.  Here, however, we leave more involved
simulations as future work.

\section{Conclusions and future work}
The key idea of FInAT, that an element library can provide structure
and not just numerical values, has made significant progress possible
within the Firedrake code stack.  While earlier work on FInAT focused
on vector and tensor product structure, our current work has enabled a
wide range of useful finite elements previously inaccessible to
automated code generation systems.  Our numerical results indicate
that these elements are indeed viable for many interesting problems.
Automation now makes Argyris and other ``difficult'' finite elements of
comparable cost to deploy as any Lagrange elements.

In the future, we hope to extend this work in many directions.  In
particular, we need to extend the transformations and hence code
infrastructure to support embedded manifolds~\citep{rognes2013automating}.  Also, we
believe that the extensions we have developed in FInAT will provide a
starting point to consider macro-element techniques for splines and
other more complex elements~\citep{LaiSch07}.

\section*{Code availability}

For reproducibility, we cite archives of the exact software versions
used to produce the results in this paper.  All major Firedrake
components have been archived on
Zenodo~\citep{zenodo/Firedrake-20190830.0}.  This record collates DOIs for
the components, as well as containing the data and scripts used to
produce the results in this paper.  An installation of Firedrake with
components matching those used to produce the results can be obtained
following the instructions at
\url{www.firedrakeproject.org/download.html} with
\begin{verbatim}
export PETSC_CONFIGURE_OPTIONS="--download-pastix --download-ptscotch"
python3 firedrake-install --doi 10.5281/zenodo.3381901 --slepc --install \
  git+https://github.com/thomasgibson/scpc.git@3a1173ebb3610dcdf5090088294a542274bbb73a
\end{verbatim}
A \texttt{README} file in the archived record contains more detailed
information.

\begin{acks}
  This work was supported by the \grantsponsor{EPSRC}{Engineering and
    Physical Sciences Research Council}{https://www.epsrc.ac.uk/}
  [grant number \grantnum{EPSRC}{EP/L000407/1}]; and the
  \grantsponsor{NSF}{National Science
    Foundation}{https://www.nsf.gov}, award number
  \grantnum{NSF}{1525697}.
  \Cref{fig:els,fig:affmap,fig:hermrefandphys,fig:morleypushforward,fig:morleybridge}
  are adapted from \citet{kirby-zany}, licensed under \href{https://creativecommons.org/licenses/by-nc-nd/4.0/}{CC-BY-NC-ND}.
\end{acks}
\appendix
\bibliographystyle{ACM-Reference-Format}
\ifarxiv
\input{paper.bbl}

\else
\bibliography{references,zenodo}
\fi
\end{document}

%% file: code/pictures/tikzmesh.tex
\begin{tikzpicture}[scale=3.0]
\draw (0.000000e+00, 0.000000e+00) -- (0.000000e+00, 1.250000e-01) -- (1.250000e-01, 0.000000e+00) -- cycle;
\draw (0.000000e+00, 1.250000e-01) -- (1.250000e-01, 0.000000e+00) -- (1.367188e-01, 1.132812e-01) -- cycle;
\draw (1.250000e-01, 0.000000e+00) -- (1.367188e-01, 1.132812e-01) -- (2.500000e-01, 0.000000e+00) -- cycle;
\draw (0.000000e+00, 1.250000e-01) -- (1.367188e-01, 1.132812e-01) -- (0.000000e+00, 2.500000e-01) -- cycle;
\draw (1.367188e-01, 1.132812e-01) -- (2.500000e-01, 0.000000e+00) -- (2.665728e-01, 1.084272e-01) -- cycle;
\draw (1.367188e-01, 1.132812e-01) -- (0.000000e+00, 2.500000e-01) -- (1.415728e-01, 2.334272e-01) -- cycle;
\draw (2.500000e-01, 0.000000e+00) -- (2.665728e-01, 1.084272e-01) -- (3.750000e-01, 0.000000e+00) -- cycle;
\draw (1.367188e-01, 1.132812e-01) -- (2.665728e-01, 1.084272e-01) -- (1.415728e-01, 2.334272e-01) -- cycle;
\draw (0.000000e+00, 2.500000e-01) -- (1.415728e-01, 2.334272e-01) -- (0.000000e+00, 3.750000e-01) -- cycle;
\draw (2.665728e-01, 1.084272e-01) -- (3.750000e-01, 0.000000e+00) -- (3.867188e-01, 1.132812e-01) -- cycle;
\draw (2.665728e-01, 1.084272e-01) -- (1.415728e-01, 2.334272e-01) -- (2.734375e-01, 2.265625e-01) -- cycle;
\draw (1.415728e-01, 2.334272e-01) -- (0.000000e+00, 3.750000e-01) -- (1.367188e-01, 3.632812e-01) -- cycle;
\draw (3.750000e-01, 0.000000e+00) -- (3.867188e-01, 1.132812e-01) -- (5.000000e-01, 0.000000e+00) -- cycle;
\draw (2.665728e-01, 1.084272e-01) -- (3.867188e-01, 1.132812e-01) -- (2.734375e-01, 2.265625e-01) -- cycle;
\draw (1.415728e-01, 2.334272e-01) -- (2.734375e-01, 2.265625e-01) -- (1.367188e-01, 3.632812e-01) -- cycle;
\draw (0.000000e+00, 3.750000e-01) -- (1.367188e-01, 3.632812e-01) -- (0.000000e+00, 5.000000e-01) -- cycle;
\draw (3.867188e-01, 1.132812e-01) -- (5.000000e-01, 0.000000e+00) -- (5.000000e-01, 1.250000e-01) -- cycle;
\draw (3.867188e-01, 1.132812e-01) -- (2.734375e-01, 2.265625e-01) -- (3.915728e-01, 2.334272e-01) -- cycle;
\draw (2.734375e-01, 2.265625e-01) -- (1.367188e-01, 3.632812e-01) -- (2.665728e-01, 3.584272e-01) -- cycle;
\draw (1.367188e-01, 3.632812e-01) -- (0.000000e+00, 5.000000e-01) -- (1.250000e-01, 5.000000e-01) -- cycle;
\draw (5.000000e-01, 0.000000e+00) -- (5.000000e-01, 1.250000e-01) -- (6.250000e-01, 0.000000e+00) -- cycle;
\draw (3.867188e-01, 1.132812e-01) -- (5.000000e-01, 1.250000e-01) -- (3.915728e-01, 2.334272e-01) -- cycle;
\draw (2.734375e-01, 2.265625e-01) -- (3.915728e-01, 2.334272e-01) -- (2.665728e-01, 3.584272e-01) -- cycle;
\draw (1.367188e-01, 3.632812e-01) -- (2.665728e-01, 3.584272e-01) -- (1.250000e-01, 5.000000e-01) -- cycle;
\draw (0.000000e+00, 5.000000e-01) -- (1.250000e-01, 5.000000e-01) -- (0.000000e+00, 6.250000e-01) -- cycle;
\draw (5.000000e-01, 1.250000e-01) -- (6.250000e-01, 0.000000e+00) -- (6.132812e-01, 1.367188e-01) -- cycle;
\draw (5.000000e-01, 1.250000e-01) -- (3.915728e-01, 2.334272e-01) -- (5.000000e-01, 2.500000e-01) -- cycle;
\draw (3.915728e-01, 2.334272e-01) -- (2.665728e-01, 3.584272e-01) -- (3.867188e-01, 3.632812e-01) -- cycle;
\draw (2.665728e-01, 3.584272e-01) -- (1.250000e-01, 5.000000e-01) -- (2.500000e-01, 5.000000e-01) -- cycle;
\draw (1.250000e-01, 5.000000e-01) -- (0.000000e+00, 6.250000e-01) -- (1.132812e-01, 6.367188e-01) -- cycle;
\draw (6.250000e-01, 0.000000e+00) -- (6.132812e-01, 1.367188e-01) -- (7.500000e-01, 0.000000e+00) -- cycle;
\draw (5.000000e-01, 1.250000e-01) -- (6.132812e-01, 1.367188e-01) -- (5.000000e-01, 2.500000e-01) -- cycle;
\draw (3.915728e-01, 2.334272e-01) -- (5.000000e-01, 2.500000e-01) -- (3.867188e-01, 3.632812e-01) -- cycle;
\draw (2.665728e-01, 3.584272e-01) -- (3.867188e-01, 3.632812e-01) -- (2.500000e-01, 5.000000e-01) -- cycle;
\draw (1.250000e-01, 5.000000e-01) -- (2.500000e-01, 5.000000e-01) -- (1.132812e-01, 6.367188e-01) -- cycle;
\draw (0.000000e+00, 6.250000e-01) -- (1.132812e-01, 6.367188e-01) -- (0.000000e+00, 7.500000e-01) -- cycle;
\draw (6.132812e-01, 1.367188e-01) -- (7.500000e-01, 0.000000e+00) -- (7.334272e-01, 1.415728e-01) -- cycle;
\draw (6.132812e-01, 1.367188e-01) -- (5.000000e-01, 2.500000e-01) -- (6.084272e-01, 2.665728e-01) -- cycle;
\draw (5.000000e-01, 2.500000e-01) -- (3.867188e-01, 3.632812e-01) -- (5.000000e-01, 3.750000e-01) -- cycle;
\draw (3.867188e-01, 3.632812e-01) -- (2.500000e-01, 5.000000e-01) -- (3.750000e-01, 5.000000e-01) -- cycle;
\draw (2.500000e-01, 5.000000e-01) -- (1.132812e-01, 6.367188e-01) -- (2.334272e-01, 6.415728e-01) -- cycle;
\draw (1.132812e-01, 6.367188e-01) -- (0.000000e+00, 7.500000e-01) -- (1.084272e-01, 7.665728e-01) -- cycle;
\draw (7.500000e-01, 0.000000e+00) -- (7.334272e-01, 1.415728e-01) -- (8.750000e-01, 0.000000e+00) -- cycle;
\draw (6.132812e-01, 1.367188e-01) -- (7.334272e-01, 1.415728e-01) -- (6.084272e-01, 2.665728e-01) -- cycle;
\draw (5.000000e-01, 2.500000e-01) -- (6.084272e-01, 2.665728e-01) -- (5.000000e-01, 3.750000e-01) -- cycle;
\draw (3.867188e-01, 3.632812e-01) -- (5.000000e-01, 3.750000e-01) -- (3.750000e-01, 5.000000e-01) -- cycle;
\draw (2.500000e-01, 5.000000e-01) -- (3.750000e-01, 5.000000e-01) -- (2.334272e-01, 6.415728e-01) -- cycle;
\draw (1.132812e-01, 6.367188e-01) -- (2.334272e-01, 6.415728e-01) -- (1.084272e-01, 7.665728e-01) -- cycle;
\draw (0.000000e+00, 7.500000e-01) -- (1.084272e-01, 7.665728e-01) -- (0.000000e+00, 8.750000e-01) -- cycle;
\draw (7.334272e-01, 1.415728e-01) -- (8.750000e-01, 0.000000e+00) -- (8.632812e-01, 1.367188e-01) -- cycle;
\draw (7.334272e-01, 1.415728e-01) -- (6.084272e-01, 2.665728e-01) -- (7.265625e-01, 2.734375e-01) -- cycle;
\draw (6.084272e-01, 2.665728e-01) -- (5.000000e-01, 3.750000e-01) -- (6.132812e-01, 3.867188e-01) -- cycle;
\draw (5.000000e-01, 3.750000e-01) -- (3.750000e-01, 5.000000e-01) -- (5.000000e-01, 5.000000e-01) -- cycle;
\draw (3.750000e-01, 5.000000e-01) -- (2.334272e-01, 6.415728e-01) -- (3.632812e-01, 6.367188e-01) -- cycle;
\draw (2.334272e-01, 6.415728e-01) -- (1.084272e-01, 7.665728e-01) -- (2.265625e-01, 7.734375e-01) -- cycle;
\draw (1.084272e-01, 7.665728e-01) -- (0.000000e+00, 8.750000e-01) -- (1.132812e-01, 8.867188e-01) -- cycle;
\draw (8.750000e-01, 0.000000e+00) -- (8.632812e-01, 1.367188e-01) -- (1.000000e+00, 0.000000e+00) -- cycle;
\draw (7.334272e-01, 1.415728e-01) -- (8.632812e-01, 1.367188e-01) -- (7.265625e-01, 2.734375e-01) -- cycle;
\draw (6.084272e-01, 2.665728e-01) -- (7.265625e-01, 2.734375e-01) -- (6.132812e-01, 3.867188e-01) -- cycle;
\draw (5.000000e-01, 3.750000e-01) -- (6.132812e-01, 3.867188e-01) -- (5.000000e-01, 5.000000e-01) -- cycle;
\draw (3.750000e-01, 5.000000e-01) -- (5.000000e-01, 5.000000e-01) -- (3.632812e-01, 6.367188e-01) -- cycle;
\draw (2.334272e-01, 6.415728e-01) -- (3.632812e-01, 6.367188e-01) -- (2.265625e-01, 7.734375e-01) -- cycle;
\draw (1.084272e-01, 7.665728e-01) -- (2.265625e-01, 7.734375e-01) -- (1.132812e-01, 8.867188e-01) -- cycle;
\draw (0.000000e+00, 8.750000e-01) -- (1.132812e-01, 8.867188e-01) -- (0.000000e+00, 1.000000e+00) -- cycle;
\draw (8.632812e-01, 1.367188e-01) -- (1.000000e+00, 0.000000e+00) -- (1.000000e+00, 1.250000e-01) -- cycle;
\draw (8.632812e-01, 1.367188e-01) -- (7.265625e-01, 2.734375e-01) -- (8.584272e-01, 2.665728e-01) -- cycle;
\draw (7.265625e-01, 2.734375e-01) -- (6.132812e-01, 3.867188e-01) -- (7.334272e-01, 3.915728e-01) -- cycle;
\draw (6.132812e-01, 3.867188e-01) -- (5.000000e-01, 5.000000e-01) -- (6.250000e-01, 5.000000e-01) -- cycle;
\draw (5.000000e-01, 5.000000e-01) -- (3.632812e-01, 6.367188e-01) -- (5.000000e-01, 6.250000e-01) -- cycle;
\draw (3.632812e-01, 6.367188e-01) -- (2.265625e-01, 7.734375e-01) -- (3.584272e-01, 7.665728e-01) -- cycle;
\draw (2.265625e-01, 7.734375e-01) -- (1.132812e-01, 8.867188e-01) -- (2.334272e-01, 8.915728e-01) -- cycle;
\draw (1.132812e-01, 8.867188e-01) -- (0.000000e+00, 1.000000e+00) -- (1.250000e-01, 1.000000e+00) -- cycle;
\draw (8.632812e-01, 1.367188e-01) -- (1.000000e+00, 1.250000e-01) -- (8.584272e-01, 2.665728e-01) -- cycle;
\draw (7.265625e-01, 2.734375e-01) -- (8.584272e-01, 2.665728e-01) -- (7.334272e-01, 3.915728e-01) -- cycle;
\draw (6.132812e-01, 3.867188e-01) -- (7.334272e-01, 3.915728e-01) -- (6.250000e-01, 5.000000e-01) -- cycle;
\draw (5.000000e-01, 5.000000e-01) -- (6.250000e-01, 5.000000e-01) -- (5.000000e-01, 6.250000e-01) -- cycle;
\draw (3.632812e-01, 6.367188e-01) -- (5.000000e-01, 6.250000e-01) -- (3.584272e-01, 7.665728e-01) -- cycle;
\draw (2.265625e-01, 7.734375e-01) -- (3.584272e-01, 7.665728e-01) -- (2.334272e-01, 8.915728e-01) -- cycle;
\draw (1.132812e-01, 8.867188e-01) -- (2.334272e-01, 8.915728e-01) -- (1.250000e-01, 1.000000e+00) -- cycle;
\draw (1.000000e+00, 1.250000e-01) -- (8.584272e-01, 2.665728e-01) -- (1.000000e+00, 2.500000e-01) -- cycle;
\draw (8.584272e-01, 2.665728e-01) -- (7.334272e-01, 3.915728e-01) -- (8.632812e-01, 3.867188e-01) -- cycle;
\draw (7.334272e-01, 3.915728e-01) -- (6.250000e-01, 5.000000e-01) -- (7.500000e-01, 5.000000e-01) -- cycle;
\draw (6.250000e-01, 5.000000e-01) -- (5.000000e-01, 6.250000e-01) -- (6.367188e-01, 6.132812e-01) -- cycle;
\draw (5.000000e-01, 6.250000e-01) -- (3.584272e-01, 7.665728e-01) -- (5.000000e-01, 7.500000e-01) -- cycle;
\draw (3.584272e-01, 7.665728e-01) -- (2.334272e-01, 8.915728e-01) -- (3.632812e-01, 8.867188e-01) -- cycle;
\draw (2.334272e-01, 8.915728e-01) -- (1.250000e-01, 1.000000e+00) -- (2.500000e-01, 1.000000e+00) -- cycle;
\draw (8.584272e-01, 2.665728e-01) -- (1.000000e+00, 2.500000e-01) -- (8.632812e-01, 3.867188e-01) -- cycle;
\draw (7.334272e-01, 3.915728e-01) -- (8.632812e-01, 3.867188e-01) -- (7.500000e-01, 5.000000e-01) -- cycle;
\draw (6.250000e-01, 5.000000e-01) -- (7.500000e-01, 5.000000e-01) -- (6.367188e-01, 6.132812e-01) -- cycle;
\draw (5.000000e-01, 6.250000e-01) -- (6.367188e-01, 6.132812e-01) -- (5.000000e-01, 7.500000e-01) -- cycle;
\draw (3.584272e-01, 7.665728e-01) -- (5.000000e-01, 7.500000e-01) -- (3.632812e-01, 8.867188e-01) -- cycle;
\draw (2.334272e-01, 8.915728e-01) -- (3.632812e-01, 8.867188e-01) -- (2.500000e-01, 1.000000e+00) -- cycle;
\draw (1.000000e+00, 2.500000e-01) -- (8.632812e-01, 3.867188e-01) -- (1.000000e+00, 3.750000e-01) -- cycle;
\draw (8.632812e-01, 3.867188e-01) -- (7.500000e-01, 5.000000e-01) -- (8.750000e-01, 5.000000e-01) -- cycle;
\draw (7.500000e-01, 5.000000e-01) -- (6.367188e-01, 6.132812e-01) -- (7.665728e-01, 6.084272e-01) -- cycle;
\draw (6.367188e-01, 6.132812e-01) -- (5.000000e-01, 7.500000e-01) -- (6.415728e-01, 7.334272e-01) -- cycle;
\draw (5.000000e-01, 7.500000e-01) -- (3.632812e-01, 8.867188e-01) -- (5.000000e-01, 8.750000e-01) -- cycle;
\draw (3.632812e-01, 8.867188e-01) -- (2.500000e-01, 1.000000e+00) -- (3.750000e-01, 1.000000e+00) -- cycle;
\draw (8.632812e-01, 3.867188e-01) -- (1.000000e+00, 3.750000e-01) -- (8.750000e-01, 5.000000e-01) -- cycle;
\draw (7.500000e-01, 5.000000e-01) -- (8.750000e-01, 5.000000e-01) -- (7.665728e-01, 6.084272e-01) -- cycle;
\draw (6.367188e-01, 6.132812e-01) -- (7.665728e-01, 6.084272e-01) -- (6.415728e-01, 7.334272e-01) -- cycle;
\draw (5.000000e-01, 7.500000e-01) -- (6.415728e-01, 7.334272e-01) -- (5.000000e-01, 8.750000e-01) -- cycle;
\draw (3.632812e-01, 8.867188e-01) -- (5.000000e-01, 8.750000e-01) -- (3.750000e-01, 1.000000e+00) -- cycle;
\draw (1.000000e+00, 3.750000e-01) -- (8.750000e-01, 5.000000e-01) -- (1.000000e+00, 5.000000e-01) -- cycle;
\draw (8.750000e-01, 5.000000e-01) -- (7.665728e-01, 6.084272e-01) -- (8.867188e-01, 6.132812e-01) -- cycle;
\draw (7.665728e-01, 6.084272e-01) -- (6.415728e-01, 7.334272e-01) -- (7.734375e-01, 7.265625e-01) -- cycle;
\draw (6.415728e-01, 7.334272e-01) -- (5.000000e-01, 8.750000e-01) -- (6.367188e-01, 8.632812e-01) -- cycle;
\draw (5.000000e-01, 8.750000e-01) -- (3.750000e-01, 1.000000e+00) -- (5.000000e-01, 1.000000e+00) -- cycle;
\draw (8.750000e-01, 5.000000e-01) -- (1.000000e+00, 5.000000e-01) -- (8.867188e-01, 6.132812e-01) -- cycle;
\draw (7.665728e-01, 6.084272e-01) -- (8.867188e-01, 6.132812e-01) -- (7.734375e-01, 7.265625e-01) -- cycle;
\draw (6.415728e-01, 7.334272e-01) -- (7.734375e-01, 7.265625e-01) -- (6.367188e-01, 8.632812e-01) -- cycle;
\draw (5.000000e-01, 8.750000e-01) -- (6.367188e-01, 8.632812e-01) -- (5.000000e-01, 1.000000e+00) -- cycle;
\draw (1.000000e+00, 5.000000e-01) -- (8.867188e-01, 6.132812e-01) -- (1.000000e+00, 6.250000e-01) -- cycle;
\draw (8.867188e-01, 6.132812e-01) -- (7.734375e-01, 7.265625e-01) -- (8.915728e-01, 7.334272e-01) -- cycle;
\draw (7.734375e-01, 7.265625e-01) -- (6.367188e-01, 8.632812e-01) -- (7.665728e-01, 8.584272e-01) -- cycle;
\draw (6.367188e-01, 8.632812e-01) -- (5.000000e-01, 1.000000e+00) -- (6.250000e-01, 1.000000e+00) -- cycle;
\draw (8.867188e-01, 6.132812e-01) -- (1.000000e+00, 6.250000e-01) -- (8.915728e-01, 7.334272e-01) -- cycle;
\draw (7.734375e-01, 7.265625e-01) -- (8.915728e-01, 7.334272e-01) -- (7.665728e-01, 8.584272e-01) -- cycle;
\draw (6.367188e-01, 8.632812e-01) -- (7.665728e-01, 8.584272e-01) -- (6.250000e-01, 1.000000e+00) -- cycle;
\draw (1.000000e+00, 6.250000e-01) -- (8.915728e-01, 7.334272e-01) -- (1.000000e+00, 7.500000e-01) -- cycle;
\draw (8.915728e-01, 7.334272e-01) -- (7.665728e-01, 8.584272e-01) -- (8.867188e-01, 8.632812e-01) -- cycle;
\draw (7.665728e-01, 8.584272e-01) -- (6.250000e-01, 1.000000e+00) -- (7.500000e-01, 1.000000e+00) -- cycle;
\draw (8.915728e-01, 7.334272e-01) -- (1.000000e+00, 7.500000e-01) -- (8.867188e-01, 8.632812e-01) -- cycle;
\draw (7.665728e-01, 8.584272e-01) -- (8.867188e-01, 8.632812e-01) -- (7.500000e-01, 1.000000e+00) -- cycle;
\draw (1.000000e+00, 7.500000e-01) -- (8.867188e-01, 8.632812e-01) -- (1.000000e+00, 8.750000e-01) -- cycle;
\draw (8.867188e-01, 8.632812e-01) -- (7.500000e-01, 1.000000e+00) -- (8.750000e-01, 1.000000e+00) -- cycle;
\draw (8.867188e-01, 8.632812e-01) -- (1.000000e+00, 8.750000e-01) -- (8.750000e-01, 1.000000e+00) -- cycle;
\draw (1.000000e+00, 8.750000e-01) -- (8.750000e-01, 1.000000e+00) -- (1.000000e+00, 1.000000e+00) -- cycle;
\end{tikzpicture}

%% file: paper.bbl
%%% -*-BibTeX-*-
%%% Do NOT edit. File created by BibTeX with style
%%% ACM-Reference-Format-Journals [18-Jan-2012].